\documentclass[aps,prb,twocolumn,10pt,superscriptaddress,showpacs]{revtex4-2}
\usepackage{amsmath,amssymb,graphics,epsfig,epstopdf,color,verbatim,ulem,braket,tabularx,multirow,array,diagbox,colortbl,hhline}
\usepackage[colorlinks,linkcolor=blue,citecolor=blue,urlcolor=blue]{hyperref}
\usepackage{physics}
\usepackage{bm}
\usepackage{blindtext}
\usepackage{booktabs}
\usepackage{csquotes}
\usepackage{slashed}
\usepackage{siunitx}
\graphicspath{{Figures/}}

\begin{document}


\title{Finite-temperature quantum Monte Carlo study of Dirac quantum criticality: Algorithmic advances and comprehensive analysis}

\author{Hou-Min Du}
\thanks{These authors contributed equally to this work.}
\affiliation{Institute of Modern Physics, Northwest University, Xi'an 710127, China}
\affiliation{Hefei National Laboratory, Hefei 230088, China}

\author{Qian Sun}
\thanks{These authors contributed equally to this work.}
\affiliation{School of Physics, Northwest University, Xi'an 710127, China}

\author{Yuan-Yao He}
\email{heyuanyao@nwu.edu.cn}
\affiliation{Institute of Modern Physics, Northwest University, Xi'an 710127, China}
\affiliation{Shaanxi Key Laboratory for Theoretical Physics Frontiers, Xi'an 710127, China}
\affiliation{Fundamental Discipline Research Center for Quantum Science and Technology of Shaanxi Province, Xi'an 710127, China}
\affiliation{Hefei National Laboratory, Hefei 230088, China}

\begin{abstract}
We present a systematic study of the quantum phase transition from a Dirac semimetal to an antiferromagnetic Mott insulator in the honeycomb-lattice Hubbard model, using finite-temperature auxiliary-field quantum Monte Carlo (AFQMC) simulations. Given the dynamical critical exponent $z=1$ at criticality, we scale the inverse temperature $\beta$ with the linear system size $L$, i.e., $\beta t=L$ (with $t$ as the hopping amplitude), to approach the thermodynamic and zero-temperature limits simultaneously. On the algorithmic front, we generalize the fast Fourier transform in path propagation to multi-sublattice systems and employ the delayed update technique in configuration updates to accelerate the AFQMC simulations. These advances enable us to access unprecedented system sizes of up to $3528$ sites (with $L=42$ and $\beta t = 42$). We additionally use twist-averaged boundary conditions to reduce the finite-size effects in computed observables. On the physical side, we determine the Dirac quantum criticality via finite-size scaling analysis of the spin correlations and off-diagonal single-particle Green's function, and locate the critical point at $U_c/t=3.731(4)$ with the critical exponents $\nu=1.097(7)$, $\eta_{\phi}=0.72(2)$ and $\eta_{\psi}=0.155(6)$. We further corroborate the critical point by examining several relevant quantities, including the quasiparticle weight, Fermi-liquid parameter, charge compressibility, $U$-derivative of double occupancy, and fidelity susceptibility. Our work establishes an independent benchmark for the chiral Heisenberg Gross-Neveu-Yukawa criticality in the honeycomb Hubbard model, offering a complementary perspective to previous ground-state simulations.
\end{abstract}

\date{\today}

\maketitle

\section{Introduction}
\label{sec:intro}

The study of quantum phase transitions in correlated electron systems has long been a central focus of condensed matter physics~\cite{Sondhi1997,Matthias2003,Sachdev2011,Sachdev2018,Sachdev2019,Castro2009,Kotov2012}, intertwining physics of band topology, strong correlations, and quantum criticality. A paradigmatic example is the interaction-driven transition from a Dirac semimetal (DSM) to an antiferromagnetic Mott insulator (AFMI) on the honeycomb lattice~\cite{Herbut2006,Herbut2009,Janssen2014,Tang2018}, where massless Dirac fermions acquire a gap via spontaneous symmetry breaking. This transition, which can be described by the half-filled Hubbard model~\cite{Sorella1992,Paiva2005,Meng2010,Sorella2012,Assaad2013,Otsuka2016,Lang2025,Wang2026}, has attracted sustained theoretical attention. It is not only a prototypical setting for investigating correlation-induced gap opening in a relativistic kinetic dispersion, but also a potential platform for realizing emergent quantum critical phenomena in two spatial dimensions.

Precision theoretical studies of the honeycomb-lattice Hubbard model at half filling rely predominantly on various quantum many-body numerical approaches~\cite{Wuwei2010,Ansgar2011,Rong2012,Wuwei2013,QingXiao2015,Marcin2020,Janssen2014,Knorr2018,Buividovich2018,Buividovich2019,Ostmeyer2020,Sorella1992,Paiva2005,Meng2010,Sorella2012,Assaad2013,Otsuka2016,Lang2025,Wang2026,Toldin2015,Yu2026}. Among these, the unbiased auxiliary-field quantum Monte Carlo (AFQMC) method plays a central role~\cite{Sorella1992,Paiva2005,Meng2010,Sorella2012,Assaad2013,Otsuka2016,Lang2025,Wang2026,Toldin2015,Yu2026}, since it is free of the fermion sign problem for this model and thereby enables large-scale simulations. Early-stage AFQMC studies, limited to relatively small systems~\cite{Sorella1992,Paiva2005}, identified a single quantum phase transition between DSM and AFMI at an interaction strength $U/t=4\sim5$. In 2010, Meng {\it et al.}~\cite{Meng2010} instead reported the possible existence of an intermediate quantum spin-liquid phase for $U/t=3.5\sim4.3$, based on simulations up to linear system size $L=18$. This scenario was subsequently ruled out by larger-scale AFQMC simulations reaching $L=36$~\cite{Sorella2012} and by AFQMC calculations using a pinning-order technique~\cite{Assaad2013}, thus establishing a direct continuous DSM-AFMI transition. Theoretically, the low-energy physics of the honeycomb Hubbard model maps onto a relativistic Gross-Neveu model of quantum field theory~\cite{Gross1974,Herbut2006,Herbut2009}, which places the DSM-AFMI transition in the $(2+1)$D chiral Heisenberg Gross-Neveu-Yukawa universality class with $N=8$ Dirac fermion components~\cite{Zerf2017,Gracey2018,Otsuka2020,Ladovrechis2023,Tolosa2025}. Subsequent AFQMC studies~\cite{Toldin2015,Otsuka2016,Lang2025} focused primarily on the location of the DSM-AFMI quantum critical point (QCP) and its critical exponents. Most recently, the Dirac quantum criticality has been further examined employing improved AFQMC techniques, including simulations on systems up to $L=72$~\cite{Wang2026} and via the nonequilibrium dynamics in imaginary time~\cite{Yu2026}. Despite broad agreement on the universality class, significant discrepancies persist in the reported QCP location and critical exponents~\cite{Toldin2015,Otsuka2016,Lang2025,Wang2026}. For instance, the estimated values of the QCP $U_c/t$ and the bosonic anomalous dimension $\eta_{\phi}$ differ notably, from $[3.85(2),0.49(3)]$~\cite{Otsuka2016} to $[3.664(5),0.79(2)]$~\cite{Wang2026}. While these discrepancies are often attributed solely to finite-size effects, the roles played by numerical methodology and the underlying critical physics have not been systematically disentangled.

From a numerical perspective, all prior AFQMC studies of the Dirac quantum criticality discussed above employed the zero-temperature AFQMC algorithm~\cite{White1989,Assaad2008,Sugiyama1986,Sorella1989,Duhao2025}. However, this algorithm is known to suffer from the infinite variance problem~\cite{Shihao2016Inf,Wan2025} even in sign-problem-free simulations, which renders statistical error bars unreliable and can result in biased finite-sampling estimates of physical observables. This issue may also contribute to the quantitative discrepancies among previous estimates of the critical properties~\cite{Toldin2015,Otsuka2016,Lang2025,Wang2026}. Furthermore, the algorithm has additional limitations when applied to the honeycomb Hubbard model. First, it obtains the ground state $|\Psi_g\rangle$ via the imaginary-time projection from a trial wave function $|\psi_T\rangle$, i.e., $|\Psi_g\rangle=\lim_{\Theta\to\infty}e^{-\Theta\hat{H}}|\psi_T\rangle$, with $\Theta$ as the projection length. To ensure the convergence, $\Theta$ must scale inversely with the finite-size gap~\cite{Assaad2008}, which, near the QCP of the honeycomb Hubbard model, implies $\Theta\propto L$~\cite{Otsuka2016,Wang2026}. In practice, performing systematic convergence tests for large system sizes is computationally demanding, often necessitating an empirical choice of $\Theta$ that may leave residual projection errors in computed observables. This issue is particularly relevant because the largest system sizes typically play the dominant role in determining the criticality. Second, the trial wave function $|\psi_T\rangle$ is commonly chosen as the ground state of the noninteracting Hamiltonian, supplemented by a tiny perturbation, such as an infinitesimal twisted boundary phase~\cite{Assaad2008,Meng2010} or small random hoppings~\cite{Wang2026}, to lift the energy level degeneracy at the Dirac points. Such a perturbation necessarily breaks the $C_6$ rotational symmetry. An alternative choice of $|\psi_T\rangle$ is the mean-field ground state~\cite{Otsuka2016}, which breaks the spin SU(2) symmetry. These symmetry-breaking effects may influence the imaginary-time projection and auxiliary-field sampling, potentially affecting the quantitative accuracy of the simulation.

In this work, we revisit the Dirac quantum criticality in the honeycomb Hubbard model by instead employing the finite-temperature AFQMC algorithm~\cite{White1989,Assaad2008,Blankenbecler1981,Hirsch1983,White1989,Scalettar1991,He2019B,He2019L,Sun2024,Song2025L,*Song2025B,Yuanyao2026}, to avoid the issues of projection and trial wave function associated with the zero-temperature algorithm. Exploiting the dynamical critical exponent $z=1$~\cite{Herbut2001} at criticality, we simultaneously approach the thermodynamic and zero-temperature limits by setting the inverse temperature $\beta$ according to $\beta t=L$ in the calculations. By implementing efficient algorithmic improvements and twist-averaged boundary conditions (TABC), our AFQMC simulations access large systems up to $L=42$ and enable a high-precision determination of the criticality. We estimate the QCP and critical exponents via finite-size scaling analyses of spin-spin correlations and fermionic Green's function, obtaining results that are broadly consistent with prior studies. Furthermore, we corroborate the extracted QCP location by examining several independent physical quantities, providing further insights into finite-size effects and associated critical behaviors.

The remainder of this paper is organized as follows. In Sec.~\ref{sec:modelmethod}, we introduce the honeycomb Hubbard model, review the finite-temperature AFQMC algorithm, and define the physical quantities we compute in this work. In Sec.~\ref{sec:NewMethod}, we present the algorithmic improvements and describe the application of TABC in the AFQMC simulations. In Sec.~\ref{sec:HalfResults}, we focus on the numerical results, including the determination of the criticality and analyses of several complementary physical quantities. Finally, in Sec.~\ref{sec:Summary}, we summarize our findings and discuss possible directions for future research. The Appendixes provide additional results on the speedup by FFT, the comparison between PBC and TABC, and benchmark data of the total energy and double occupancy. 

\begin{figure}[t]
\includegraphics[width=0.780\columnwidth]{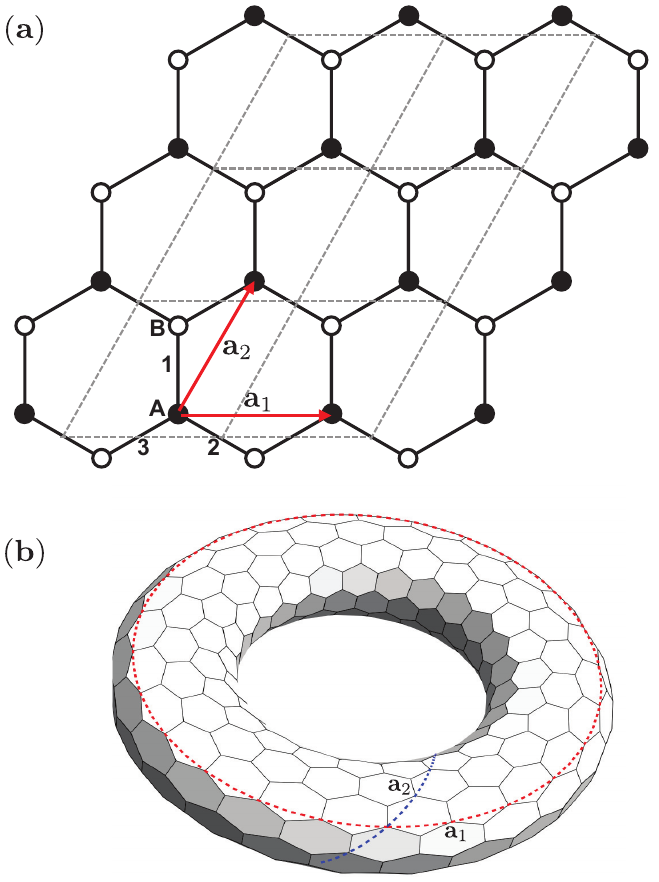}
\caption{
(a) Illustration of the honeycomb lattice with primitive lattice vectors $\mathbf{a}_1=(\sqrt{3},0)$ and $\mathbf{a}_2=(\sqrt{3}/2,3/2)$, where the lattice constant is set to unity. (b) Schematic illustration of the honeycomb lattice topologically wrapped onto a torus, where $\mathbf{a}_1$ and $\mathbf{a}_2$ directions are represented by the red and blue dashed loops, respectively.
}
\label{fig:fig01honeycomb}
\end{figure}

\section{Model, Method, and physical observables}
\label{sec:modelmethod}

\subsection{The half-filled honeycomb Hubbard model}
\label{sec:TheModel}

We study the honeycomb-lattice Hubbard model at half filling described by the following Hamiltonian
\begin{equation}\begin{aligned}
\label{eq:HubbModel}
\hat{H}= -t \sum_{\langle ij\rangle\sigma} 
(c_{i\sigma}^+ c_{j\sigma}^{} + h.c.)
+ U \sum_i \big(
\hat{n}_{i\uparrow}\hat{n}_{i\downarrow}-\frac{\hat{n}_{i\uparrow}+\hat{n}_{i\downarrow}}{2}\big),
\end{aligned}\end{equation}
where $\hat{n}_{i\sigma}=c_{i\sigma}^+c_{i\sigma}^{}$ and $\hat{n}_i=\hat{n}_{i\uparrow}+\hat{n}_{i\downarrow}$ are the density operators, with $i,j$ as the indices of the lattice sites and $\sigma$ ($=\uparrow$ or $\downarrow$) denoting spin. The notation $\langle ij\rangle$ represents the nearest-neighbor lattice site pair, with $t$ as the corresponding hopping amplitude. Throughout this work, we set $t=1$ as the energy unit, and focus on the repulsive interaction ($U>0$). As shown in Fig.~\ref{fig:fig01honeycomb}(a), the honeycomb lattice is a bipartite lattice consisting of two sublattices $A$ and $B$, with primitive lattice vectors $\mathbf{a}_1=(\sqrt{3},0)$ and $\mathbf{a}_2=(\sqrt{3}/2,3/2)$ (with the lattice constant set to unity). Then the primitive lattice vectors in the reciprocal lattice are $\mathbf{b}_1=(2\pi/\sqrt{3},-2\pi/3)$ and $\mathbf{b}_2=(0,4\pi/3)$. The linear system size is denoted by $L$, such that the numbers of unit cells and lattice sites are $N_c=L^2$ and $N_s=2N_c=2L^2$, respectively. Then the momentum is accordingly defined as $\mathbf{k}=(l_1\mathbf{b}_1+l_2\mathbf{b}_2)/L$ ($l_1,l_2$ as integers). In Fig.~\ref{fig:fig01honeycomb}(b), we schematically plot the torus geometry of the honeycomb lattice with periodic boundary conditions (PBC), which provides the basis for introducing the twisted boundary conditions (TBC) in Sec.~\ref{sec:TABC}. Under PBC, the noninteracting part $\hat{H}_0$ in the model~(\ref{eq:HubbModel}) can be transformed into reciprocal space and expressed as
\begin{equation}\begin{aligned}
\hat{H}_0 = \sum_{\mathbf{k}\sigma}
\begin{pmatrix}
a_{\mathbf{k}\sigma}^+ &
b_{\mathbf{k}\sigma}^+
\end{pmatrix}
\mathbf{h}(\mathbf{k})
\begin{pmatrix}
a_{\mathbf{k}\sigma}^{}\\
b_{\mathbf{k}\sigma}^{}
\end{pmatrix},
\end{aligned}\end{equation}
where $a_{\mathbf{k}\sigma}^+$ ($a_{\mathbf{k}\sigma}^{}$) and $b_{\mathbf{k}\sigma}^+$ ($b_{\mathbf{k}\sigma}^{}$) are the operators for A and B sublattices, respectively, and $\mathbf{h}(\mathbf{k})$ is a $2\times 2$ momentum-resolved Hamiltonian matrix reading
\begin{equation}\begin{aligned}
\label{eq:h0kMat}
\mathbf{h}(\mathbf{k}) = 
\begin{pmatrix}
0 & f(\mathbf{k}) \\
[f(\mathbf{k})]^* & 0
\end{pmatrix},
\end{aligned}\end{equation}
with $f(\mathbf{k})=-t[1+e^{+{\rm i}\mathbf{k}\cdot(\mathbf{a}_1-\mathbf{a}_2)}+e^{-{\rm i}\mathbf{k}\cdot\mathbf{a}_2}]$. Via diagonalizing the $\mathbf{h}(\mathbf{k})$ matrix, we obtain the energy dispersion $\varepsilon_{\mathbf{k},\pm}=\pm|f(\mathbf{k})|$.

The model~(\ref{eq:HubbModel}) preserves particle-hole symmetry, which guarantees half filling and renders both zero- and finite-temperature AFQMC simulations free of the fermion sign problem~\cite{Congjun2005}. It undergoes a continuous quantum phase transition from a DSM to an AFMI with increasing $U/t$. In the noninteracting limit, we can express the Hamiltonian in Eq.~(\ref{eq:HubbModel}) as a continuum massless Dirac Hamiltonian at the vicinity of Dirac points as 
\begin{equation}\begin{aligned}
\hat{H}_{eff}=v_{\mathrm{F}}(\pm q_x\tau_x+q_y\tau_y),
\end{aligned}\end{equation}
where $v_{\mathrm{F}}$ is the Fermi velocity, $\tau_x$ and $\tau_y$ are Pauli matrices acting in the sublattice space, $q_x$ and $q_y$ are the momenta measured from the corresponding Dirac point, and the $\pm$ sign corresponds to the two valleys (or Dirac points). The Hamiltonian is identical for the two spin species. Upon including the local quartic interaction, the effective low-energy Dirac theory is described by the chiral Heisenberg Gross-Neveu model, whose Lagrangian takes the form $\mathcal{L}=i\bar{\psi}^{i}\slashed{\partial}\psi^{i}+\frac{g^2}{2}(\bar{\psi}^{i}\sigma^{a}\psi^{i})^2$~\cite{Gross1974,Herbut2006,Herbut2009}, with the flavor index $i$ running over the eight fermionic degrees of freedom associated with the spin, valley, and sublattice, and $\sigma^{a}$ (with $a=1,2,3$) representing the Pauli matrices operating on spin space. From this low-energy theory and previous studies~\cite{Toldin2015,Otsuka2016,Lang2025,Wang2026}, the DSM-AFMI transition in the model~(\ref{eq:HubbModel}) should belong to the (2+1)D chiral Heisenberg Gross-Neveu-Yukawa universality class with $N=8$ Dirac fermion components~\cite{Zerf2017,Gracey2018,Otsuka2020,Ladovrechis2023,Tolosa2025}.

\subsection{The formalism of AFQMC algorithm}
\label{sec:AFQMCmethod}

We employ the unbiased finite-temperature AFQMC (also known as Determinant Quantum Monte Carlo) algorithm~\cite{White1989,Assaad2008,Blankenbecler1981,Hirsch1983,White1989,Scalettar1991,He2019B,He2019L,Sun2024,Song2025L,*Song2025B,Yuanyao2026} to numerically solve the honeycomb Hubbard model in Eq.~(\ref{eq:HubbModel}). In this subsection, we briefly review the essential ingredients of the algorithm and establish the notation and formalism used in Sec.~\ref{sec:NewMethod}, where the algorithmic improvements implemented in this work are presented.

We first decompose the model Hamiltonian in Eq.~(\ref{eq:HubbModel}) as $\hat{H}=\hat{H}_0+\hat{H}_U$, where $\hat{H}_0$ and $\hat{H}_U$ denote the hopping and interaction terms, respectively. The AFQMC algorithm begins by discretizing the inverse temperature as $\beta = M\Delta\tau$, such that the partition function can be written as $Z={\rm Tr}(e^{-\beta\hat{H}})={\rm Tr}[(e^{-\Delta\tau\hat{H}})^M]$. The Trotter-Suzuki (TS) decomposition is then applied to separate the $\hat{H}_0$ and $\hat{H}_U$ terms in $e^{-\Delta\tau\hat{H}}$, such as the second-order formula $e^{-\Delta\tau\hat{H}}=e^{-\Delta\tau\hat{H}_U}e^{-\Delta\tau\hat{H}_0}+\mathcal{O}[(\Delta\tau)^2]$. Next, the Hubbard-Stratonovich (HS) transformation~\cite{Hirsch1983,Assaad1998,Shihao2013,WangDa2014,Xie2025} is used to decouple two-body interactions in $e^{-\Delta\tau\hat{H}_U}$ into free fermions coupled to auxiliary fields. There are two commonly used HS transformations for the Hubbard repulsion ($U>0$)~\cite{Hirsch1983,Xie2025}: one into the spin-$\hat{s}^z$ channel (denoted HS-$\hat{s}^z$) as
\begin{equation}\begin{aligned}
\label{eq:SzHS}
e^{-\Delta\tau U \big(\hat{n}_{i\uparrow} \hat{n}_{i\downarrow} - \frac{\hat{n}_{i\uparrow} + \hat{n}_{i\downarrow}}{2}\big) } = \frac{1}{2} \sum\limits_{x_i = \pm 1} e^{\gamma_{\mathrm{s}} x_i (\hat{n}_{i\uparrow} - \hat{n}_{i\downarrow})},
\end{aligned}\end{equation}
and the other into the charge channel (denoted HS-$\hat{n}$) as
\begin{equation}\begin{aligned}
\label{eq:DensityHS}
e^{-\Delta\tau U \big(\hat{n}_{i\uparrow} \hat{n}_{i\downarrow} - \frac{\hat{n}_{i\uparrow} + \hat{n}_{i\downarrow}}{2}\big) } = \frac{e^{\Delta\tau U/2}}{2} \sum\limits_{x_i = \pm 1} e^{\gamma_{\mathrm{c}} x_i (\hat{n}_{i\uparrow} + \hat{n}_{i\downarrow} - 1)},
\end{aligned}\end{equation}
where the coupling constants $\gamma_{\mathrm{s}}$ and $\gamma_{\mathrm{c}}$ read
\begin{equation}\begin{aligned}
\label{eq:coupling}
\gamma_{\mathrm{s}} &= \cosh^{-1}(e^{+\Delta\tau U/2}),\\
\gamma_{\mathrm{c}} &= {\rm i}\cos^{-1}(e^{-\Delta\tau U/2}).
\end{aligned}\end{equation}
Combining the TS decomposition and HS transformation, the $e^{-\Delta\tau\hat{H}}$ operator at the $\ell$-th imaginary-time slice (with $\ell=1,2,\cdots,M$) can be expressed as
\begin{equation}\begin{aligned}
\label{eq:ExpDtH}
e^{-\Delta\tau\hat{H}(\ell)} 
\simeq \sum_{\mathbf{x}_{\ell}}p(\mathbf{x}_{\ell})\hat{B}(\mathbf{x}_{\ell}),
\end{aligned}\end{equation}
where $\hat{B}(\mathbf{x}_{\ell})=\hat{B}_I(\mathbf{x}_{\ell})e^{-\Delta\tau\hat{H}_0}$ is a single-particle propagator with the auxiliary fields $\mathbf{x}_{\ell}=(x_{\ell,1},x_{\ell,2},\cdots,x_{\ell,N_s})$, and the ``$\simeq$'' symbol indicates Trotter error. For the HS-$\hat{s}^z$ transformation in Eq.~(\ref{eq:SzHS}), the $\hat{B}_I(\mathbf{x}_{\ell})$ operator takes the form
\begin{equation}\begin{aligned}
\label{eq:BIOpSz}
\hat{B}_I(\mathbf{x}_{\ell})
= {\rm exp}\Big[\gamma_{\rm s}\sum_{i=1}^{N_s} x_{\ell,i}(\hat{n}_{i\uparrow}-\hat{n}_{i\downarrow})\Big],
\end{aligned}\end{equation}
with $p(\mathbf{x}_{\ell})=1/2^{N_s}$. For the HS-$\hat{n}$ transformation in Eq.~(\ref{eq:DensityHS}), $\hat{B}_I(\mathbf{x}_{\ell})$ instead reads
\begin{equation}\begin{aligned}
\label{eq:BIOpnT}
\hat{B}_I(\mathbf{x}_{\ell})
= {\rm exp}\Big[\gamma_{\rm c}\sum_{i=1}^{N_s} x_{\ell,i}(\hat{n}_{i\uparrow}+\hat{n}_{i\downarrow})\Big],
\end{aligned}\end{equation}
with $p(\mathbf{x}_{\ell})=(e^{\Delta\tau UN_s/2}/2^{N_s})\times e^{-\gamma_{\rm c}\sum_{i}x_{\ell,i}}$. After above procedures, the trace in the partition function only contains single-particle propagators and thus can be calculated to yield $Z\simeq\sum_{\mathbf{X}}W(\mathbf{X})$ with the auxiliary-field configuration weight $W(\mathbf{X})$ given by~\cite{Assaad2008}
\begin{equation}\begin{aligned}
\label{eq:ConfgWeight}
W(\mathbf{X}) 
&= P(\mathbf{X})\times {\rm Tr}\big[\hat{B}(\mathbf{x}_M)\cdots\hat{B}(\mathbf{x}_{\ell})\cdots\hat{B}(\mathbf{x}_1)\big] \\
&= P(\mathbf{X})\prod_{\sigma=\uparrow,\downarrow}{\rm det}\big(\mathbf{1}_{N_s}+\mathbf{B}_{M}^{\sigma}\cdots\mathbf{B}_{\ell}^{\sigma}\cdots\mathbf{B}_{1}^{\sigma}\big),
\end{aligned}\end{equation}
where $\mathbf{X}=(\mathbf{x}_1,\mathbf{x}_2,\cdots,\mathbf{x}_M)$ is a complete auxiliary-field configuration (or a full path), and $P(\mathbf{X})=\prod_{\ell=1}^M p(\mathbf{x}_{\ell})$ is a probability density. In Eq.~(\ref{eq:ConfgWeight}), we have applied the spin-decoupled condition as indicated by the $\hat{H}_0$ term and HS transformations in Eqs.~(\ref{eq:SzHS}) and~(\ref{eq:DensityHS}). The propagator matrix $\mathbf{B}_{\ell}^{\sigma}$ is expressed as $\mathbf{B}_{\ell}^{\sigma}=e^{\mathbf{H}_I^{\sigma}(\mathbf{x}_{\ell})}e^{-\Delta\tau\mathbf{H}_0^{\sigma}}$, where $\mathbf{H}_0^{\sigma}$ is the hopping matrix defined from the noninteracting Hamiltonian as $\hat{H}_0=\sum_{ij,\sigma}(\mathbf{H}_0^{\sigma})_{ij}c_{i\sigma}^+c_{j\sigma}^{}$, and $\mathbf{H}_I^{\sigma}(\mathbf{x}_{\ell})$ is the interaction-related matrix in the $\hat{B}_I(\mathbf{x}_{\ell})$ operator formulated as $\hat{B}_I(\mathbf{x}_{\ell})=\exp\{\sum_{ij,\sigma}[\mathbf{H}_I^{\sigma}(\mathbf{x}_{\ell})]_{ij}c_{i\sigma}^+c_{j\sigma}^{}\}$. Here in this work, $\mathbf{B}_{\ell}^{\sigma}$, $\mathbf{H}_0^{\sigma}$, and $\mathbf{H}_I^{\sigma}(\mathbf{x})$ are $N_s\times N_s$ matrices. Moreover, comparing with Eqs.~(\ref{eq:BIOpSz}) and~(\ref{eq:BIOpnT}), $\mathbf{H}_I^{\sigma}(\mathbf{x})$ is a diagonal matrix for the Hubbard interaction.

We then proceed to compute physical observables using $\langle\hat{O}\rangle={\rm Tr}(\hat{O}e^{-\beta\hat{H}})/Z$, which based on above framework can be reformulated as
\begin{equation}\begin{aligned}
\label{eq:GSObsQMC}
\langle\hat{O}\rangle
= \frac{\sum_{\mathbf{X}}W(\mathbf{X})O(\mathbf{X})}{\sum_{\mathbf{X}^{\prime}}W(\mathbf{X}^{\prime})}
= \sum_{\mathbf{X}}\omega(\mathbf{X})O(\mathbf{X}),
\end{aligned}\end{equation}
where $\omega(\mathbf{X})=W(\mathbf{X})/\sum_{\mathbf{X}^{\prime}}W(\mathbf{X}^{\prime})$ denotes the normalized probability density adopted for the importance sampling, and $O(\mathbf{X})$ is the measurement of $\hat{O}$ given by
\begin{equation}\begin{aligned}
O(\mathbf{X})
= \frac{{\rm Tr}(\hat{B}_{M}\cdots\hat{B}_{\ell+1}\hat{O}\hat{B}_{\ell}\cdots\hat{B}_1)}{{\rm Tr}(\hat{B}_{M}\cdots\hat{B}_{\ell+1}\hat{B}_{\ell}\cdots\hat{B}_1)}.
\end{aligned}\end{equation}
Specifically, for $\hat{O}=c_{i\sigma}^+c_{j\sigma}^{}$, this equation yields the element of equal-time single-particle Green's function matrix defined as $\mathbf{G}^{\sigma}(\tau,\tau)=\{G_{ij}^{\sigma}=\langle c_{i\sigma}^{} c_{j\sigma}^+\rangle_{\tau}\}$ ($\tau=\ell\Delta\tau$), which possesses the following matrix formula~\cite{Assaad2008}
\begin{equation}\begin{aligned}
\label{eq:GrFMat}
\mathbf{G}^{\sigma}(\tau,\tau)
=(\mathbf{1}_{N_s}+\mathbf{R}^{\sigma}\mathbf{L}^{\sigma})^{-1},
\end{aligned}\end{equation}
with $\mathbf{L}^{\sigma}=\mathbf{B}_{M}^{\sigma}\mathbf{B}_{M-1}^{\sigma}\cdots\mathbf{B}_{\ell+1}^{\sigma}$ and $\mathbf{R}^{\sigma}=\mathbf{B}_{\ell}^{\sigma}\cdots\mathbf{B}_2^{\sigma}\mathbf{B}_1^{\sigma}$ as $N_s\times N_s$ propagation matrices, and $\mathbf{1}_{N_s}$ being the $N_s\times N_s$ identity matrix.

We further discuss several implementation details relevant to practical AFQMC simulations. {\it First}, the Trotter error induced by TS decomposition can be eliminated via extrapolating $\Delta\tau t\to0$. In practice, we perform a modest set of tests and choose the value of $\Delta\tau t$ at which computed observables converge within statistical errors. Formally, we use the second-order TS decomposition for its simplicity to demonstrate the algorithmic details. But we actually apply the third-order formula in the simulations, i.e., $e^{-\Delta\tau\hat{H}}=e^{-\frac{\Delta\tau}{2}\hat{H}_0}e^{-\Delta\tau\hat{H}_U}e^{-\frac{\Delta\tau}{2}\hat{H}_0}+\mathcal{O}[(\Delta\tau)^3]$, which ensures faster convergence with decreasing $\Delta\tau t$~\cite{Song2025B}. In this work, we adopt $\Delta\tau t=0.05$ for the involved interaction strength $3.50\le U/t\le 4.50$, for which we verify that the residual Trotter error is negligible. {\it Second}, the stable calculation of $\mathbf{G}^{\sigma}$ matrix in Eq.~(\ref{eq:GrFMat})~\cite{He2019B} requires periodically applying a numerical stabilization procedure to $\mathbf{L}^{\sigma}$ and $\mathbf{R}^{\sigma}$ matrices, for instance, via the QR algorithm. {\it Third}, the HS-$\hat{s}^z$ and HS-$\hat{n}$ transformations in Eqs.~(\ref{eq:SzHS}) and~(\ref{eq:DensityHS}) explicitly break the spin and charge SU(2) symmetries~\cite{Cornelia2020}, respectively, which typically introduces significant fluctuations in observables associated with the corresponding symmetry channel~\cite{Song2025B,Xie2025}. Therefore, to achieve high-precision results, we adopt HS-$\hat{n}$ to compute spin-related properties (such as spin-spin correlations), and instead use HS-$\hat{s}^z$ for the density-related quantities (such as $U$-derivative of double occupancy). {\it Fourth}, the absence of the sign problem for model~(\ref{eq:HubbModel}) is established by applying a partial particle-hole transformation, i.e., $c_{i\downarrow}^{}\to(-1)^ic_{i\downarrow}^+$, $c_{i\downarrow}^+\to(-1)^ic_{i\downarrow}^{}$. This leads to $\mathbf{H}_0^{\downarrow}=\mathbf{H}_0^{\uparrow}$ (both matrices are real) and $\mathbf{H}_I^{\downarrow}(\mathbf{x}_{\ell})=[\mathbf{H}_I^{\uparrow}(\mathbf{x}_{\ell})]^*$, which together yield the relation $\mathbf{B}_{\ell}^{\downarrow}=(\mathbf{B}_{\ell}^{\uparrow})^*$. It then follows that the spin-up and spin-down determinants in Eq.~(\ref{eq:ConfgWeight}) are complex conjugate, ensuring their product is real and nonnegative. In this situation, we only need to perform the simulation for the spin-up channel and compute $\mathbf{G}^{\uparrow}$ matrix, while $\mathbf{G}^{\downarrow}$ can be readily obtained through the above particle-hole transformation as $\mathbf{G}^{\downarrow}=\mathbf{1}_{N_s}-(\mathbf{G}^{\uparrow})^+$. {\it Fifth}, for a given configuration, the measurements of all static single- and two-particle observables are obtained either directly from $\mathbf{G}^{\sigma}(\tau,\tau)$ or via the Wick decomposition~\cite{Assaad2008}. The Monte Carlo averages are computed using Eq.~(\ref{eq:GSObsQMC}), and the corresponding statistical errors are estimated via standard binning analysis. 

\subsection{Physical observables}
\label{sec:AFQMCObs}

In this work, we focus primarily on the physical observables associated with the QCP, including the AFM spin correlations and several related quantities. In this subsection, we outline the definitions and computation formulas for all these observables.

To characterize the magnetic properties, we measure the intra-sublattice spin-spin correlation function as
\begin{equation}\begin{aligned}
\label{eq:SzSzCrF}
C_{zz}(\mathbf{r})
= \frac{1}{2N_c}\sum_{i}\big\langle\hat{s}^{z}(\mathbf{r}_i)\hat{s}^{z}(\mathbf{r}_i+\mathbf{r})\big\rangle,
\end{aligned}\end{equation}
where $\hat{s}^{z}(\mathbf{r}_i)=(\hat{n}_{i\uparrow}-\hat{n}_{i\downarrow})/2$ is the spin-$\hat{s}^z$ operator, and the distance vector $\mathbf{r}$ is restricted to the Bravais lattice vectors $\mathbf{R}=n_1\mathbf{a}_1+n_2\mathbf{a}_2$ (with $n_1,n_2$ as integers) of the honeycomb lattice. Using $C_{zz}(\mathbf{r})$, the spin structure factor is expressed as
\begin{equation}\begin{aligned}
\label{eq:SpinStrucF}
S_{zz}(\mathbf{k})=\sum_{\mathbf{r}}C_{zz}(\mathbf{r})e^{{\rm i}\mathbf{k}\cdot\mathbf{r}}.
\end{aligned}\end{equation}
Then the mean-squared magnetization for the N\'{e}el AFM order reads $m^2=S_{zz}(\boldsymbol{\Gamma})/(2N_c)$ with $\boldsymbol{\Gamma}=(0,0)$. The QCP location $U_c/t$ and the critical exponents $\nu$ and $\eta_{\phi}$ (bosonic anomalous dimension) can be extracted via the finite-size scaling analysis of $m^2$~\cite{Toldin2015,Otsuka2016,Lang2025,Wang2026}. We also compute the correlation ratio for the N\'{e}el order using
\begin{equation}\begin{aligned}
\label{eq:CrfRatio}
R_{m^2}=1-\frac{S_{zz}(\boldsymbol{\Gamma}+\delta\mathbf{k})}{S_{zz}(\boldsymbol{\Gamma})},
\end{aligned}\end{equation}
with $\delta\mathbf{k}=\mathbf{b}_1/L$ or $\mathbf{b}_2/L$ as the smallest momentum on the lattice. 

We further examine several physical observables associated with the QCP, including both single-particle and two-body quantities. For the former, we focus on the fermionic momentum distribution $\mathbf{n}(\mathbf{k})$, which is measured as a $2\times 2$ matrix (for spin-up or -down channel)
\begin{equation}\begin{aligned}
\label{eq:nkMatrix}
\mathbf{n}(\mathbf{k})=
\begin{pmatrix}
n_{\rm AA}(\mathbf{k}) & n_{\rm AB}(\mathbf{k})\\
n_{\rm BA}(\mathbf{k}) & n_{\rm BB}(\mathbf{k})
\end{pmatrix},
\end{aligned}\end{equation}
where A and B denote the sublattices. The off-diagonal element $n_{\rm AB}(\mathbf{k})=\langle a_{\mathbf{k}\uparrow}^{+}b_{\mathbf{k}\uparrow}^{}\rangle$ [or $n_{\rm BA}(\mathbf{k})=\langle b_{\mathbf{k}\uparrow}^{+}a_{\mathbf{k}\uparrow}^{}\rangle$] can be used to calculate the fermionic anomalous dimension $\eta_{\psi}$~\cite{Otsuka2016,Lang2025,Wang2026}. In addition, we use $n_{+}(\mathbf{k})$ and $n_{-}(\mathbf{k})$ to denote the eigenvalues of $\mathbf{n}(\mathbf{k})$ matrix corresponding to $\varepsilon_{\mathbf{k},-}$ and $\varepsilon_{\mathbf{k},+}$ bands, respectively, from which we construct the energy-resolved momentum distribution $n(\varepsilon_{\mathbf{k}})$ by assigning $n(\varepsilon_{\mathbf{k}}=\varepsilon_{\mathbf{k},-})=n_{+}(\mathbf{k})$ and $n(\varepsilon_{\mathbf{k}}=\varepsilon_{\mathbf{k},+})=n_{-}(\mathbf{k})$ ($\varepsilon_{\mathbf{k},\pm}$ as the kinetic energy dispersions, see Sec.~\ref{sec:TheModel}). Based on $n(\varepsilon_{\mathbf{k}})$, we can compute the finite-size quasiparticle weight~\cite{Otsuka2016}
\begin{equation}\begin{aligned}
\label{eq:QpWeight}
Z_{L}=n(\varepsilon_{\mathbf{k}}=0^{-})-n(\varepsilon_{\mathbf{k}}=0^{+}),
\end{aligned}\end{equation}
and the Fermi-liquid parameter~\cite{Rossi2018}
\begin{equation}\begin{aligned}
\label{eq:FLP}
Q_{\mathrm{FL}}=\big(\big|\nabla_{\mathbf{k}}n(\varepsilon_\mathbf{k})\big|_{\mathbf{k}=\mathbf{K}}\big)^{-1}=\left[-v_{\mathrm{F}}\frac{\partial n(\varepsilon_\mathbf{k})}{\partial\varepsilon_{\mathbf{k}}}\bigg|_{\varepsilon_{\mathbf{k}}=0}\right]^{-1},
\end{aligned}\end{equation}
where $v_{\mathrm{F}}=|\nabla_{\mathbf{k}}\varepsilon_{\mathbf{k}}|_{\mathbf{k}=\mathbf{K}}=3t/2$ is the bare Fermi velocity with $\mathbf{K}$ being a Dirac point. For the two-body observables, we concentrate on those that are most relevant to the Dirac quantum criticality. They include the charge compressibility~\cite{Song2025B,Xie2025,Lu2026}
\begin{equation}\begin{aligned}
\label{eq:ChiCharge}
\chi_e = \frac{\beta}{N_s}\sum_{ij}\big(\langle \hat{n}_{i} \hat{n}_{j} \rangle - \langle \hat{n}_{i} \rangle\langle \hat{n}_{j} \rangle\big),
\end{aligned}\end{equation}
the double occupancy $D=N_{s}^{-1}\sum_{i}\langle\hat{n}_{i\uparrow}\hat{n}_{i\downarrow}\rangle$ as well as its derivative over $U$ as~\cite{Song2025L,*Song2025B}
\begin{equation}\begin{aligned}
\label{eq:DouOccDerive}
\frac{\partial D}{\partial U} =
-\frac{2}{N_s}\int_0^{\beta/2} C_{\hat{H}_I}(\tau) d\tau,
\end{aligned}\end{equation}
and the fidelity susceptibility $\chi_{\rm F}$ as~\cite{You2007,Venuti2007,Gu2009,Schwandt2009,Albuquerque2010,WangLei2015,HuangLi2016}
\begin{equation}\begin{aligned}
\label{eq:fidelitysusp}
\chi_{\rm F} = \int_{0}^{\beta/2} \tau C_{\hat{H}_I}(\tau) d\tau,
\end{aligned}\end{equation}
where $C_{\hat{H}_{I}}(\tau)=\langle\hat{H}_{I}(\tau)\hat{H}_{I}(0)\rangle-\langle\hat{H}_{I}(\tau)\rangle\langle\hat{H}_{I}(0)\rangle$ with the operator $\hat{H}_{I}=\sum_{i}[\hat n_{i\uparrow}\hat n_{i\downarrow}-(\hat n_{i\uparrow}+\hat n_{i\downarrow})/2]$ is a dynamical correlation function that can be directly measured in AFQMC simulations. For Eqs.~(\ref{eq:DouOccDerive}) and~(\ref{eq:fidelitysusp}), we practically first perform polynomial fit for $C_{\hat{H}_{I}}(\tau)$ versus $\tau$, then compute the integral analytically via the fitting curve, and further estimate the error bars using the bootstrapping technique~\cite{Song2024}.

\section{Algorithmic improvements and TABC application in AFQMC}
\label{sec:NewMethod}

In this section, we present the algorithmic improvements implemented in our AFQMC simulations, including the fast Fourier transform (FFT) for path (imaginary-time) propagation (Sec.~\ref{sec:FFT}) and the delayed update technique for configuration updates (Sec.~\ref{sec:Update}). We also discuss the application of TABC in AFQMC simulations (Sec.~\ref{sec:TABC}).

\subsection{FFT in the path propagation}
\label{sec:FFT}

The path (imaginary-time) propagation is one of the key components of the AFQMC algorithm. Specifically, from the $(\ell-1)$-th to $\ell$-th time slice, we need to propagate the $\mathbf{R}^{\sigma}$, $\mathbf{L}^{\sigma}$, and $\mathbf{G}^{\sigma}$ matrices [see Eq.~(\ref{eq:GrFMat})] using $\mathbf{R}^{\sigma}\to\mathbf{B}_{\ell}^{\sigma}\mathbf{R}^{\sigma}$, $\mathbf{L}^{\sigma}\to\mathbf{L}^{\sigma}(\mathbf{B}_{\ell}^{\sigma})^{-1}$, and $\mathbf{G}^{\sigma}\to\mathbf{B}_{\ell}^{\sigma}\mathbf{G}^{\sigma}(\mathbf{B}_{\ell}^{\sigma})^{-1}$, with the propagator matrix $\mathbf{B}_{\ell}^{\sigma}=e^{\mathbf{H}_I^{\sigma}(\mathbf{x}_{\ell})}e^{-\Delta\tau\mathbf{H}_0^{\sigma}}$. Conventionally, these propagations are performed by standard matrix-matrix multiplications (DGEMM or ZGEMM), especially for the term $e^{-\Delta\tau\mathbf{H}_0^{\sigma}}$, leading to the computational cost of $O(N_s^3)$. The application of FFT is based on the observation that~\cite{Yuanyao2026} $\mathbf{H}_0^{\sigma}$ is diagonal (or block diagonal) when transformed into momentum ($\mathbf{k}$) space, whereas $\mathbf{H}_I^{\sigma}(\mathbf{x}_{\ell})$ is diagonal in real ($\mathbf{r}$) space for the Hubbard interaction after the HS-$\hat{s}^z$ and HS-$\hat{n}$ transformations [Eqs.~(\ref{eq:SzHS}) and~(\ref{eq:DensityHS})]. Consequently, the propagation of $\mathbf{B}_{\ell}^{\sigma}$ matrix can be carried out by alternating multiplications of diagonal (or block-diagonal) matrices in $\mathbf{r}$ and $\mathbf{k}$ space with FFT operations connecting the two representations. This procedure reduces the computational complexity from $O(N_s^3)$ to $O(N_s^2\ln N_s)$. 

The implementation of FFT relies on the Fourier transformation matrix $\mathbf{U}$ from the $\mathbf{r}$ space to $\mathbf{k}$ space, whose matrix element reads $U_{\mathbf{k},\mathbf{r}}=e^{-{\rm i}\mathbf{k}\cdot\mathbf{r}}/L$. Then we can reformulate $\mathbf{B}_{\ell}^{\sigma}=e^{\mathbf{H}_I^{\sigma}(\mathbf{x}_{\ell})}e^{-\Delta\tau\mathbf{H}_0^{\sigma}}$ as
\begin{equation}\begin{aligned}
\label{eq:NewBlMat}
\mathbf{B}_{\ell}^{\sigma}
= e^{\mathbf{H}_I^{\sigma}(\mathbf{x}_{\ell})}\mathbf{U}^{+}\mathbf{U}e^{-\Delta\tau\mathbf{H}_0^{\sigma}}\mathbf{U}^{+}\mathbf{U}
= e^{\mathbf{H}_I^{\sigma}(\mathbf{x}_{\ell})}\mathbf{U}^{+} e^{-\Delta\tau\boldsymbol{\Lambda}}\mathbf{U},
\end{aligned}\end{equation}
where $e^{-\Delta\tau\boldsymbol{\Lambda}}=\mathbf{U}e^{-\Delta\tau\mathbf{H}_0^{\sigma}}\mathbf{U}^{+}$ with $\boldsymbol{\Lambda}=\mathbf{U}\mathbf{H}_0^{\sigma}\mathbf{U}^{+}$ as the $\mathbf{k}$-space Hamiltonian matrix for $\hat{H}_0$. Notably, $\boldsymbol{\Lambda}$ is diagonal for the simple (single-sublattice) lattice (i.e., the square and triangular lattice), whereas it becomes block diagonal for multi-sublattice systems (such as the honeycomb lattice). Based on Eq.~(\ref{eq:NewBlMat}), calculating the propagation $\mathbf{B}_{\ell}^{\sigma}\mathbf{R}^{\sigma}=e^{\mathbf{H}_I^{\sigma}(\mathbf{x}_{\ell})}\mathbf{U}^{+} e^{-\Delta\tau\boldsymbol{\Lambda}}\mathbf{U}\mathbf{R}^{\sigma}$ can be divided into four steps (note $\mathbf{R}^{\sigma}$ is a dense matrix in the $\mathbf{r}$-space representation): (i) compute $\mathbf{R}_1=\mathbf{U}\mathbf{R}^{\sigma}$ by applying an FFT from $\mathbf{r}$ to $\mathbf{k}$ space; (ii) compute $\mathbf{R}_2=e^{-\Delta\tau\boldsymbol{\Lambda}}\mathbf{R}_1$ by multiplying a diagonal (or block-diagonal) matrix; (iii) compute $\mathbf{R}_3=\mathbf{U}^{+}\mathbf{R}_2$ by applying the inverse FFT from $\mathbf{k}$ to $\mathbf{r}$ space; (iv) compute $e^{\mathbf{H}_I^{\sigma}(\mathbf{x}_{\ell})}\mathbf{R}_3$ by multiplying a diagonal matrix. Similar procedures apply for the calculations of $\mathbf{L}^{\sigma}(\mathbf{B}_{\ell}^{\sigma})^{-1}$ and $\mathbf{B}_{\ell}^{\sigma}\mathbf{G}^{\sigma}(\mathbf{B}_{\ell}^{\sigma})^{-1}$. The FFT operations are implemented using subroutines in the FFTW library~\cite{Frigo2005}. The above steps (i)-(iii) correspond to the FFT-related calculations combined as $(\mathbf{U}^{+} e^{-\Delta\tau\boldsymbol{\Lambda}}\mathbf{U}\mathbf{R}^{\sigma})$, and they are practically performed sequentially for each column of $\mathbf{R}^{\sigma}$. For the special case of simple lattices, two FFT operations plus one diagonal matrix multiplication by $e^{-\Delta\tau\boldsymbol{\Lambda}}$ are required for each column, which features the computational cost of $O(2N_s\ln N_s+N_s)$. Repeating such calculations for all $N_s$ columns results in an overall computational complexity of $O[2N_s^2\ln N_s+N_s^2]$.

For the honeycomb lattice, $\boldsymbol{\Lambda}$ is block diagonal and it is composed of $2\times2$ matrix $\mathbf{h}(\mathbf{k})$ [see Eq.~(\ref{eq:h0kMat})] at all $\mathbf{k}$ points, i.e., $\boldsymbol{\Lambda}={\rm Diag}[\mathbf{h}(\mathbf{k}_1),\mathbf{h}(\mathbf{k}_2),\cdots,\mathbf{h}(\mathbf{k}_{N_c})]$. Therefore, the $e^{-\Delta\tau\boldsymbol{\Lambda}}$ matrix is evaluated as
\begin{equation}\begin{aligned}
\label{eq:ExpLambda}
e^{-\Delta\tau\boldsymbol{\Lambda}}={\rm Diag}[e^{-\Delta\tau\mathbf{h}(\mathbf{k}_1)},e^{-\Delta\tau\mathbf{h}(\mathbf{k}_2)},\cdots,e^{-\Delta\tau\mathbf{h}(\mathbf{k}_{N_c})}],
\end{aligned}\end{equation}
where $e^{-\Delta\tau\mathbf{h}(\mathbf{k})}$ is a $2\times2$ matrix and can be precomputed at the beginning of the simulation. Here we label all the A-sublattice sites as $i=1,2,\cdots,N_c$ and the B-sublattice sites as $i=N_c+1,N_c+2,\cdots,N_s$. The Fourier transformation matrix $\mathbf{U}$ also becomes block diagonal as
\begin{equation}\begin{aligned}
\label{eq:FourierMat}
\mathbf{U}=
\begin{pmatrix}
\mathbf{U}_{\rm A} & \mathbf{0}\\
\mathbf{0} & \mathbf{U}_{\rm B}
\end{pmatrix},
\end{aligned}\end{equation}
with $\mathbf{U}_{\rm A}$ and $\mathbf{U}_{\rm B}$ as the $N_c\times N_c$ Fourier transformation matrices for A and B sublattices, respectively. The matrix elements of $\mathbf{U}_{\rm A}$ read $(\mathbf{U}_{\rm A})_{\mathbf{k},\mathbf{r}}=e^{-{\rm i}\mathbf{k}\cdot\mathbf{r}}/L$, while those of $\mathbf{U}_{\rm B}$ can, up to a gauge transformation, be chosen as either $(\mathbf{U}_{\rm B})_{\mathbf{k},\mathbf{r}}=e^{-{\rm i}\mathbf{k}\cdot\mathbf{r}}/L$ or $(\mathbf{U}_{\rm B})_{\mathbf{k},\mathbf{r}}=e^{-{\rm i}\mathbf{k}\cdot(\mathbf{r}+\boldsymbol{\tau}_{\rm AB})}/L$, as discussed in Sec.~\ref{sec:TABC}. Then the propagation $\mathbf{B}_{\ell}^{\sigma}\mathbf{R}^{\sigma}=e^{\mathbf{H}_I^{\sigma}(\mathbf{x}_{\ell})}\mathbf{U}^{+} e^{-\Delta\tau\boldsymbol{\Lambda}}\mathbf{U}\mathbf{R}^{\sigma}$ using FFT proceeds in four steps analogous to those for simple lattices, with several specific modifications. In the steps (i)-(iii), the calculations for $(\mathbf{U}^{+} e^{-\Delta\tau\boldsymbol{\Lambda}}\mathbf{U}\mathbf{R}^{\sigma})$ are still carried out sequentially for each column of $\mathbf{R}^{\sigma}$. For instance, for the $j$-th column of $\mathbf{R}^{\sigma}$ matrix denoted $\mathbf{p}^{j}$, we perform the FFT to calculate $\mathbf{q}^{j}=\mathbf{U}\mathbf{p}^{j}$ as
\begin{equation}\begin{aligned}
\label{eq:ColumnDem}
\mathbf{U}\mathbf{p}^{j} =
\begin{pmatrix}
\mathbf{U}_{\rm A} & \mathbf{0}\\
\mathbf{0} & \mathbf{U}_{\rm B}
\end{pmatrix}
\begin{pmatrix}
\mathbf{p}_{\rm A}\\
\mathbf{p}_{\rm B}
\end{pmatrix}
=
\begin{pmatrix}
\mathbf{U}_{\rm A}\mathbf{p}_{\rm A}\\
\mathbf{U}_{\rm B}\mathbf{p}_{\rm B}
\end{pmatrix}
= 
\begin{pmatrix}
\mathbf{q}_{\rm A}\\
\mathbf{q}_{\rm B}
\end{pmatrix},
\end{aligned}\end{equation}
where $\mathbf{p}_{\rm A}$ and $\mathbf{p}_{\rm B}$ are $N_c$-dimensional column vectors in $\mathbf{r}$-space representation that together form $\mathbf{p}^{j}$, and the $\mathbf{k}$-space vector $\mathbf{q}^{j}$ consists of $\mathbf{q}_{\rm A}=\mathbf{U}_{\rm A}\mathbf{p}_{\rm A}$ and $\mathbf{q}_{\rm B}=\mathbf{U}_{\rm B}\mathbf{p}_{\rm B}$ that are obtained through two separate FFT operations. Then, the $j$-th column of $\mathbf{R}_2=e^{-\Delta\tau\boldsymbol{\Lambda}}(\mathbf{U}\mathbf{R}^{\sigma})$ matrix [in step (ii)] is obtained by multiplying $\mathbf{q}^{j}$ by $e^{-\Delta\tau\boldsymbol{\Lambda}}$. Note that the combined $\mathbf{k}$ and sublattice basis ordering of $\mathbf{q}^{j}$ differs from that of $e^{-\Delta\tau\boldsymbol{\Lambda}}$ [see Eq.~(\ref{eq:ExpLambda})]. Therefore, the latter must be reshaped before performing the multiplication. Next, we perform the inverse FFT for the $j$-th column of $\mathbf{R}_2$ matrix using a similar formula as Eq.~(\ref{eq:ColumnDem}), i.e., two separate FFT operations for A- and B-sublattice components. Consequently, to compute each column of $\mathbf{R}_3=\mathbf{U}^{+} e^{-\Delta\tau\boldsymbol{\Lambda}}\mathbf{U}\mathbf{R}^{\sigma}$ matrix, it requires four FFT operations for $N_c$-dimensional vector and one block diagonal matrix multiplication, featuring the computational cost of $O(4N_c\ln N_c + 4N_c)$. Considering all $N_s=2N_c$ columns in $\mathbf{R}^{\sigma}$, the overall computational complexity for steps (i)-(iii) is $O[8N_c^2(\ln N_c+1)]$ (or equivalently $O[8L^4(2\ln L + 1)]$ using $N_c=L^2$). Next, the step (iv) of computing $e^{\mathbf{H}_I^{\sigma}(\mathbf{x}_{\ell})}\mathbf{R}_3$ is the same as that for simple lattices, since $e^{\mathbf{H}_I^{\sigma}(\mathbf{x}_{\ell})}$ is still diagonal in $\mathbf{r}$-space representation.

\begin{figure}[t]
\includegraphics[width=0.913\columnwidth]{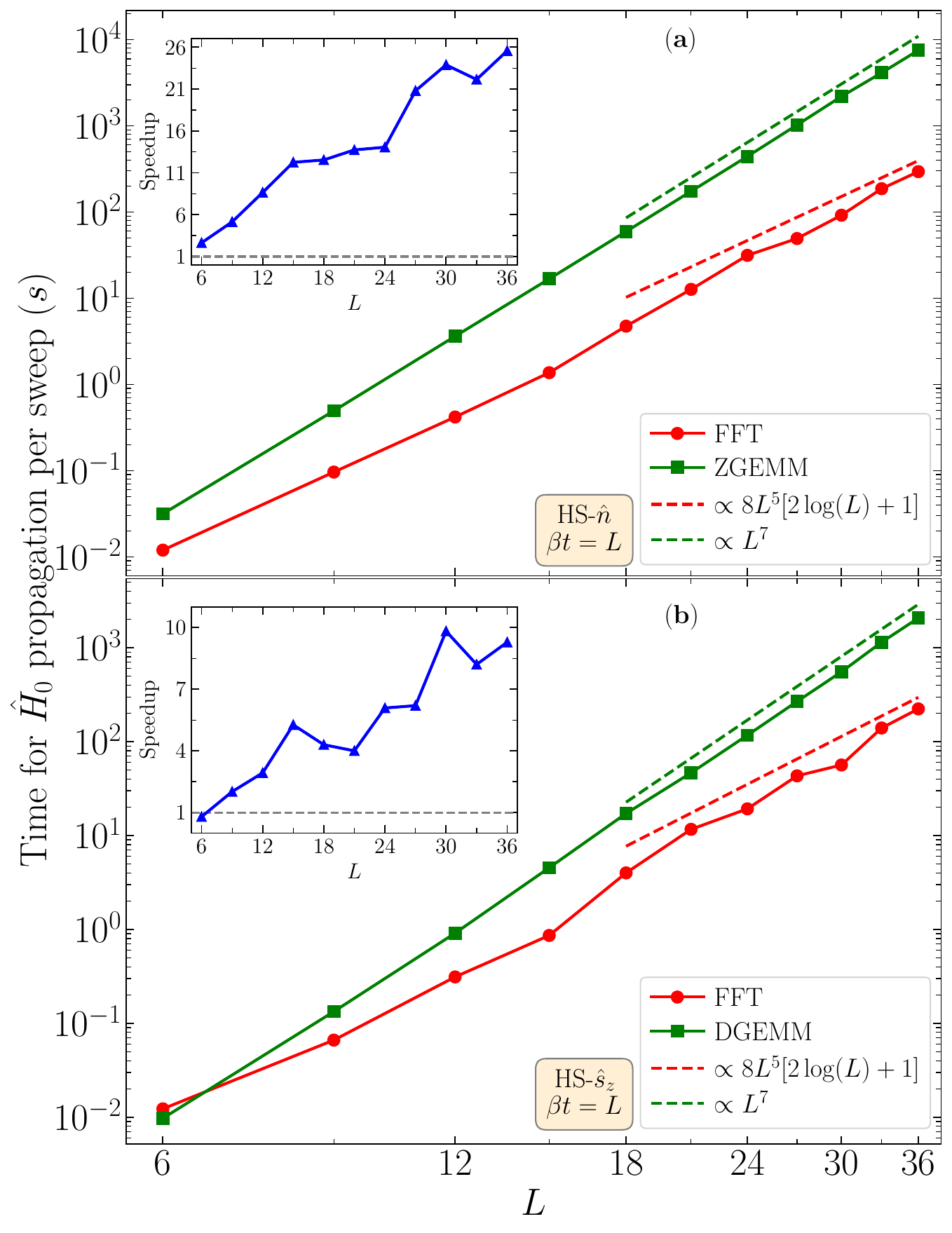}
\caption{
Comparison of the average time for propagating the noninteracting part per sweep (in seconds) between the FFT and ZGEMM/DGEMM methods in AFQMC simulations of the honeycomb Hubbard model with $U/t=1$ and $\beta t=L$. Panels (a) and (b) show the results applying HS-$\hat{n}$ [Eq.~(\ref{eq:DensityHS})] and HS-$\hat{s}^z$ [Eq.~(\ref{eq:SzHS})], respectively. The dashed lines plot the corresponding theoretical scalings. The insets in both panels illustrate the corresponding speedups achieved by the FFT compared to ZGEMM/DGEMM for the propagation.
}
\label{fig:fig02FFTGEMM}
\end{figure}

The application of FFT in the path propagation relies on the existence of a well-defined momentum-space representation, which is available only when PBC or TBC is imposed in the simulation. This technique reduces the computational cost of propagating the noninteracting part by nearly an order of magnitude in $N_s$, i.e., from $O(N_s^3)$ to $O(N_s^2)$ considering $\ln N_s\ll N_s$. An alternative route to achieving comparable acceleration is to employ Trotter breakups to the $e^{-\Delta\tau\hat{H}_0}$ operator, which involves factorizing $e^{-\Delta\tau\mathbf{H}_0^{\sigma}}$ into a sequential product of $2\times 2$ block matrices~\cite{Lang2025,Wang2026}. While this approach can surpass the FFT method in efficiency and extends naturally to open boundary conditions, it comes at the cost of introducing additional Trotter error. Nevertheless, the FFT-based strategy provides easy access to the single-particle Green's function in momentum space, from which the pairing matrix and condensate fraction can be readily computed to characterize superconductivity and superfluidity~\cite{Yang1962,Boronat2005,Fukushima2007,Shihao2015,He2019L,YuanYao2022,Yuanyao2026}. Moreover, the FFT procedure described above can be extended to more sophisticated multi-band or multi-orbital Hubbard models~\cite{He2016,Hongxia2019,Ettore2019,Adam2020,Hao2020,Luo2023,Zhang2026,Ma2026,Rubem2026}, and is equally applicable to the ground-state AFQMC algorithm~\cite{White1989,Assaad2008,Sugiyama1986,Sorella1989,Duhao2025,Wang2026}.

In practical simulations, the FFT-based propagation of the noninteracting part, i.e., $e^{-\Delta\tau\mathbf{H}_0^{\sigma}}\mathbf{R}^{\sigma}$, can achieve substantial speedup over the conventional approach of dense matrix-matrix multiplication. In Fig.~\ref{fig:fig02FFTGEMM}, we present the average time per sweep consumed by these two methods in AFQMC simulations applying both HS-$\hat{n}$ [Eq.~(\ref{eq:DensityHS})] and HS-$\hat{s}^z$ [Eq.~(\ref{eq:SzHS})] transformations, for which we set $U/t=1$ and $\beta t=L$. For $U>0$, AFQMC simulations are performed within the complex-number framework for HS-$\hat{n}$, while for HS-$\hat{s}^z$ all involved matrices ($\mathbf{L}^{\sigma}$, $\mathbf{R}^{\sigma}$, and $\mathbf{B}_{\ell}^{\sigma}$) are purely real (with PBC). The conventional propagation is thus performed accordingly with ZGEMM (complex) or DGEMM (real). Note that an AFQMC sweep refers to a complete forward and backward scan over all auxiliary fields across $\ell = 1$ to $\ell = M$ imaginary-time slices. Given $\beta t=L$, the theoretical computational scaling is $O(L^7)$ for the ZGEMM/DGEMM method, whereas for the FFT method it scales as $O[8L^5(2\ln L + 1)]$. These scalings are well confirmed by the results in Fig.~\ref{fig:fig02FFTGEMM}. The corresponding speedups achieved by the FFT, shown in the insets, exhibit an overall enhancement with increasing system size, reaching $\sim26$ for the complex case and $\sim10$ for the real case at $L=36$. The lower speedup in the real case is due to the fact that its FFT operations involve real-to-complex and complex-to-real conversions, which introduce additional computational overhead. We also observe similar speedups in simulations with fixed $\beta t$ (see Appendix~\ref{sec:AppFFTSpeedUp}). These results illustrate that the FFT-based propagation is robust and highly efficient, and its advantage becomes increasingly pronounced for larger system sizes.

\subsection{The delayed update}
\label{sec:Update}

In AFQMC, the auxiliary-field configuration $\mathbf{X}$ is typically updated to a new configuration $\mathbf{X}^{\prime}$ according to the Metropolis algorithm. Imposing detailed balance condition then yields the acceptance probability as
\begin{equation}\begin{aligned}
\label{eq:DetailBalance}
\mathcal{A}(\mathbf{X}\to\mathbf{X}^{\prime})
=\min\Big\{1,\frac{W(\mathbf{X}^{\prime})}{W(\mathbf{X})}\Big\}.
\end{aligned}\end{equation}
According to Eq.~(\ref{eq:ConfgWeight}), the weight ratio can be expressed as $W(\mathbf{X}^{\prime})/W(\mathbf{X})=P(\mathbf{X}^{\prime})/P(\mathbf{X})\times r_{\uparrow}r_{\downarrow}$, with $r_{\sigma}$ denoting the determinant ratio in spin-$\sigma$ channel. The term $P(\mathbf{X}^{\prime})/P(\mathbf{X})$ is easily obtained with the HS transformation, leaving the evaluation of $r_{\sigma}$ as the central task in determining whether to accept the update. If the update is accepted, the single-particle Green's function matrix, $\mathbf{G}^{\sigma}=(\mathbf{1}_{N_s}+\mathbf{R}^{\sigma}\mathbf{L}^{\sigma})^{-1}$ in Eq.~(\ref{eq:GrFMat}), needs to be upgraded. In the following, we briefly review the general update scheme, the local update and delayed update algorithms in finite-temperature AFQMC method. 

We consider the case where $\mathbf{X}$ and $\mathbf{X}^{\prime}$ differ only at the $\ell$-th time slice, i.e., $\mathbf{x}_{\ell}$ in $\mathbf{X}$ and $\mathbf{x}_{\ell}^{\prime}$ in $\mathbf{X}^{\prime}$. Then the ratio $r_{\sigma}$ is given by 
\begin{equation}\begin{aligned}
\label{eq:DRatio}
r_{\sigma}
&= \frac{\det[\mathbf{1}_{N_s}+\mathbf{B}_{M}^{\sigma}\cdots\mathbf{B}_{\ell+1}^{\sigma}(\mathbf{B}_{\ell}^{\sigma})'\mathbf{B}_{\ell-1}^{\sigma}\cdots\mathbf{B}_1^{\sigma}]}{{\rm det}(\mathbf{1}_{N_s}+\mathbf{B}_{M}^{\sigma}\cdots\mathbf{B}_{\ell+1}^{\sigma}\mathbf{B}_{\ell}^{\sigma}\mathbf{B}_{\ell-1}^{\sigma}\cdots\mathbf{B}_1^{\sigma})}\\
&= \frac{\det[\mathbf{1}_{N_s}+\mathbf{L}^{\sigma}(\mathbf{1}_{N_s}+\boldsymbol{\Delta}^{\sigma})\mathbf{R}^{\sigma}]}{{\rm det}(\mathbf{1}_{N_s}+\mathbf{L}^{\sigma}\mathbf{R}^{\sigma})} \\
&= \det[\mathbf{1}_{N_s}+\boldsymbol{\Delta}^{\sigma}(\mathbf{I}_{N_s}-\mathbf{G}^{\sigma})],
\end{aligned}\end{equation}
with $(\mathbf{B}_{\ell}^{\sigma})^{\prime}=e^{\mathbf{H}_I^{\sigma}(\mathbf{x}_{\ell}^{\prime})}e^{-\Delta\tau\mathbf{H}_0^{\sigma}}$, $\mathbf{B}_{\ell}^{\sigma}=e^{\mathbf{H}_I^{\sigma}(\mathbf{x}_{\ell})}e^{-\Delta\tau\mathbf{H}_0^{\sigma}}$, and $\mathbf{L}^{\sigma}=\mathbf{B}_{M}^{\sigma}\mathbf{B}_{M-1}^{\sigma}\cdots\mathbf{B}_{\ell+1}^{\sigma}$, $\mathbf{R}^{\sigma}=\mathbf{B}_{\ell}^{\sigma}\cdots\mathbf{B}_2^{\sigma}\mathbf{B}_1^{\sigma}$. We introduce the diagonal matrix $\boldsymbol{\Delta}^{\sigma}=e^{\mathbf{H}_I^{\sigma}(\mathbf{x}_{\ell}^{\prime})}e^{-\mathbf{H}_I^{\sigma}(\mathbf{x}_{\ell})}-\mathbf{1}_{N_s}$ to rewrite $(\mathbf{B}_{\ell}^{\sigma})^{\prime}$ as $(\mathbf{B}_{\ell}^{\sigma})^{\prime}=(\mathbf{1}_{N_s}+\boldsymbol{\Delta}^{\sigma})\mathbf{B}_{\ell}^{\sigma}$. From Eq.~(\ref{eq:DRatio}), only $\boldsymbol{\Delta}^{\sigma}$ and $\mathbf{G}^{\sigma}$ are involved in the calculation of $r_{\sigma}$. Once the update is accepted, we need to upgrade $\mathbf{G}^{\sigma}$ as
\begin{equation}\begin{aligned}
\label{eq:GFupdate}
(\mathbf{G}^{\sigma})^{\prime}
&= [\mathbf{1}_{N_s}+(\mathbf{1}_{N_s}+\boldsymbol{\Delta}^{\sigma})\mathbf{R}^{\sigma}\mathbf{L}^{\sigma}]^{-1} \\
&= \mathbf{G}^{\sigma}[\mathbf{1}_{N_s}+\boldsymbol{\Delta}^{\sigma}(\mathbf{1}_{N_s}-\mathbf{G}^{\sigma})]^{-1}.
\end{aligned}\end{equation}
Equations~(\ref{eq:DRatio}) and~(\ref{eq:GFupdate}) constitute the key results for the configuration update process.

The {\it local update} algorithm updates the auxiliary fields $\mathbf{x}_{\ell}=(x_{\ell,1},x_{\ell,2},\cdots,x_{\ell,N_s})$ sequentially, one element at a time. For instance, for the $i$-th lattice site, the single-step update flips $x_{\ell,i}$ (from $+1$ to $-1$, or vice versa), and computes $r_{\sigma}$ to determine the acceptance. Accordingly, the $\boldsymbol{\Delta}^{\sigma}$ matrix in Eqs.~(\ref{eq:DRatio}) and~(\ref{eq:GFupdate}) only has one nonzero element, i.e., $\Delta_{ii}^{\sigma}=e^{\gamma f_{\sigma}(x_{\ell,i}^{\prime}-x_{\ell,i})}-1$ with $\gamma=\gamma_{\rm s},f_{\uparrow}=+1,f_{\downarrow}=-1$ for HS-$\hat{s}^z$ [Eq.~(\ref{eq:SzHS})] and $\gamma=\gamma_{\rm c},f_{\uparrow(\downarrow)}=+1$ for HS-$\hat{n}$ [Eq.~(\ref{eq:DensityHS})]. To facilitate the derivation, we define the column vector $\mathbf{v}_i=(\boldsymbol{\Delta}^{\sigma})_{i{\rm-col}}$ (``$i$-col'' as the $i$-th column) and the row vector $\mathbf{y}^{\rm T}=(\mathbf{1}_{N_s}-\mathbf{G}^{\sigma})_{i{\rm-row}}$ (``$i$-row'' as the $i$-th row, and the superscript ``T'' meaning transpose). Then the $r_{\sigma}$ in Eq.~(\ref{eq:DRatio}) can be simplified as
\begin{equation}\begin{aligned}
\label{eq:rRatioLocal}
r_{\sigma}
&= {\rm \det}(\mathbf{1}_{N_s}+\mathbf{v}_i\mathbf{y}^{\rm T})
 = 1 + \mathbf{y}^{\rm T}\mathbf{v}_i \\
&= 1 + \Delta_{ii}^{\sigma}[1-(\mathbf{G}^{\sigma})_{ii}].
\end{aligned}\end{equation}
Once this update is accepted, the $\mathbf{G}^{\sigma}$ matrix in Eq.~(\ref{eq:GFupdate}) is upgraded using the Sherman-Morrison formula~\cite{Duhao2025}, proceeding as
\begin{equation}\begin{aligned}
\label{eq:GupdateLocal}
(\mathbf{G}^{\sigma})^{\prime}
&= \mathbf{G}^{\sigma}(\mathbf{1}_{N_s}+\mathbf{v}_i\mathbf{y}^{\rm T})^{-1}
 = \mathbf{G}^{\sigma} - \mathbf{G}^{\sigma}\mathbf{v}_i\mathbf{y}^{\rm T}/r_{\sigma} \\
&= \mathbf{G}^{\sigma} - \mathbf{x}\mathbf{y}^{\rm T},
\end{aligned}\end{equation}
with $\mathbf{x}=\mathbf{G}^{\sigma}\mathbf{v}_i/r_{\sigma}=(\mathbf{G}^{\sigma})_{{i{\rm-col}}}\Delta_{ii}^{\sigma}/r_{\sigma}$ as a column vector. The calculation of $\mathbf{x}\mathbf{y}^{\rm T}$ involves a vector-vector outer product (ZGERU), which is the leading computational complexity $\mathcal{O}(N_s^2)$ of the local update process. Then repeating the above procedure and sweeping from $i=1$ to $i=N_s$ constitute the {\it local update} for the entire $\ell$-th time slice, featuring a computational cost of $\mathcal{O}(N_s^3)$. Nevertheless, the vector operation involved in computing $\mathbf{x}\mathbf{y}^{\rm T}$ in Eq.~(\ref{eq:GupdateLocal}) suffers from low efficiency and becomes the performance bottleneck of the {\it local update} algorithm.

The {\it delayed update} technique builds on the above {\rm local update} algorithm and is specially designed to replace the inefficient vector-vector outer products (in upgrading $\mathbf{G}^{\sigma}$) with matrix-matrix multiplications~\cite{Sun2024}. Practically, it computes the ratio $r_{\sigma}$ in a similar way, but delays the upgrade of $\mathbf{G}^{\sigma}$ until the number of accepted local updates reaches a predetermined delay rank $n_d$. 

We use $(\mathbf{G}^{\sigma})_{[0]}$ and $(\mathbf{G}^{\sigma})_{[k]}$ (with $k\ge1$) to denote the initial $\mathbf{G}^{\sigma}$ matrix before the update and the new $\mathbf{G}^{\sigma}$ matrix after the $k$-th accepted update, respectively. The target of the delayed update is to achieve the relation
\begin{equation}\begin{aligned}
\label{eq:DelayG0}
(\mathbf{G}^{\sigma})_{[k]}=(\mathbf{G}^{\sigma})_{[0]}-\sum_{m=1}^{k}\mathbf{x}_{m}\mathbf{y}_{m}^{\mathrm{T}},
\end{aligned}\end{equation}
for which we need to derive the explicit expressions of the column vector $\mathbf{x}_m$ and the row vector $\mathbf{y}_m^{\rm T}$ in terms of $\boldsymbol{\Delta}^{\sigma}$ and $(\mathbf{G}^{\sigma})_{[0]}$. For the special case of $k=1$, the delayed upgrade of $\mathbf{G}^{\sigma}$ in Eq.~(\ref{eq:DelayG0}) becomes the local update as in Eq.~(\ref{eq:GupdateLocal}). We then focus on the $k$-th accepted move ($k>1$), which involves flipping the auxiliary field at the $i_k$-th lattice site (denoted as $x_{\ell,i_k}$) and upgrading $\mathbf{G}^{\sigma}$ from $(\mathbf{G}^{\sigma})_{[k-1]}$ to $(\mathbf{G}^{\sigma})_{[k]}$. Applying Eq.~(\ref{eq:rRatioLocal}), the ratio $r_{i_k,\sigma}$ for this update move reads $r_{i_k,\sigma}=1+\mathbf{y}_k^{\rm T}\mathbf{v}_{i_k}$, with the row vector $\mathbf{y}_k^{\rm T}=[\mathbf{1}_{N_s}-(\mathbf{G}^{\sigma})_{[k-1]}]_{i_k{\rm-row}}$ and the column vector $\mathbf{v}_{i_k}=(\boldsymbol{\Delta}^{\sigma})_{i_k{\rm-col}}$. Substituting the relation of $(\mathbf{G}^{\sigma})_{[k-1]}=(\mathbf{G}^{\sigma})_{[0]}-\sum_{m=1}^{k-1}\mathbf{x}_{m}\mathbf{y}_{m}^{\mathrm{T}}$ from Eq.~(\ref{eq:DelayG0}) into $\mathbf{y}_k^{\rm T}$, we can reach a recursive formula for $r_{i_k,\sigma}$ as
\begin{equation}\begin{aligned}
\label{eq:DelayRatio}
r_{i_k,\sigma} 
= 1 + \Delta_{i_ki_k}^{\sigma}\Big\{1-[(\mathbf{G}^{\sigma})_{[0]}]_{i_ki_k}
+ \sum_{m=1}^{k-1}(\mathbf{x}_m)_{i_k}(\mathbf{y}_m^{\rm T})_{i_k}\Big\},
\end{aligned}\end{equation}
with $(\mathbf{x}_m)_{i_k}$ and $(\mathbf{y}_m^{\rm T})_{i_k}$ representing the $i_k$-th element of $\mathbf{x}_m$ and $\mathbf{y}_m^{\rm T}$, respectively. We then calculate the configuration weight ratio as $W(\mathbf{X}^{\prime})/W(\mathbf{X})=[P(\mathbf{X}^{\prime})/P(\mathbf{X})]\times r_{i_k,\uparrow}r_{i_k,\downarrow}$ and decide whether to accept this update move using Eq.~(\ref{eq:DetailBalance}). In analogy with that in the local update, we define the column vector $\mathbf{x}_k=(\mathbf{G}^{\sigma})_{[k-1]}\mathbf{v}_{i_k}/r_{i_k,\sigma}=[(\mathbf{G}^{\sigma})_{[k-1]}]_{{i_k{\rm-col}}}\Delta_{i_ki_k}^{\sigma}/r_{i_k,\sigma}$. If this update is accepted, we need to compute and store the vectors $\mathbf{x}_k$ and $\mathbf{y}_k^{\rm T}$. Based on the relation between $(\mathbf{G}^{\sigma})_{[k-1]}$ and $(\mathbf{G}^{\sigma})_{[0]}$, the recursive expressions for $\mathbf{x}_k$ and $\mathbf{y}_k^{\rm T}$ are given by
\begin{equation}\begin{aligned}
\label{eq:DelayRgenk}
\mathbf{x}_k 
&= \frac{\Delta_{i_ki_k}^{\sigma}}{r_{i_k,\sigma}}\Big\{ [(\mathbf{G}^{\sigma})_{[0]}]_{i_k{\rm-col}} - \sum_{m=1}^{k-1}\mathbf{x}_m (\mathbf{y}_m^{\rm T})_{i_k} \Big\}, \\
\mathbf{y}_k^{\rm T} 
&= [\mathbf{1}_{N_s}-(\mathbf{G}^{\sigma})_{[0]}]_{i_k{\rm-row}} + \sum_{m=1}^{k-1}(\mathbf{x}_m)_{i_k}\mathbf{y}_m^{\rm T}.
\end{aligned}\end{equation}
On the other hand, if this update for $x_{\ell,i_k}$ is rejected, we simply proceed to examine the update of $x_{\ell,i_k+1}$, during which Eq.~(\ref{eq:DelayRatio}) is used to compute $r_{i_k+1,\sigma}$. Upon accumulating $n_d$ accepted updates, the $\mathbf{G}^{\sigma}$ matrix can be upgraded using
\begin{equation}\begin{aligned}
\label{eq:FinalDelay}
(\mathbf{G}^{\sigma})_{[n_d]} = (\mathbf{G}^{\sigma})_{[0]} - \sum_{m=1}^{n_d} \mathbf{x}_m\mathbf{y}_m^{\rm T} = (\mathbf{G}^{\sigma})_{[0]} - \mathbb{X}\mathbb{Y},
\end{aligned}\end{equation}
where $\mathbb{X}=(\mathbf{x}_1|\mathbf{x}_2|\cdots|\mathbf{x}_{n_d})$ is an $N_s\times n_d$ matrix, and $\mathbb{Y}=(\mathbf{y}_1|\mathbf{y}_2|\cdots|\mathbf{y}_{n_d})^{\rm T}$ is an $n_d\times N_s$ matrix. By applying the above procedure, the $n_d$ separate vector-vector outer products $\mathbf{x}_m\mathbf{y}_m^{\rm T}$ are replaced by a single matrix-matrix multiplication $\mathbb{X}\mathbb{Y}$. The latter is generally much more efficient, and thus accelerates the conventional local update. More specifically, the computational cost for the delayed update of $\mathbf{G}^{\sigma}$ using Eq.~(\ref{eq:FinalDelay}) scales as $\mathcal{O}(n_dN_s^2)$, which is the same as the local update process. The acceleration therefore originates from a substantially reduced prefactor, rather than a change in the asymptotic scaling. Beyond the leading cost of upgrading $\mathbf{G}^{\sigma}$, the {\it delayed update} requires additional effort in computing $r_{i_k,\sigma}$, $\mathbf{x}_k$, and $\mathbf{y}_k^{\rm T}$, compared to the local update. For $r_{i_k,\sigma}$, Eq.~(\ref{eq:DelayRatio}) can be rewritten as $r_{i_k,\sigma} = 1 + \Delta_{i_ki_k}^{\sigma}\{1-[(\mathbf{G}^{\sigma})_{[k-1]}]_{i_ki_k}\}$, which only involves the diagonal elements of $(\mathbf{G}^{\sigma})_{[k-1]}$. Hence, in practical simulations, we typically upgrade the diagonal elements of $\mathbf{G}^{\sigma}$ if the local update move is accepted, using $[(\mathbf{G}^{\sigma})_{[k]}]_{ii} = [(\mathbf{G}^{\sigma})_{[k-1]}]_{ii} - (\mathbf{x}_m)_{i}(\mathbf{y}_m^{\rm T})_{i}$ for $i=1,2,\cdots,N_s$, and subsequently upgrade the full $\mathbf{G}^{\sigma}$ matrix using Eq.~(\ref{eq:FinalDelay}) after every $n_d$ accepted moves. For $\mathbf{x}_k$ and $\mathbf{y}_k^{\rm T}$ in Eq.~(\ref{eq:DelayRgenk}), the evaluation of $\sum_m \mathbf{x}_m (\mathbf{y}_m^{\rm T})_{i_k}$ and $\sum_m (\mathbf{x}_m)_{i_k}\mathbf{y}_m^{\rm T}$ contribute a subleading computational cost of $\mathcal{O}(n_d(n_d-1)N_s)$.

\begin{figure}[t]
\includegraphics[width=0.976\columnwidth]{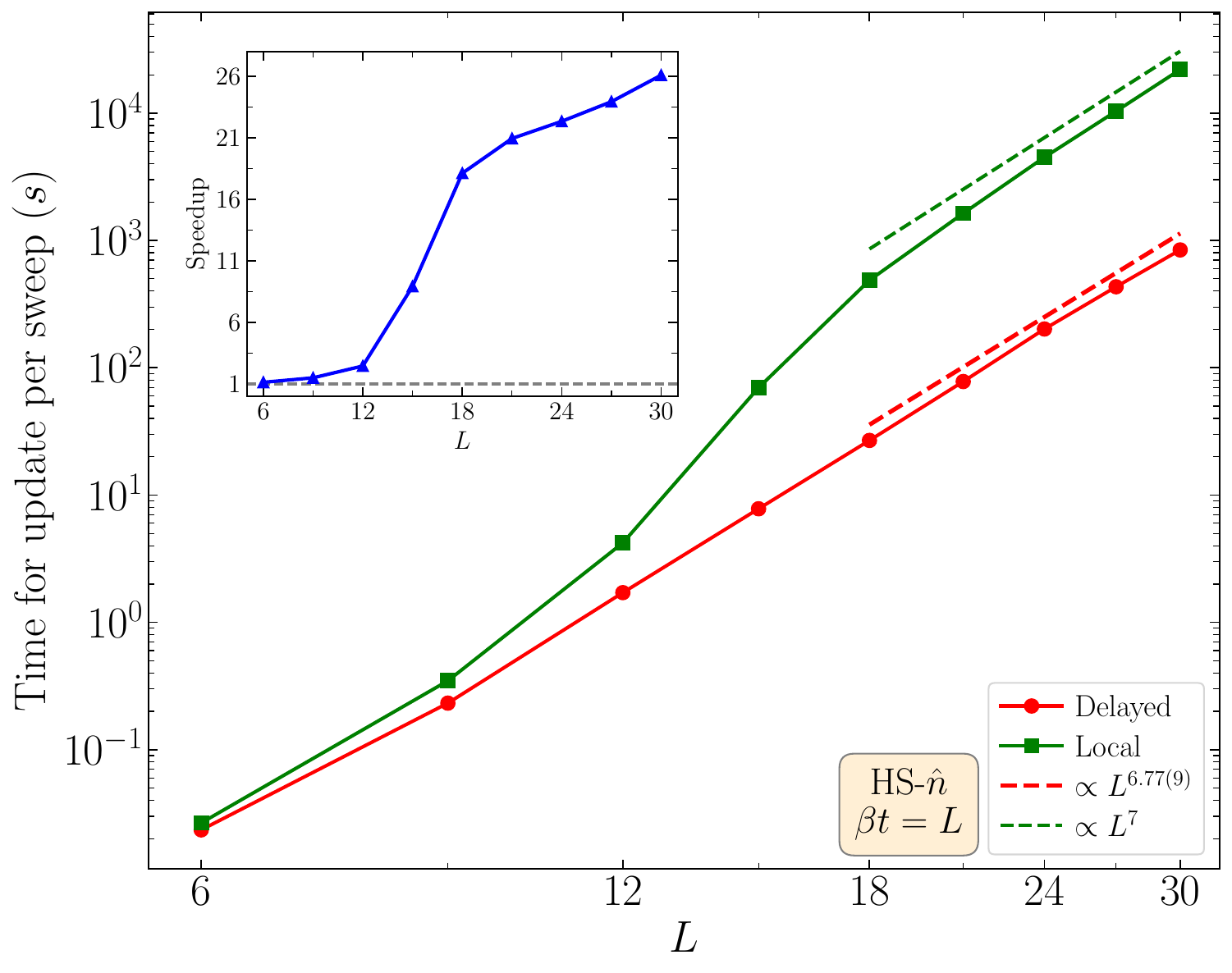}
\caption{Comparison of the average time for update per sweep (in seconds) between the local and delayed updates, as a function of $L$ in AFQMC simulations of the model~(\ref{eq:HubbModel}) with $U/t=3.70$ and $\beta t=L$. The HS-$\hat{n}$ transformation [Eq.~(\ref{eq:DensityHS})] is applied. The green dashed line plots the theoretical computational complexity ($\propto L^7$), while the red dashed line shows algebraic fit ($\propto L^{\alpha}$) to the consumed time of delayed update, yielding fitted exponents $\alpha$ slightly below $7$. The inset illustrates the speedup achieved by the delayed update compared to the local update.} 
\label{fig:fig03Delay}
\end{figure}

We refer the above process, which contains $n_d$ accepted local update moves, as a single-step {\it delayed update}. The update of all auxiliary fields on a given time slice (i.e., $\mathbf{x}_\ell$) can be divided into multiple successive delayed-update steps, satisfying $\sum n_d$$\sim$$N_s$ and thus leading to the overall computational complexity of $\mathcal{O}(N_s^3)$. Besides, the optimal value of $n_d$, which yields the largest speedup over the local update, can be determined through moderate simulation tests~\cite{Duhao2025}. If $n_d$ is too small, the efficiency gain of using matrix-matrix multiplication for $\mathbb{X}\mathbb{Y}$ in Eq.~(\ref{eq:FinalDelay}) is limited, as $\mathbb{X}$ and $\mathbb{Y}$ remain highly rectangular. As $n_d$ increases, however, the subleading cost $\mathcal{O}(n_d(n_d-1)N_s)$ constantly grows, and eventually becomes comparable to the leading cost $\mathcal{O}(n_dN_s^2)$, thereby offsetting the advantage of the {\it delayed update}. For the model~(\ref{eq:HubbModel}), our test simulations reveal its optimal value as $n_d\simeq\sqrt{N_s}$~\cite{LocalDelayed}. 

The applicability of the {\it delayed update} algorithm described above is not limited to on-site interactions. It can be readily extended to off-site interaction models~\cite{Sun2024} and spin-orbit coupled systems~\cite{Song2024,Duhao2025}. Building on this technique, a more advanced submatrix update algorithm has been proposed~\cite{Fanjie2025,Wang2026} to further accelerate the delayed update calculations in AFQMC, though it is not employed in the present work. A similar delayed update scheme has recently been developed for the ground-state AFQMC~\cite{Duhao2025} as well.

In Fig.~\ref{fig:fig03Delay}, we illustrate the speedup of the delayed update over the local update by comparing the average time consumed per sweep by these two update schemes for the model~(\ref{eq:HubbModel}). We set $U/t=3.7$ and $\beta t=L$, and apply the HS-$\hat{n}$ transformation [Eq.~(\ref{eq:DensityHS})]. As $L$ increases, the consumed time by the local update conforms well with the theoretical scaling $\mathcal{O}(L^7)$ (since $N_s=2L^2$ and $\beta t=L$). For the delayed update, an algebraic fit to the numerical data shows that the actual computational complexity is slightly slower than $L^7$. As shown in the inset, the corresponding speedup steadily grows with $L$, reaching $\sim$$26$ for $L=30$. We also find that applying the HS-$\hat{s}^z$ transformation for $U>0$ [Eq.~(\ref{eq:SzHS})], i.e., the real case, displays similar speedup (not shown). These results demonstrate that the delayed update technique provides a substantial and scalable speedup, making it particularly advantageous for large-scale AFQMC calculations.

The path propagation and auxiliary-field update are the two key components of the AFQMC algorithm. In conventional implementations, the matrix-matrix multiplication for the former and the local update for the latter typically account for more than $90\%$ of the total computational time~\cite{Sun2024}. The combination of the FFT and delayed update techniques developed in this work effectively addresses this bottleneck, achieving an overall speedup of more than an order of magnitude in AFQMC simulations for the model~(\ref{eq:HubbModel}) within $L\ge18$. As a result, we are able to efficiently simulate the $\beta t= L =42$ system with a computational cost of $\sim$$10^4$ CPU hours, while achieving a relative error of $\sim$$0.6\%$ for the mean-squared magnetization $m^2$. This extends the boundary of cutting-edge AFQMC simulations for the honeycomb Hubbard model.

\subsection{Twist-averaged boundary conditions}
\label{sec:TABC}

Finite-size effects pose a major challenge in resolving the Dirac quantum criticality in the model~(\ref{eq:HubbModel}), due to the gapless nature near the QCP. These effects consist of both single-particle and many-body contributions. While the many-body part is interaction-dependent and difficult to handle, the single-particle component can be significantly reduced by TABC~\cite{Song2025B,Lin2001,Qin2016b,Vitali2016,Haoxu2024}. Since the QCP is located at an intermediate interaction $U_c/t<4$~\cite{Toldin2015,Otsuka2016,Lang2025,Wang2026}, the single-particle finite-size effect arising from the noninteracting energy spectrum is expected to be sizable. Hence, TABC offers a practical and efficient route to suppress this effect and facilitates a reliable extraction of the critical properties. However, this TABC technique has not been utilized in previous ground-state AFQMC studies of the Dirac quantum criticality~\cite{Sorella1992,Paiva2005,Meng2010,Sorella2012,Assaad2013,Otsuka2016,Lang2025,Wang2026,Toldin2015,Yu2026}.

We first discuss the implementation of twisted boundary conditions (TBC) for the model~(\ref{eq:HubbModel}) in AFQMC simulations, focusing specifically on how the noninteracting Hamiltonian $\hat{H}_0$ is modified. Under TBC, when an electron hops across boundaries of the system, its wave function $\psi(\mathbf{r})$, satisfying $\hat{H}_0\psi(\mathbf{r})=E\psi(\mathbf{r})$, gains a phase as
\begin{equation}\begin{aligned}
\label{eq:TwistWvFct}
\psi(\mathbf{r} + L_{\alpha}\mathbf{a}_{\alpha}) = e^{{\rm i}\Theta_{\alpha}}\psi(\mathbf{r}),
\end{aligned}\end{equation}
where $L_{\alpha}$ is the linear system size along the $\mathbf{a}_{\alpha}$ direction (with $\alpha=1,2$) of the honeycomb lattice (see Fig.~\ref{fig:fig01honeycomb}), and $\Theta_{\alpha}=2\pi\theta_{\alpha}$ with $\theta_{\alpha}\in[0,1)$ is the twist angle. Here we apply a general $L_1\times L_2$ lattice for the discussion. This condition is equivalent to introducing fluxes $\theta_1$ and $\theta_2$ (in units of the flux quantum) through the two noncontractible loops of the torus [see Fig.~\ref{fig:fig01honeycomb}(b)]. Specifically, $\theta_1$ ($\theta_2$) corresponds to the flux threading the hole associated with the $\mathbf{a}_2$ ($\mathbf{a}_1$) direction. For convenience, a pure and uniform gauge with its vector potential $\mathbf{A}$ as constant, i.e., $\nabla\times\mathbf{A}=0$, is typically adopted to account for the flux, where the line integral of $\mathbf{A}$ around the entire torus along the $\mathbf{a}_{\alpha}$ direction equals the twist angle as
\begin{equation}\begin{aligned}
\label{eq:VecATBC}
\oint_{\alpha} \mathbf{A}(\mathbf{r}) \cdot d\bm{\ell} 
= \mathbf{A}\cdot L_{\alpha}\mathbf{a}_{\alpha}
= \Theta_{\alpha}
= 2\pi\theta_{\alpha}.
\end{aligned}\end{equation}
This equation leads to the relation $\mathbf{A}\cdot\mathbf{a}_{\alpha}=2\pi\theta_{\alpha}/L_{\alpha}$. Here we absorb the electron charge $e$ and $\hbar$ into the definition of $\mathbf{A}$. Although the $\psi(\mathbf{r})$ in Eq.~(\ref{eq:TwistWvFct}) satisfying TBC does not obey the conventional PBC, a local gauge transformation $\psi^{\prime}(\mathbf{r})=e^{-{\rm i}\mathbf{A}\cdot\mathbf{r}}\psi(\mathbf{r})$ transforms it into a periodic wave function as $\psi^{\prime}(\mathbf{r} + L_{\alpha}\mathbf{a}_{\alpha})=\psi^{\prime}(\mathbf{r})$. Accordingly, the eigenvalue equation becomes $\hat{H}_0^{\prime}\psi^{\prime}(\mathbf{r})=E\psi^{\prime}(\mathbf{r})$, with $\hat{H}_0^{\prime}=e^{-{\rm i}\mathbf{A}\cdot\mathbf{r}}\hat{H}_0e^{{\rm i}\mathbf{A}\cdot\mathbf{r}}$. Using $e^{-{\rm i}\mathbf{A}\cdot\mathbf{r}}\mathbf{p}e^{{\rm i}\mathbf{A}\cdot\mathbf{r}}=\mathbf{p}+\mathbf{A}$ (with $\mathbf{p}$ as the momentum), this transformation is equivalent to the substitution $\mathbf p\rightarrow\mathbf p+\mathbf A$ in the Hamiltonian. Upon projection onto Wannier orbitals and the second quantization, this yields the Peierls substitution~\cite{Peierls1933,Kohn1959,Wannier1962,Nenciu1991} for the hopping term as $t_{ij}c_{i\sigma}^+c_{j\sigma}^{}\to t_{ij}e^{-{\rm i}\phi_{ij}}c_{i\sigma}^+c_{j\sigma}^{}$ (with $t_{ij}$ as the hopping amplitude), where the phase factor $\phi_{ij}$ reads
\begin{equation}\begin{aligned}
\label{eq:PhaseFact}
\phi_{ij} 
= \int_{\mathbf{r}_i}^{\mathbf{r}_j}\mathbf{A}\cdot d\bm{\ell}
= \mathbf{A}\cdot (\mathbf{r}_j - \mathbf{r}_i).
\end{aligned}\end{equation}
Therefore, the remaining task is to calculate the Peierls phase factor $\phi_{ij}$ for various hopping processes.

For simple lattices, the distance vector $\delta\mathbf{r}=\mathbf{r}_j - \mathbf{r}_i$ in Eq.~(\ref{eq:PhaseFact}) can be naturally expressed as a specific superposition of the primitive lattice vectors $\mathbf{a}_{\alpha}$, and hence allows $\phi_{ij}$ to be directly obtained from Eq.~(\ref{eq:VecATBC}). On the other hand, for multi-sublattice geometry, e.g., the honeycomb lattice, the evaluation of $\phi_{ij}$ depends on how the coordinate vectors $\mathbf{r}$ of different sublattice sites are defined. There are typically two choices. The first neglects the relative displacements between different sublattices within the same unit cell and simply assigns a common $\mathbf{r}$ to all sublattices. For instance, the $f(\mathbf{k})$ function in the $\mathbf{h}(\mathbf{k})$ matrix in Eq.~(\ref{eq:h0kMat}) is obtained based on this choice, under which the Fourier transformation between $\mathbf{r}$ and $\mathbf{k}$ spaces is defined using the same expression for A- and B-sublattice operators as $c_{\mathbf{k}\sigma}^+ = N_c^{-1/2}\sum_{I}e^{+{\rm i}\mathbf{k}\cdot\mathbf{r}_I}c_{I\sigma}^+$ (with $c=a$ or $c=b$), where $I=1,2,\cdots,N_c$ labels the unit cells, with $\mathbf{r}_I$ denoting the coordinate of the $I$-th unit cell. The second choice, in contrast, accounts for the relative displacements across sublattices and uses the actual $\mathbf{r}$ for each sublattice. For this case, the Fourier transformation for $a_{\mathbf{k}\sigma}^+$ keeps unchanged whereas that for $b_{\mathbf{k}\sigma}^+$ is modified as $b_{\mathbf{k}\sigma}^+ = N_c^{-1/2}\sum_{I}e^{+{\rm i}\mathbf{k}\cdot(\mathbf{r}_I+\boldsymbol{\tau}_{\rm AB})}b_{\mathbf{r}_I,\sigma}^+$, with $\boldsymbol{\tau}_{\rm AB}=(0,1)$ as the relative displacement between A and B sublattices. Using this transformation, we arrive at the $f(\mathbf{k})$ function in Eq.~(\ref{eq:h0kMat}) as $f(\mathbf{k})=-te^{+{\rm i}\mathbf{k}\cdot\boldsymbol{\tau}_{\rm AB}}[1+e^{+{\rm i}\mathbf{k}\cdot(\mathbf{a}_1-\mathbf{a}_2)}+e^{-{\rm i}\mathbf{k}\cdot\mathbf{a}_2}]$. In terms of $\mathbf{h}(\mathbf{k})$, the above two choices are connected by a local gauge transformation $b_{\mathbf{k}\sigma}^+\to b_{\mathbf{k}\sigma}^+e^{+{\rm i}\mathbf{k}\cdot\boldsymbol{\tau}_{\rm AB}}$, under which the energy spectrum $\varepsilon_{\mathbf{k},\pm}=\pm|f(\mathbf{k})|$ remains unchanged. We refer to these two choices as Gauge-I and Gauge-II, respectively, in the subsequent evaluations of the Peierls phase factor.

We then focus on the three nearest-neighbor hoppings from A-sublattice sites, i.e., $-t\sum_{\eta=1}^3 e^{-{\rm i}\Phi_{\eta}}a_{\mathbf{r}_I,\sigma}^+b_{\mathbf{r}_I+\delta\mathbf{r}_{\eta},\sigma}^{}$, as labeled in Fig.~\ref{fig:fig01honeycomb}(a), and calculate the corresponding phase factor $\Phi_{\eta}$ based on Eqs.~(\ref{eq:VecATBC}) and~(\ref{eq:PhaseFact}). For Gauge-I, the distance vectors for the three hoppings are $\delta\mathbf{r}_1=\mathbf{0}$, $\delta\mathbf{r}_2=\mathbf{a}_1-\mathbf{a}_2$, and $\delta\mathbf{r}_3=-\mathbf{a}_2$. From Eq.~(\ref{eq:PhaseFact}), we readily obtain $\Phi_{\eta}=\mathbf{A}\cdot \delta\mathbf{r}_{\eta}$, which, together with Eq.~(\ref{eq:VecATBC}), gives
\begin{equation}\begin{aligned}
\label{eq:GaugeIPhase}
\Phi_1 = 0, \hspace{0.2cm}
\Phi_2 = \frac{2\pi\theta_1}{L_1} - \frac{2\pi\theta_2}{L_2}, \hspace{0.2cm}
\Phi_3 = - \frac{2\pi\theta_2}{L_2}.
\end{aligned}\end{equation}
For Gauge-II, the distance vectors are instead $\delta\mathbf{r}_1=(0,1)=(2\mathbf{a}_2-\mathbf{a}_1)/3$, $\delta\mathbf{r}_2=(\sqrt{3}/2,-1/2)=(2\mathbf{a}_1-\mathbf{a}_2)/3$, and $\delta\mathbf{r}_3=(-\sqrt{3}/2,-1/2)=-(\mathbf{a}_1+\mathbf{a}_2)/3$. Similarly, we obtain the phase factors as
\begin{equation}\begin{aligned}
\label{eq:GaugeIIPhase}
\Phi_1 &= -\frac{1}{3}\frac{2\pi\theta_1}{L_1} +\frac{2}{3}\frac{2\pi\theta_2}{L_2}, \\
\Phi_2 &= +\frac{2}{3}\frac{2\pi\theta_1}{L_1} -\frac{1}{3}\frac{2\pi\theta_2}{L_2}, \\
\Phi_3 &= -\frac{1}{3}\frac{2\pi\theta_1}{L_1} -\frac{1}{3}\frac{2\pi\theta_2}{L_2}.
\end{aligned}\end{equation}
Both sets of phase factors in Eqs.~(\ref{eq:GaugeIPhase}) and~(\ref{eq:GaugeIIPhase}) can be directly used to construct the $\mathbf{r}$-space hopping matrix $\mathbf{H}_0^{\sigma}$ under TBC, when the matrix-matrix multiplication formalism (see Sec.~\ref{sec:FFT}) is employed for the path propagation in AFQMC simulations. Furthermore, since Gauge-I and Gauge-II are connected by a gauge transformation, they must produce identical physical results in the simulations. 

\begin{figure}[t]
\includegraphics[width=1.00\columnwidth]{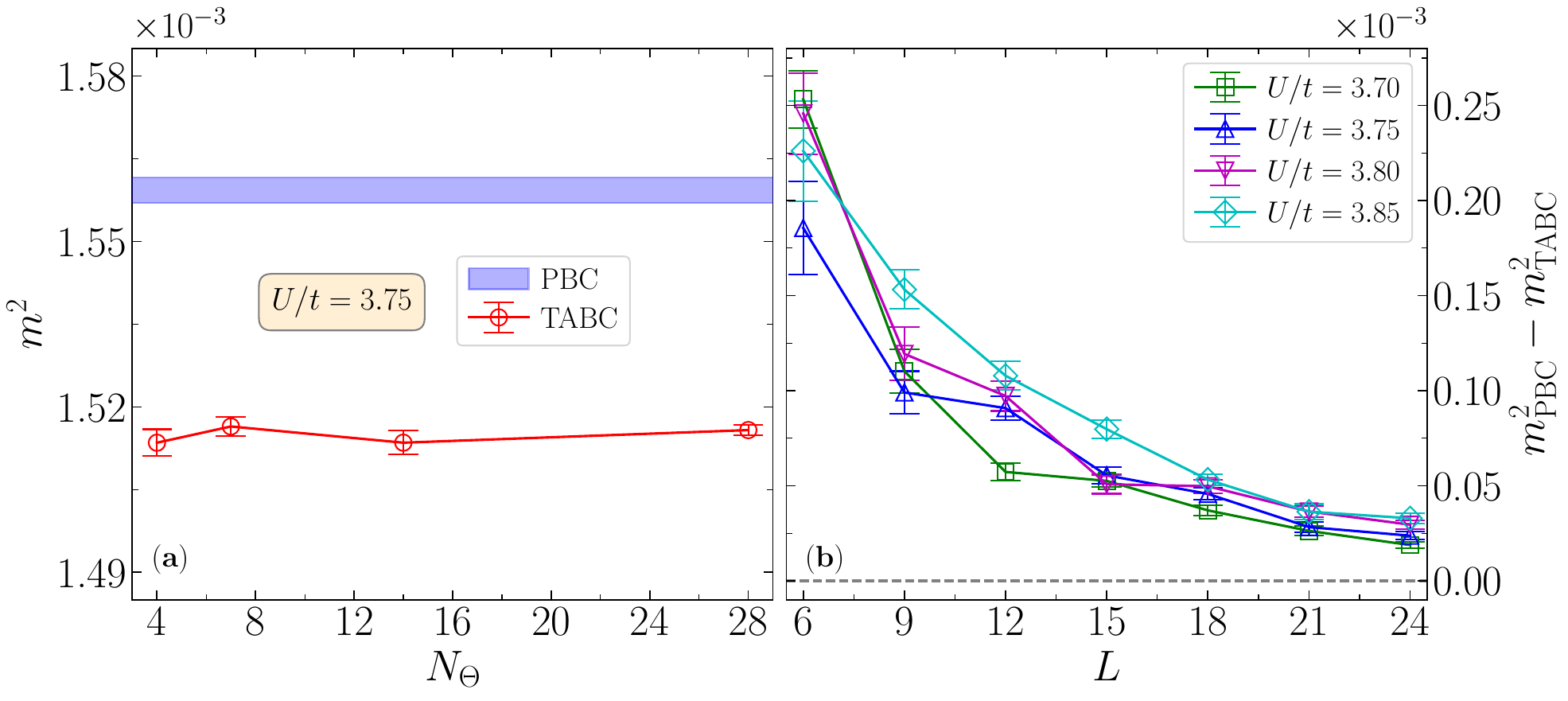}
\caption{(a) The convergence of mean-squared magnetization $m^2$ computed from AFQMC simulations with TABC using the number of twist angles $N_{\Theta}=4,7,14,28$. The horizontal blue shaded band indicates the PBC result of $m^2$ for reference. The simulation parameters are $\beta t=L=18$, and $U/t=3.75$. (b) The difference $(m_{\rm PBC}^2-m_{\rm TABC}^2)$, plotted as a function of $L$ ranging from $L=6$ to $L=24$, for $U/t=3.70,3.75,3.80,3.85$.}
\label{fig:fig04Twist}
\end{figure}

Once the TBC is applied in combination with the FFT technique, we need to evaluate $e^{-\Delta\tau\mathbf{h}(\mathbf{k})}$, which depends on the function $f(\mathbf{k})$. By performing the Fourier transformation and explicitly incorporating the phase factor $\Phi_{\eta}=\mathbf{A}\cdot\delta\mathbf{r}_{\eta}$ [from Eq.~(\ref{eq:PhaseFact})], we obtain a modified function $f^{\prime}(\mathbf{k})$, in which $\mathbf{k}$ in the original $f(\mathbf{k})$ is simply replaced by $\mathbf{k}+\mathbf{A}$, i.e., $f^{\prime}(\mathbf{k})=f(\mathbf{k}+\mathbf{A})$. Given Eq.~(\ref{eq:VecATBC}) and the equality $\mathbf{a}_{\alpha}\cdot\mathbf{b}_{\rho}=2\pi\delta_{\alpha\rho}$ (with $\alpha,\rho=1,2$), a practical solution for $\mathbf{A}$ is $\mathbf{A} = (\theta_1/L_1)\mathbf{b}_1 + (\theta_2/L_2)\mathbf{b}_2$, where $\mathbf{b}_1$ and $\mathbf{b}_2$ are $\mathbf{k}$-space primitive lattice vectors (see Sec.~\ref{sec:TheModel}). Therefore, in the FFT propagation, imposing TBC is equivalent to shifting the set of discrete $\mathbf{k}$ points from $\mathbf{k}=(l_1/L_1)\mathbf{b}_1+(l_2/L_2)\mathbf{b}_2$, where $l_1,l_2$ are integers satisfying $0\le l_1<L_1$ and $0\le l_2<L_2$, to
\begin{equation}\begin{aligned}
\label{eq:kpoints}
\mathbf{k} = \frac{l_1+\theta_1}{L_1}\mathbf{b}_1 + \frac{l_2+\theta_2}{L_2}\mathbf{b}_2,
\end{aligned}\end{equation}
while retaining the original form of $f(\mathbf{k})$. This illustrates that the effect of the Peierls phase under TBC can be absorbed into a shift of the discrete momentum grid. The same procedure applies to both the aforementioned Gauge-I and Gauge-II, for which the functions $f(\mathbf{k})$ differ only by a phase factor $e^{+{\rm i}\mathbf{k}\cdot\boldsymbol{\tau}_{\rm AB}}$. In practice, however, it is more convenient to adopt Gauge-I in the simulations, since Gauge-II requires an additional multiplication by $e^{+{\rm i}\mathbf{k}\cdot\boldsymbol{\tau}_{\rm AB}}$ for the B-sublattice component vector in the actual FFT calculations. 
 
For the honeycomb Hubbard model~(\ref{eq:HubbModel}), we apply the same set of $(\theta_1,\theta_2)$ for spin-up and spin-down channels when imposing TBC, to avoid the fermion sign problem in AFQMC simulations~\cite{Congjun2005}. The TABC calculation~\cite{Qin2016b} is implemented by performing independent TBC simulations for $N_\Theta$ sets of $(\theta_1,\theta_2)$, followed by averaging the resulting observables over all twist-angle configurations. Specifically, the ensemble of $(\theta_1,\theta_2)$ [with $\theta_1,\theta_2\in[0,1)$] values is generated using a 2D quasirandom number sequence~\cite{QuasiRandom}, which has a cumulative property and converges rapidly with $N_{\Theta}$~\cite{Qin2016b}. Theoretically, the value of $N_{\Theta}$ required to achieve satisfying statistics decreases with increasing $L$, as the effect of boundary conditions becomes progressively less important. In Fig.~\ref{fig:fig04Twist}(a), we show the convergence of the mean-squared magnetization $m^2$ computed from the TABC simulation with different $N_{\Theta}$ for $\beta t=L=18$ and $U/t=3.75$. We find that $N_{\Theta}\ge7$ is sufficient to achieve good statistics, and the difference between TABC and PBC results reveals the substantial single-particle finite-size effects. Based on such tests, we adopt $N_{\Theta}=14$ for $L\le39$ and $N_{\Theta}=8$ for $L=42$ whenever TABC is employed in this work. In Fig.~\ref{fig:fig04Twist}(b), we further plot the difference $(m_{\rm PBC}^2-m_{\rm TABC}^2)$ as a function of $L$ for four values of $U/t$, with $\beta t=L$. As expected, this difference decreases systematically with increasing $L$, but remains larger than the AFQMC error bars at $L=24$. In Appendix~\ref{sec:AppTABCPBC}, we also show that TABC simulations can effectively eliminate the oscillating results versus $L$ and thus allow for a better finite-size scaling analysis. 

Although the above calculations consider only nearest-neighbor (NN) hopping, the same procedure for computing Peierls phase factors under TBC applies to longer-range hoppings as well, e.g., the next-nearest and third-nearest-neighbor ones. When applying FFT, these hoppings similarly involve $\mathbf{k}$-space functions analogous to $f(\mathbf{k})$ for the NN hopping, and then TBC is incorporated via the substitution $\mathbf{k}\to\mathbf{k}+\mathbf{A}$. This methodology generalizes naturally to other multi-sublattice systems, e.g., the Kagome-lattice Hubbard model~\cite{Zhang2026,Ma2026,Rubem2026}. Taken together, our TABC simulations, which are built upon above TBC implementation and accelerated by the FFT technique (see Sec.~\ref{sec:FFT}) and delayed update algorithm (see Sec.~\ref{sec:Update}), provide a powerful and highly efficient AFQMC framework for the honeycomb Hubbard model.

\section{Numerical results}
\label{sec:HalfResults}

In this section, we present the AFQMC simulation results for the model defined in Eq.~(\ref{eq:HubbModel}). We begin by examining the critical properties of the Dirac quantum criticality via finite-size scaling analysis in Sec.~\ref{sec:Critical}, and subsequently corroborate the location of the QCP using independent physical observables in Sec.~\ref{sec:IndepExamQCP}. We also provide high-precision benchmark data for total energy and double occupancy around the QCP in Sec.~\ref{sec:EnergyDouble}. All simulations are performed using the finite-temperature AFQMC method, with the inverse temperature $\beta$ fixed at $\beta t=L$, and employ either TABC or PBC. 

\subsection{Critical properties from finite-size scaling}
\label{sec:Critical}

\begin{figure}[t]
\includegraphics[width=0.992\columnwidth]{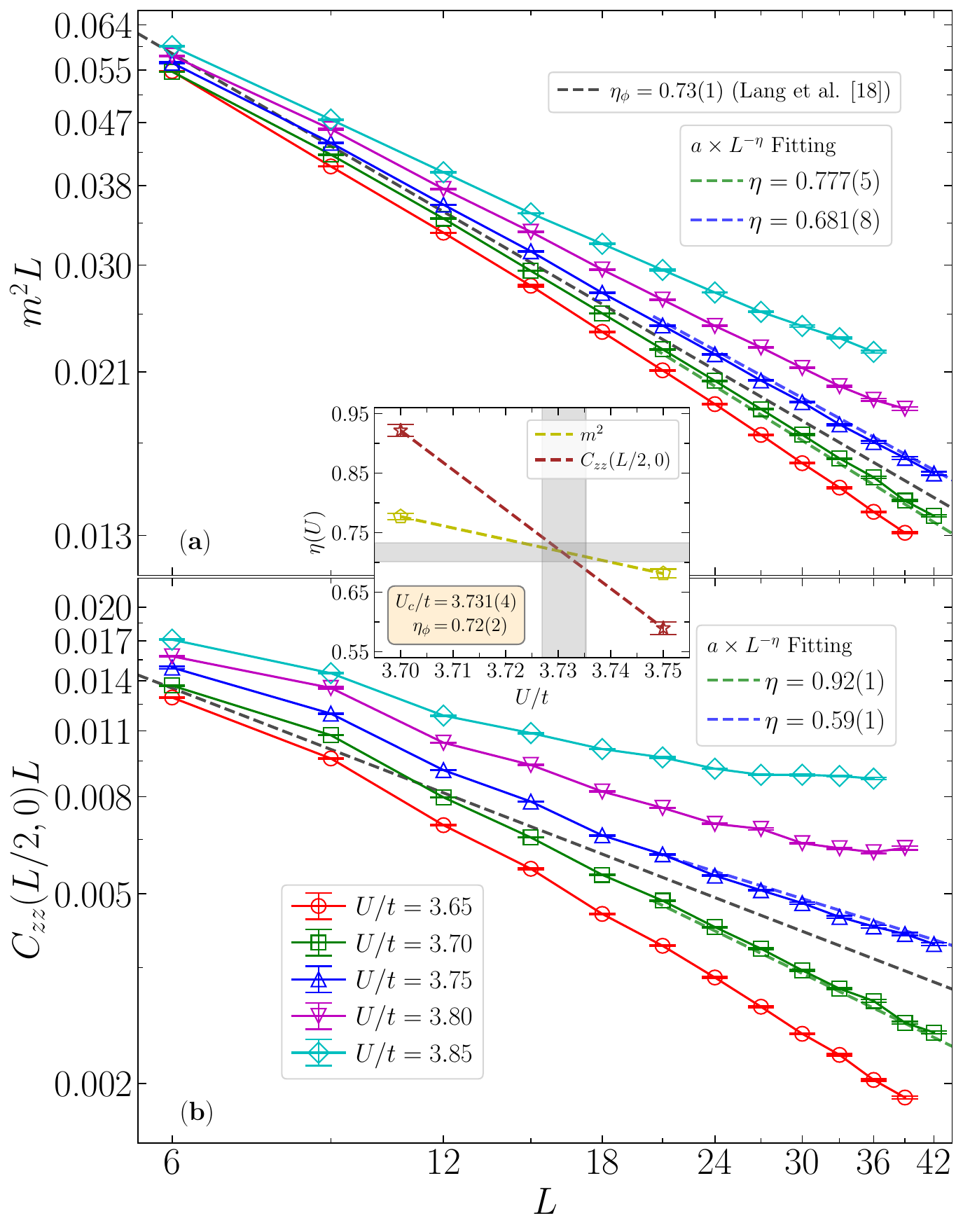}
\caption{Log-log plots of (a) the rescaled mean-squared magnetization $m^2L$, and (b) the rescaled real-space spin-spin correlation function $C_{zz}(L/2,0)L$, versus linear system size $L$, for the honeycomb Hubbard model at $U/t=3.65\sim3.85$. In both panels, green and blue dashed lines denote power-law fits (using $aL^{-\eta}$) to data with $L\geq21$ for $U/t=3.70$ and $3.75$, respectively, while the gray dashed line marks the reference bosonic anomalous dimension $\eta_{\phi}=0.73(1)$ from Ref.~\cite{Lang2025}. The inset plots $\eta$ extracted from $m^2$ and $C_{zz}(L/2,0)$ as a function of $U/t$, and via linear interpolations their intersection (gray shading) determines the QCP location $U_c/t=3.731(4)$ and the critical exponent $\eta_{\phi}=0.72(2)$. The AFQMC simulations are performed at $\beta t=L$ with TABC. }
\label{fig:fig05m2CrF}
\end{figure}


The standard approach to extract critical properties of the Dirac quantum criticality~\cite{Toldin2015,Otsuka2016,Lang2025,Wang2026} in the model~(\ref{eq:HubbModel}) is via finite-size scaling analysis of the mean-squared magnetization $m^2$ or the spin-spin correlation function $C_{zz}(\mathbf{r})$, evaluated at large and fixed $r/L$. For simplicity, we choose $\mathbf{r}=(L/2,0)$ for the latter. At the critical point $U_c/t$, both quantities should exhibit the universal scaling form $m^2\propto L^{-(1+\eta_{\phi})}$ and $C_{zz}(L/2,0)\propto L^{-(1+\eta_{\phi})}$, with $\eta_{\phi}$ being the bosonic anomalous dimension. Hence, the rescaled observables $m^2L$ and $C_{zz}(L/2,0)L$ obey the scaling relation of $\propto L^{-\eta_{\phi}}$ at $U=U_c$. 

\begin{figure}[t]
\includegraphics[width=1.00\columnwidth]{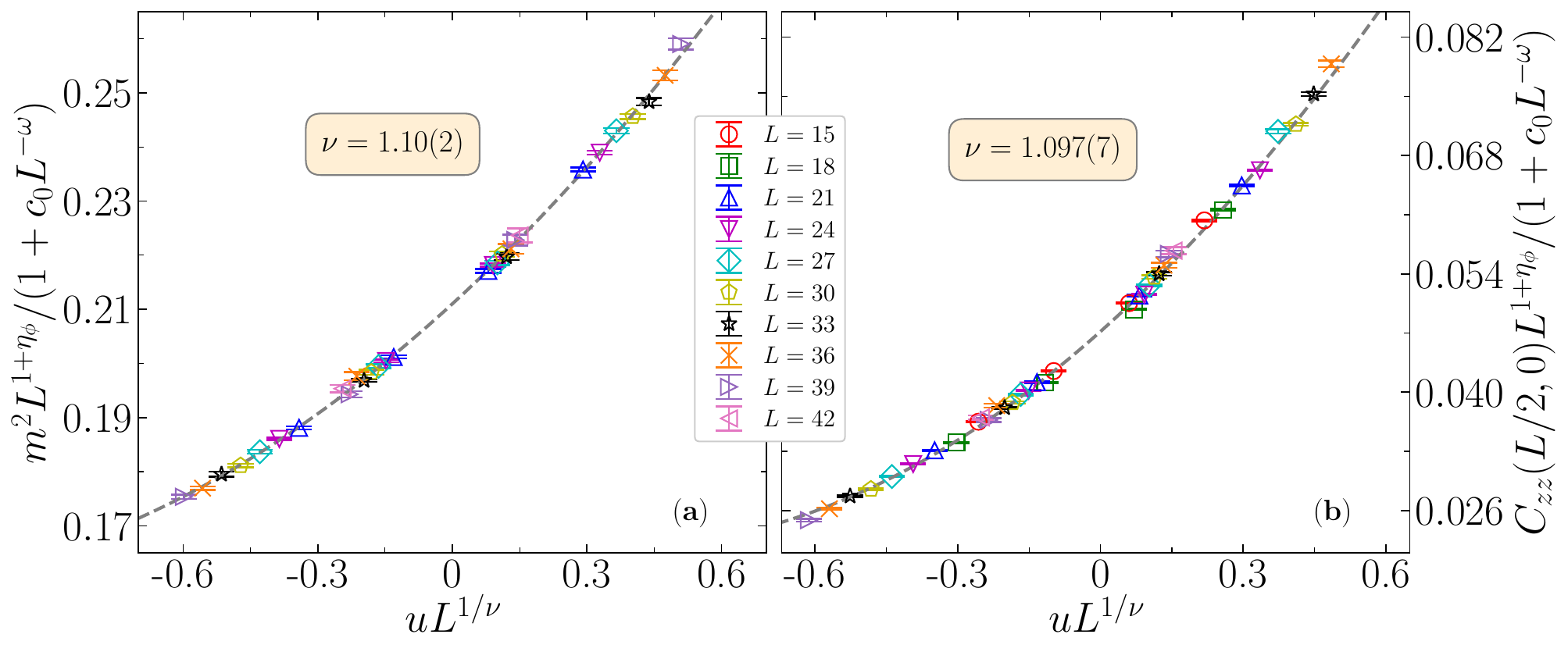}
\caption{The data collapse for (a) $m^2L^{1+\eta_{\phi}}/(1+c_0L^{-\omega})$, and (b) $C_{zz}(L/2,0)L^{1+\eta_{\phi}}/(1+c_0L^{-\omega})$, versus $uL^{1/\nu}$ according to Eq.~(\ref{eq:Collap1}), with fixed $U_c$ and $\eta_{\phi}$ from Fig.~\ref{fig:fig05m2CrF}. The fitting process uses the AFQMC data of $m^2$ and $C_{zz}(L/2,0)$ within $L=15\sim 42$ presented in Fig.~\ref{fig:fig05m2CrF}. Consistent results of the critical exponent $\nu=1.10(2)$ and $\nu=1.097(7)$ are obtained from $m^2$ and $C_{zz}(L/2,0)$, respectively. Gray dashed lines plot the function $f_s(uL^{1/\nu})$, which is approximated by a second-order polynomial in $uL^{1/\nu}$ during the fitting. }
\label{fig:fig06Collapse}
\end{figure}

In Fig.~\ref{fig:fig05m2CrF}, we present the numerical results of $m^2L$ and $C_{zz}(L/2,0)L$ as a function of $L$ near the QCP, with $U/t=3.65\sim 3.85$. Notably, we find that the data curves for $U/t=3.80$ and $3.85$ bend upward with increasing $L$, particularly evident in $C_{zz}(L/2,0)L$, which signals that these two interaction strengths lie in the AFMI regime. This observation places an upper bound $U_c/t<3.80$ on the QCP location, in contrast to $U_c/t=3.85(2)$ reported in Ref.~\cite{Otsuka2016}. Conversely, the downward curvature in the data at $U/t=3.65$ yields a lower bound $U_c/t>3.65$. At the intermediate $U/t=3.70$ and $3.75$, both $m^2L$ and $C_{zz}(L/2,0)L$ exhibit rather clean power-law decays, which actually suggests the vicinity of QCP. Given that the TABC is used, this behavior reveals strong finite-size effects of many-body nature, as perfect power-law scaling is only expected exactly at the critical point for sufficiently large $L$. We then take advantage of this behavior, and perform power-law fits (using $aL^{-\eta}$) for both sets of data. Interestingly, as shown in the inset, the $\eta$ values extracted from $m^2$ and $C_{zz}(L/2,0)$ show a reversal in their relative trend from $U/t=3.70$ to $3.75$. Since both quantities share the same $\eta=\eta_{\phi}$ at the critical point, this trend reversal essentially indicates $3.70<U_c/t<3.75$. By linearly interpolating $\eta(U)$ versus $U/t$, we identify the QCP as the crossing point of the two curves from $m^2$ and $C_{zz}(L/2,0)$, yielding $U_c/t=3.731(4)$ and $\eta_{\phi}=0.72(2)$. Hence, the two quantities offer complementary probes to examine the finite-size scaling behavior, and allow for cross-validation of the critical properties.

Compared to previous ground-state AFQMC studies, our estimates of $(U_c/t,\eta_{\phi})$ differ notably from the values $[3.85(2),0.49(3)]$ reported in Ref.~\cite{Otsuka2016}, but agree well with those extracted using $L=42\sim 72$ data in Ref.~\cite{Wang2026}. However, the latter work also presented slightly different results of $[3.664(5),0.79(2)]$ from additional extrapolations, whose reliability needs to be verified more carefully. The remaining discrepancies between our results and those previous studies likely stem from the algorithmic issues discussed in Sec.~\ref{sec:intro}, together with our use of TABC and a distinct finite-size scaling strategy. Meanwhile, our value $\eta_{\phi}=0.72(2)$ is in excellent agreement with $\eta_{\phi}=0.73(1)$ (plotted as a reference in Fig.~\ref{fig:fig05m2CrF}) reported in a very recent work~\cite{Lang2025} investigating the same universality class. That study, which employs the single Dirac cone formulation of the Hubbard model, shows substantially weaker finite-size effects and achieves well-converged result for $\eta_{\phi}$. Thus, the agreement justifies our procedure for estimating $U_c/t$ and $\eta_{\phi}$, which is reasonable for the system sizes accessed in our work. Nevertheless, we emphasize that, for sufficiently large $L$, the data at $U/t=3.70$ and $3.75$ should eventually bend downward and upward, respectively, reflecting their locations on opposite sides of the QCP. 

\begin{figure}[t]
\includegraphics[width=0.828\columnwidth]{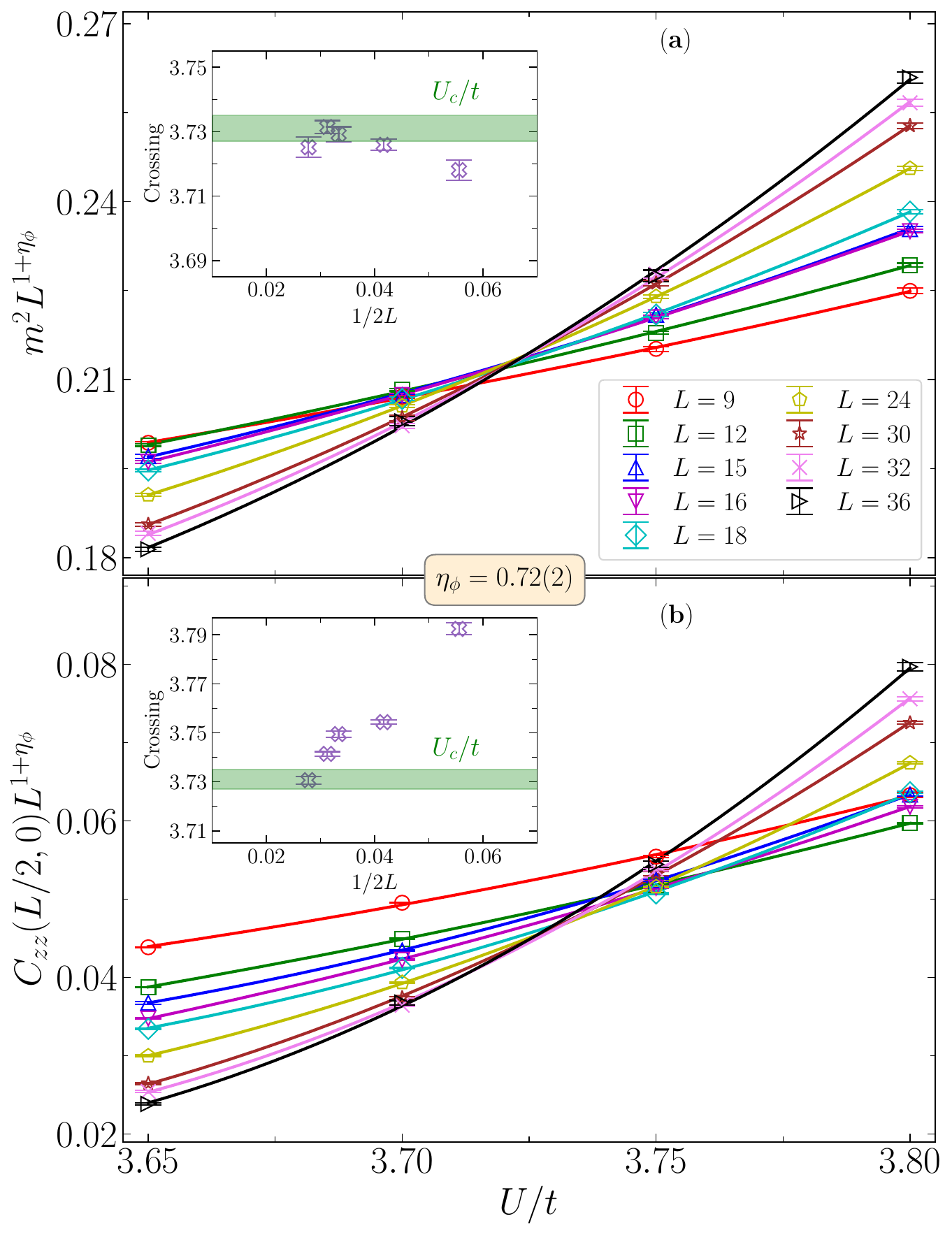}
\caption{Finite-size scaling plots for the rescaled quantities of (a) $m^2L^{1+\eta_{\phi}}$ and (b) $C_{zz}(L/2,0)L^{1+\eta_{\phi}}$, as a function of $U/t$ near the QCP, with $\eta_{\phi}=0.72(2)$. Solid curves are least-square fits by quadratic polynomials in $U/t$ to the AFQMC data presented in Fig.~\ref{fig:fig05m2CrF}. The insets plot the crossings of successive $(L,2L)$ pairs versus $1/(2L)$, with the green shaded bands marking the QCP location $U_c/t=3.731(4)$ from Fig.~\ref{fig:fig05m2CrF}.}
\label{fig:fig07Crossing}
\end{figure}

\begin{figure}[t]
\includegraphics[width=0.890\columnwidth]{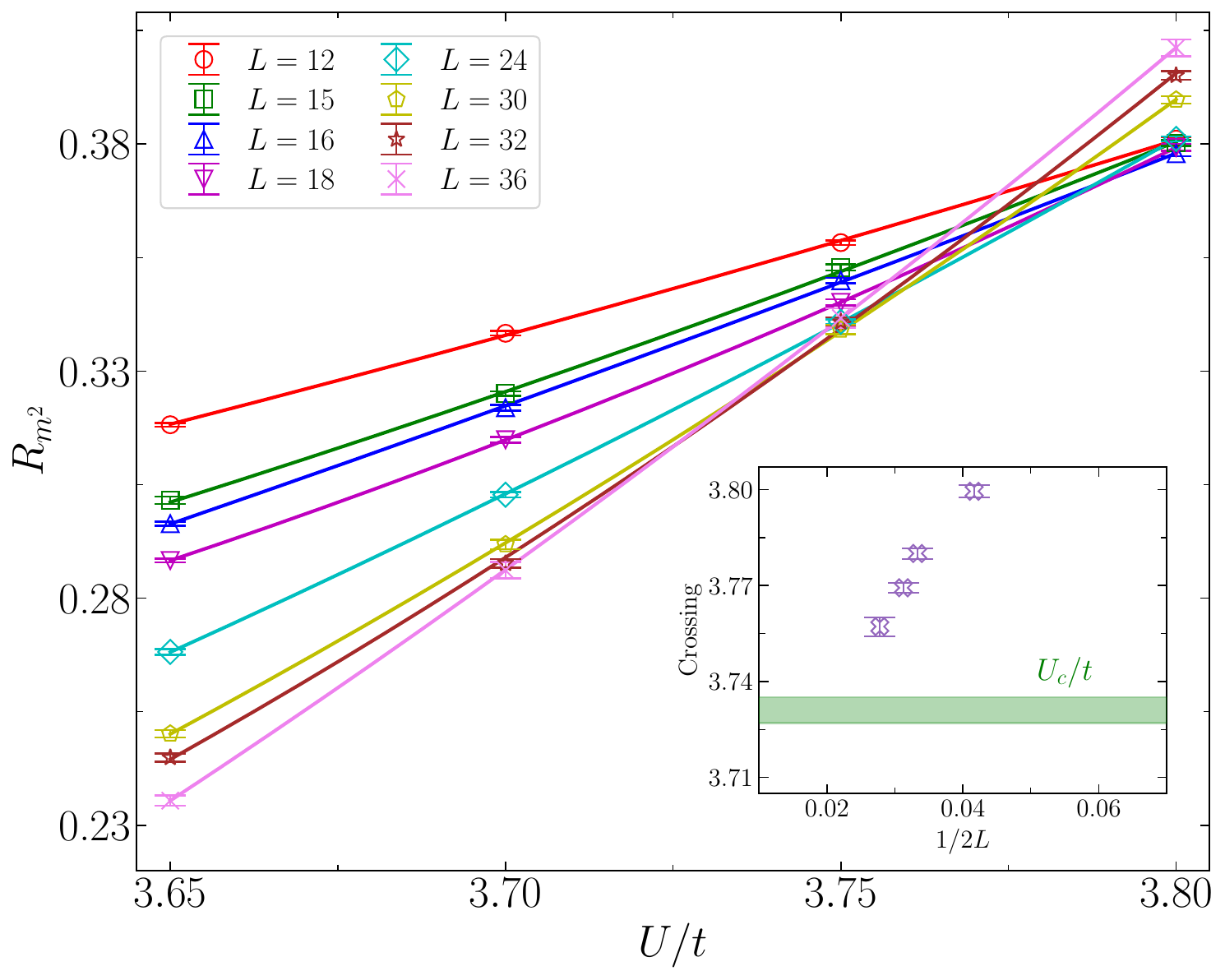}
\caption{The correlation ratio $R_{m^2}$ computed via Eq.~(\ref{eq:CrfRatio}) as a function of $U/t$ near the QCP. Solid curves are least-square fits to the AFQMC data by quadratic polynomials in $U/t$. The inset plots the crossings of successive $(L,2L)$ pairs versus $1/(2L)$, with the green shaded bands marking the QCP location $U_c/t=3.731(4)$ determined from Fig.~\ref{fig:fig05m2CrF}. The AFQMC simulations are performed at $\beta t=L$ with TABC.}
\label{fig:fig08Ratio}
\end{figure}

We next turn to the standard and widely applied approach to determine the critical properties, i.e., the data collapse analysis, which is essentially a multi-parameter fitting procedure. In the critical regime, both $m^2$ and $C_{zz}(L/2,0)$ should satisfy the following finite-size scaling ansatz~\cite{Toldin2015,Otsuka2016,Wang2026} as
\begin{equation}\begin{aligned}
\label{eq:Collap1}
O(u,L)=L^{-(1+\eta_{\phi})}(1+c_0L^{-\omega})f_s(uL^{1/\nu}),
\end{aligned}\end{equation}
where $u=(U-U_c)/U_c$, $\nu$ is the correlation length critical exponent, and $f_s(x)$ is a scaling invariant (universal) function with $f_s(x=0)$ as a finite constant. The correction term $c_0L^{-\omega}$ (typically with $\omega>0$) is included to account for the subleading finite-size effect around the critical point. In our calculations, we approximate $f_s(uL^{1/\nu})$ by a second-order polynomial in $uL^{1/\nu}$ and perform least-square fits to the AFQMC data of $m^2$ and $C_{zz}(L/2,0)$ shown in Fig.~\ref{fig:fig05m2CrF}. To effectively reduce the instability in the fitting, we fix $(U_c,\eta_{\phi})$ at our reliably estimated values $U_c/t=3.731(4)$ and $\eta_{\phi}=0.72(2)$, and leave $\nu$, $c_0$, and $\omega$ as free fitting parameters. In Fig.~\ref{fig:fig06Collapse}, we show the data collapse plots for both quantities with $L=15\sim 42$, which achieve well consistent results for $\nu$ as $\nu=1.10(2)$ from $m^2$ and $\nu=1.097(7)$ from $C_{zz}(L/2,0)$. These estimates of $\nu$ also conform well with $\nu=1.11(5)$ reported in Ref.~\cite{Wang2026}. 

\begin{figure}[t]
\includegraphics[width=1.00\columnwidth]{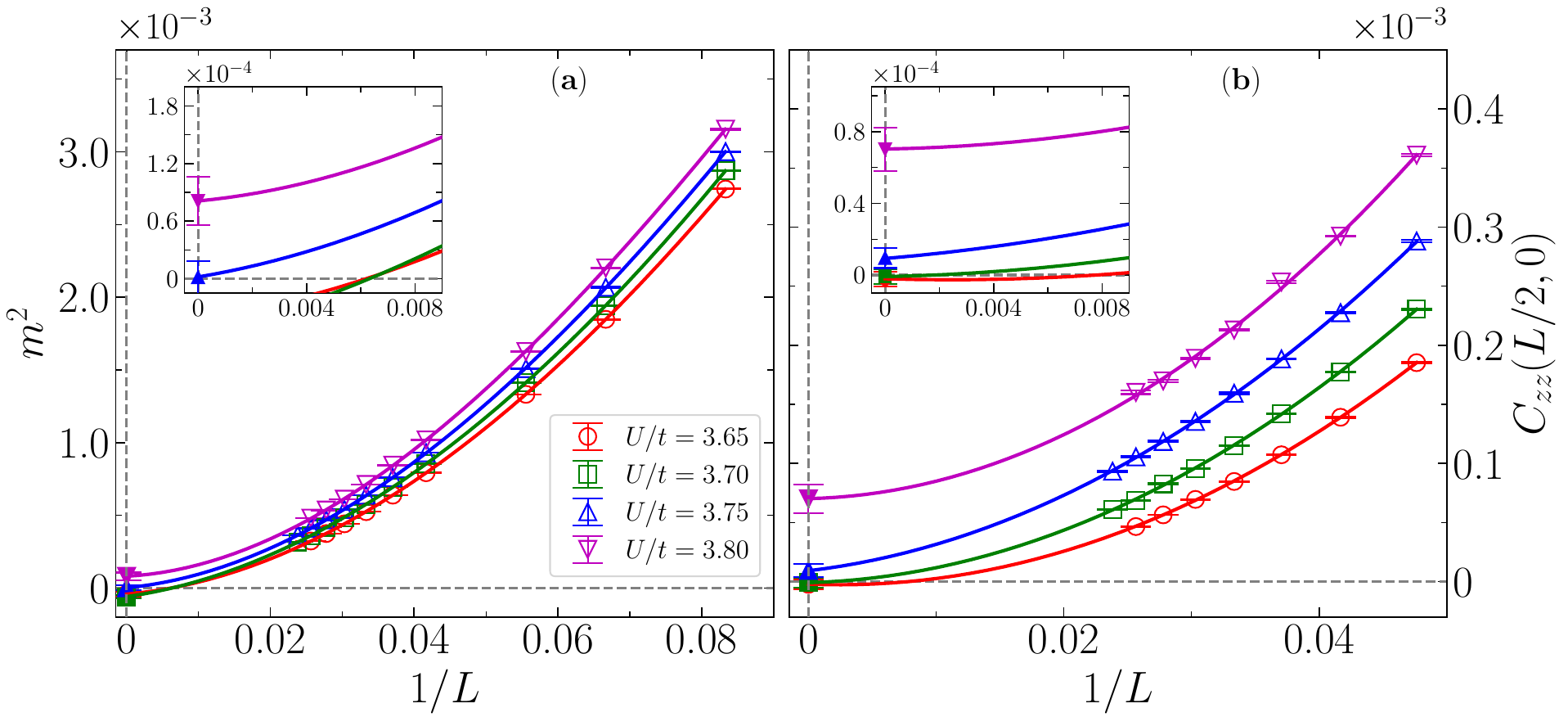}
\caption{Extrapolations of the mean-squared magnetization $m^2$ and the real-space spin-spin correlation $C_{zz}(L/2,0)$ as a function of $1/L$, for $U/t=3.65\sim 3.80$. Solid curves are least-square fits to the data by cubic (quadratic) polynomials in $1/L$. Insets show enlarged plots around $1/L\to0$. The AFQMC simulations are performed at $\beta t=L$ with TABC.}
\label{fig:fig09m2CrFExtrap}
\end{figure}

Based on the results in Figs.~\ref{fig:fig05m2CrF} and~\ref{fig:fig06Collapse}, we can further check the robustness of the extracted critical properties through several related routes. {\it First}, the rescaled quantities $m^2L^{1+\eta_{\phi}}$ and $C_{zz}(L/2,0)L^{1+\eta_{\phi}}$ versus $U/t$ should cross for different system sizes, with the crossings converging to the QCP for sufficiently large $L$. In Fig.~\ref{fig:fig07Crossing}, we plot these results near the QCP and determine the crossings of successive $(L,2L)$ pairs from polynomial fits to the AFQMC data. As illustrated in the insets, the crossing points from $m^2$ converge rapidly and saturate to our estimated $U_c/t$, while those from $C_{zz}(L/2,0)$ show clear shift with increasing $L$ and eventually approach the correct $U_c/t$ region by the $L=18$ pair. {\it Second}, the correlation ratio $R_{m^2}$ computed via Eq.~(\ref{eq:CrfRatio}) should exhibit the same scaling behavior as $m^2L^{1+\eta_{\phi}}$ and $C_{zz}(L/2,0)L^{1+\eta_{\phi}}$, and is likewise expected to show crossings when plotted versus $U/t$. In Fig.~\ref{fig:fig08Ratio}, we show the results of $R_{m^2}$ versus $U/t$, and plot the crossings versus $1/(2L)$ in the inset. Here, the crossing points exhibit a more pronounced shift toward the correct $U_c/t$ as $L$ increases, while the value from the $L=18$ pair still shows a visible deviation. This behavior reveals strong finite-size effects in $R_{m^2}$, consistent with observations in Refs.~\cite{Lang2025,Wang2026}. Nevertheless, an extrapolation of the crossing points still yields a comparable $U_c/t$. {\it Third}, the Dirac quantum criticality separates the DSM and AFMI phases, thus the QCP location is associated with the onset of a nonzero AFM order parameter $m_{\rm AFM}$ in the thermodynamic limit (TDL) for $U>U_c$. Theoretically, whether long-range AFM order exists can be determined by extrapolating $m^2$ and $C_{zz}(L/2,0)$ to TDL, where both quantities converge to the same value, i.e., $m_{\rm AFM}^2=\lim_{L\to\infty}m^2=\lim_{L\to\infty}C_{zz}(L/2,0)$. In Fig.~\ref{fig:fig09m2CrFExtrap}, we present the results of $m^2$ and $C_{zz}(L/2,0)$ versus $1/L$, and show the extrapolations to the TDL by polynomial fits. At $U/t=3.80$, the nonzero intercepts from both fits are consistent with each other, indicating the existence of long-range AFM order. At $U/t=3.65$ and $3.70$, both quantities extrapolates to zero (the negative intercepts for $m^2$ are expected to vanish once the data with larger $L$ are included), signaling the absence of AFM order. At $U/t=3.75$, however, the two fits yield slightly different results, as $C_{zz}(L/2,0)$ suggests a tiny but nonzero $m_{\rm AFM}$ while $m^2$ indicates a vanishing one. This disagreement may be attributed to the distinct finite-size effects in the two quantities, as well as the ambiguity in choosing the correct fitting ansatz. In principle, such a fitting procedure can only locate the QCP within a relatively narrow window~\cite{Otsuka2016}, e.g., $U_c/t=3.70\sim 3.80$ in Fig.~\ref{fig:fig09m2CrFExtrap}, rather than achieving a precise $U_c/t$. Collectively, although all three routes discussed above and depicted in Figs.~\ref{fig:fig07Crossing}-\ref{fig:fig09m2CrFExtrap} are related to spin-spin correlations, they adopt different analysis strategies, i.e., finite-size crossings, the correlation ratio, and TDL extrapolations, thus offering complementary validations for $(U_c/t,\eta_{\phi})$ determined in Fig.~\ref{fig:fig05m2CrF}. 

\begin{figure}[t]
\includegraphics[width=0.99\columnwidth]{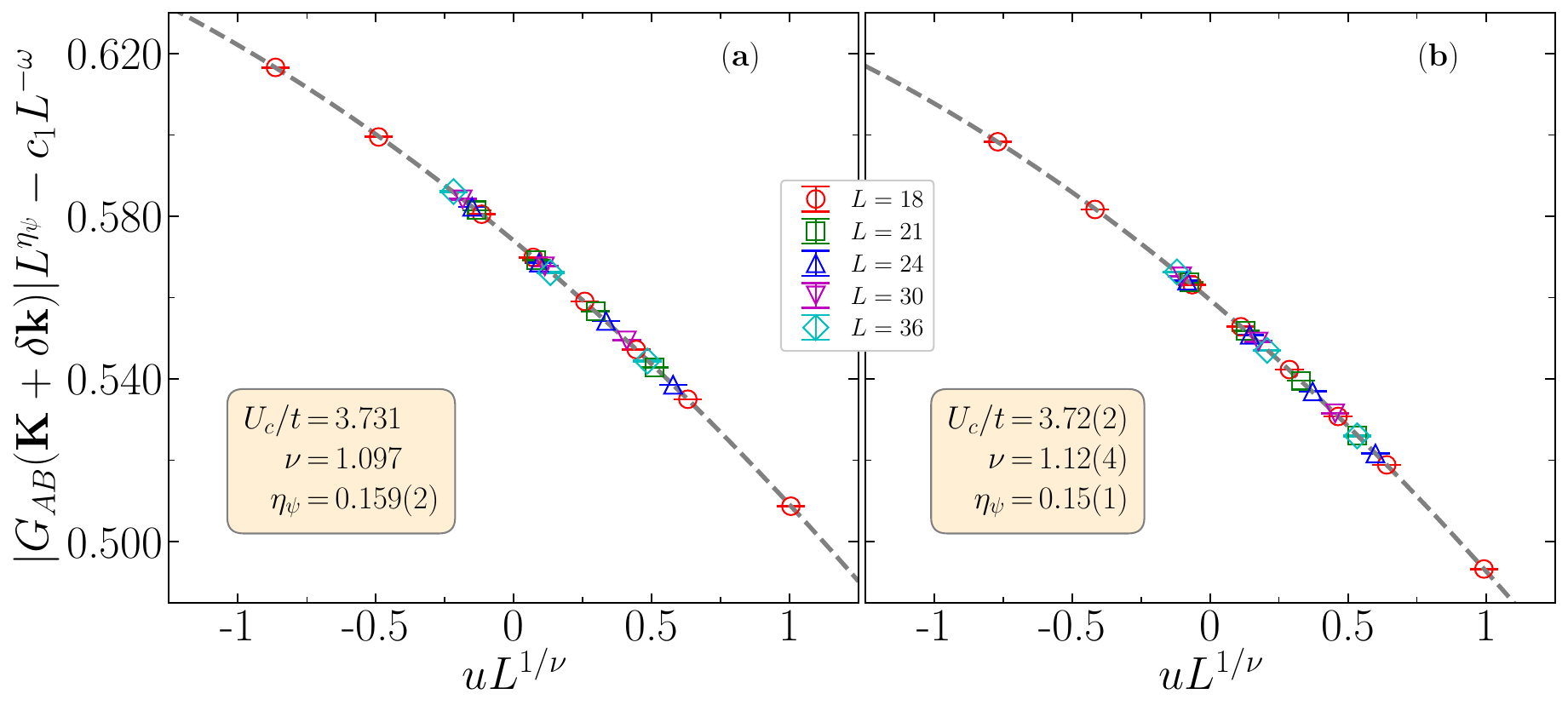}
\caption{
The data collapse for $(|G_{\rm AB}(\mathbf{K}+\delta \mathbf{k})|L^{\eta_{\psi}}-c_1L^{-\omega})$, versus $uL^{1/\nu}$ according to Eq.~(\ref{eq:Gansatz}). In panel (a), fixed $U_c=3.731$ and $\nu=1.097$ are used, while in (b) $U_c$ and $\nu$ are treated as fitting parameters. The corresponding results of $\eta_{\psi}$, $U_c/t$, and $\nu$ from the fits are also included. Gray dashed lines plot the universal function $f_g(uL^{1/\nu})$, which is approximated by a second-order polynomial in $uL^{1/\nu}$ during the fitting. The AFQMC simulations are performed at $\beta t=L$ with PBC.
}
\label{fig:fig10DataClapFermi}
\end{figure}

In addition to the aforementioned critical properties, the fermionic anomalous dimension $\eta_{\psi}$ is another key ingredient of the Dirac quantum criticality in the model~(\ref{eq:HubbModel}). Practically, it can be extracted via finite-size scaling analysis of the modulus of off-diagonal single-particle Green's function $G_{\rm AB}(\mathbf{K}+\delta\mathbf{k})=\langle a_{\mathbf{K}+\delta\mathbf{k},\sigma}^{+}
b_{\mathbf{K}+\delta\mathbf{k},\sigma}^{}\rangle$, where $\mathbf{K}$ denotes the Dirac point and $\delta\mathbf{k}$ is the minimal momentum around $\mathbf{K}$. In our calculations, we average over the two spin components and the two inequivalent Dirac points. We note that this quantity must be evaluated under PBC in the AFQMC simulations, since TBC shifts the discrete momentum grid. Moreover, $G_{\rm AB}(\mathbf{K}+\delta\mathbf{k})$ is directly connected to the jump in energy-resolved momentum distribution $n(\varepsilon_{\mathbf{k}})$ across the Fermi level, which was used to compute $\eta_{\psi}$ in Ref.~\cite{Otsuka2016}. The modulus $|G_{\rm AB}(\mathbf{K}+\delta\mathbf{k})|$ should follow the finite-size scaling ansatz as~\cite{Lang2025,Wang2026,Gansatz}
\begin{equation}\begin{aligned}
\label{eq:Gansatz}
|G_{\rm AB}(\mathbf{K}+\delta \mathbf{k})|=L^{-\eta_{\psi}}\big[f_g(uL^{1/\nu})+c_1L^{-\omega}\big],
\end{aligned}\end{equation}
where $u=(U-U_c)/U_c$, $f_g(x)$ is a scaling invariant function, and $c_1L^{-\omega}$ denotes the subleading correction term. We approximate $f_g(uL^{1/\nu})$ by a second-order polynomial in $uL^{1/\nu}$ and perform least-square fits to the AFQMC data of $|G_{\rm AB}(\mathbf{K}+\delta \mathbf{k})|$. The resulting data collapse plots for the rescaled quantity $(|G_{\rm AB}(\mathbf{K}+\delta \mathbf{k})|L^{\eta_{\psi}}-c_1L^{-\omega})$ are shown in Fig.~\ref{fig:fig10DataClapFermi}, for both cases where $(U_c/t=3.731,\nu=1.097)$ are fixed and where they are left as free fitting parameters. The first case yields $\eta_{\psi}=0.159(2)$, while the second case achieves $\eta_{\psi}=0.15(1)$, $U_c/t=3.72(2)$, and $\nu=1.12(4)$. This demonstrates that, allowing $(U_c/t,\nu)$ to vary produces $\eta_{\psi}$ consistent with the fixed case, and the fitted $(U_c/t,\nu)$ in the second case are in good agreement with those obtained in Figs.~\ref{fig:fig05m2CrF} and~\ref{fig:fig06Collapse}. We therefore take the averaged value $\eta_{\psi}=0.155(6)$ as our final estimate. In previous ground-state AFQMC studies, the computed $\eta_{\psi}$ varies considerably, ranging from $0.09(1)$ in Ref.~\cite{Lang2025} to $0.20(2)$ in Ref.~\cite{Otsuka2016}. Among these, our result $\eta_{\psi}=0.155(6)$ lies closest to the value $0.1888(40)$, as reported in the most recent work~\cite{Wang2026}. The remaining deviation may stem from the different finite-size scaling procedures employed, i.e., scaling with respect to $uL^{1/\nu}$ in our work versus $R_{m^2}$ in Ref.~\cite{Wang2026}.

Combining the results in Figs.~\ref{fig:fig05m2CrF},~\ref{fig:fig06Collapse} and~\ref{fig:fig10DataClapFermi}, our finite-temperature AFQMC calculations (applying $\beta t=L$) yield the critical properties for the honeycomb Hubbard model~(\ref{eq:HubbModel}) as $U_c/t=3.731(4)$, $\nu=1.097(7)$, $\eta_{\phi}=0.72(2)$, and $\eta_{\psi}=0.155(6)$. These results are very close to those reported in previous ground-state AFQMC studies, particularly Refs.~\cite{Otsuka2016,Lang2025,Wang2026}, where the agreements and discrepancies have been discussed in detail. We have also checked the $(U_c/t,\eta_{\phi})$ results from several perspectives related to spin-spin correlations. In next subsection, we focus on examinations of $U_c/t$ using other quantities beyond the spin sector. 

\begin{figure}[t]
\includegraphics[width=1.00\columnwidth]{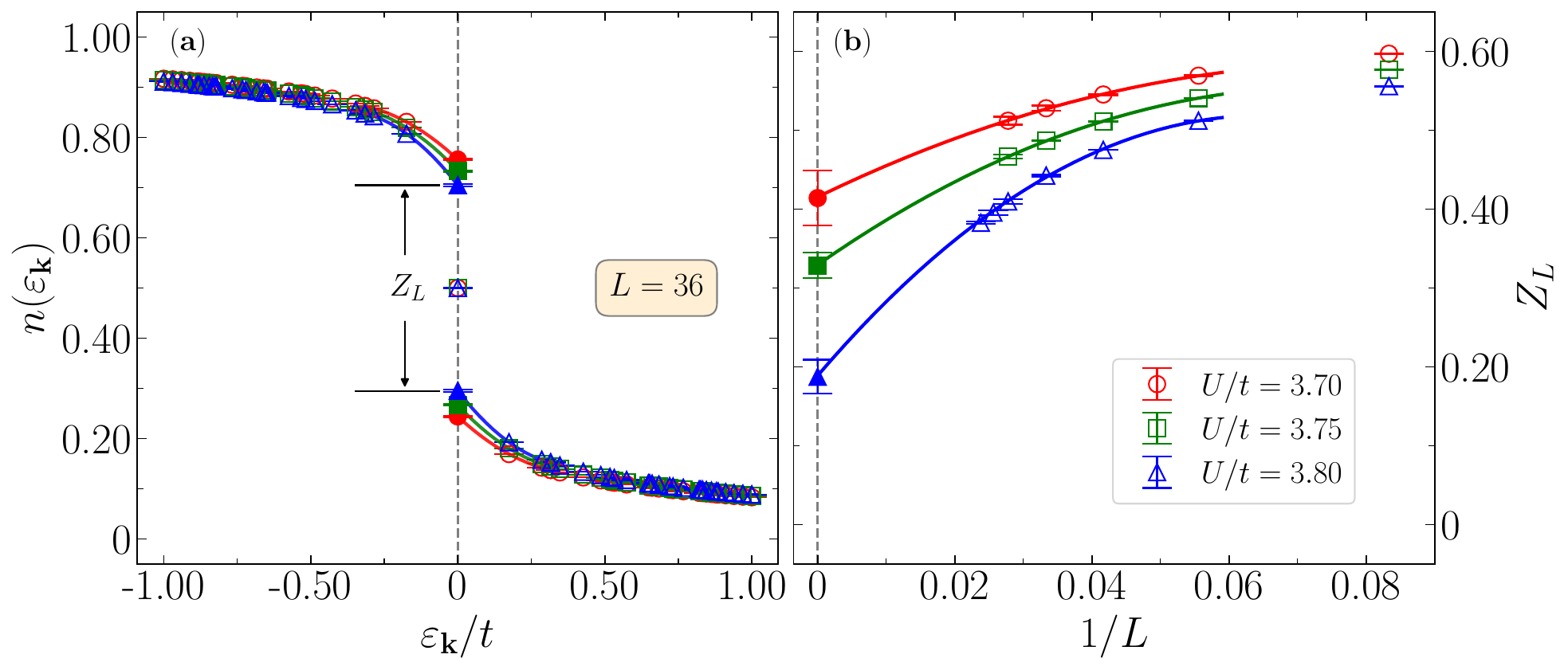}
\caption{
The quasiparticle weight $Z_L$ computed by Eq.~(\ref{eq:QpWeight}). Panel (a) illustrates the calculation of finite-size $Z_L$ by plotting the energy-resolved momentum distribution $n(\varepsilon_{\mathbf{k}})$ near the Fermi level (the gray vertical dashed line at $\varepsilon_{\mathbf{k}}=0$) for the model~(\ref{eq:HubbModel}) with $L=36$. Solid curves are least-square fits to the three data points closest to $\varepsilon_{\mathbf{k}}=0$ on both sides, with $\varepsilon_{\mathbf{k}}<0$ and $\varepsilon_{\mathbf{k}}>0$, using quadratic polynomials in $\varepsilon_{\mathbf{k}}$. $Z_L$ is marked for $U/t=3.80$. Panel (b) shows the extrapolation of $Z_{L}$ to TDL for $U/t=3.70$, $3.75$, and $3.80$, using least-square fits with quadratic polynomials in $1/L$. The AFQMC simulations are performed at $\beta t=L$ with PBC.}
\label{fig:fig11Qpwght}
\end{figure}

\subsection{Independent examinations of the QCP location}
\label{sec:IndepExamQCP}

\begin{figure}[t]
\includegraphics[width=0.99\columnwidth]{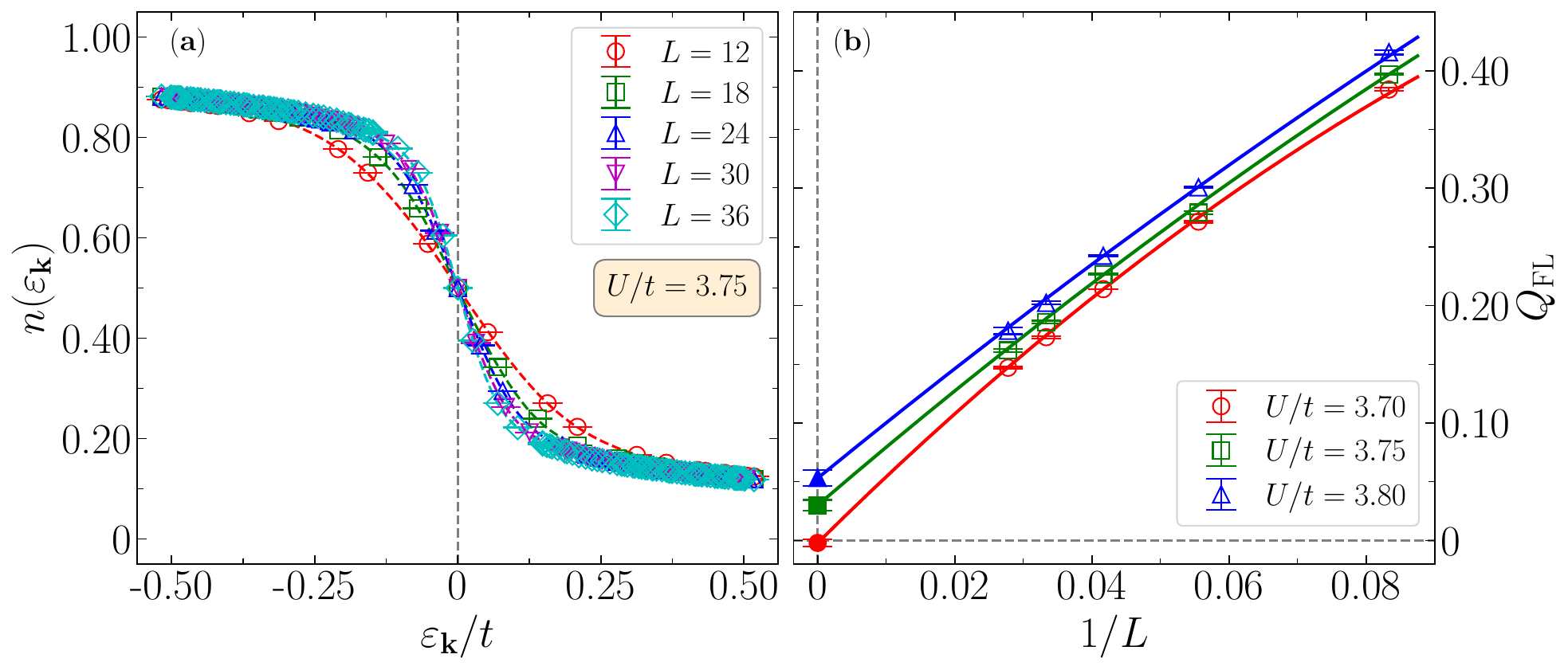}
\caption{
Fermi-liquid parameter $Q_{\rm FL}$ computed by Eq.~(\ref{eq:FLP}). Panel (a) plots the results of $n(\varepsilon_{\mathbf{k}})$, combined from PBC and TBC with two sets of specially chosen twist angles, near the Fermi level (the gray vertical dashed line at $\varepsilon_{\mathbf{k}}=0$), for the model~(\ref{eq:HubbModel}) with $U/t=3.75$ and $L=12\sim 36$. Dashed lines are the cubic-spline interpolations of the $n(\varepsilon_{\mathbf{k}})$ data, from which the finite-size $Q_{\rm FL}$ can be extracted. Panel (b) shows the extrapolation of $Q_{\mathrm{FL}}$ to the TDL for $U/t=3.70$, $3.75$, and $3.80$, using least-square fits with quadratic polynomials in $1/L$. The AFQMC simulations are performed at $\beta t=L$.}
\label{fig:fig12FLparamt}
\end{figure}

In the preceding subsection, the critical properties of the Dirac quantum criticality, especially the QCP location $U_c/t$, were determined mainly through the magnetic nature of the transition, i.e., the presence of long-range AFM order for $U>U_c$. Moreover, in terms of the charge degrees of freedom, this criticality also corresponds to a metal-Mott insulator transition, which provides independent perspectives to corroborate the QCP location. The associated physical quantities include the single-particle momentum distribution, charge compressibility, and double occupancy. Furthermore, fidelity susceptibility offers a general diagnostic for quantum phase transitions~\cite{You2007,Venuti2007,Gu2009,Schwandt2009,Albuquerque2010,WangLei2015,HuangLi2016}. In this subsection, we concentrate on these quantities and their corresponding signatures to independently examine our estimated $U_c/t$ for the honeycomb Hubbard model~(\ref{eq:HubbModel}).

We begin with the single-particle momentum distribution, which provides a sensitive probe of metal-Mott insulator transition from the fermionic sector. In the DSM phase of the model~(\ref{eq:HubbModel}), the energy-resolved momentum distribution $n(\varepsilon_{\mathbf{k}})$ (see Sec.~\ref{sec:AFQMCObs}) displays a jump across the Fermi level, which disappears in the AFMI phase. In a finite-size system under PBC, we refer to the size of this jump as the quasiparticle weight $Z_L$, compute via Eq.~(\ref{eq:QpWeight}). Hence, the critical point of the Dirac quantum criticality corresponds to the vanishing of $Z_L$ in the TDL. As illustrated in Fig.~\ref{fig:fig11Qpwght}(a), we first extrapolate the three data points of $n(\varepsilon_{\mathbf{k}})$ closest to $\varepsilon_{\mathbf{k}}=0$ on both sides (with $\varepsilon_{\mathbf{k}}<0$ and $\varepsilon_{\mathbf{k}}>0$) to $\varepsilon_{\mathbf{k}}=0$, and then identify $Z_L$ as the difference between the two extrapolated values. Subsequently, we obtain its TDL value $Z_{\infty}$ via a polynomial fit to $Z_L$, as shown in Fig.~\ref{fig:fig11Qpwght}(b). Notably, $Z_{\infty}$ remains significantly nonzero at $U/t=3.75$ and $3.80$. This seemingly contradicts our estimate $U_c/t=3.731(4)$ since both couplings fall in the AFMI phase where $Z_{\infty}$ should vanish. We attribute this apparent contradiction to strong finite-size effects in $Z_L$, as evidenced by the downward curvature of the fitting curves in Fig.~\ref{fig:fig11Qpwght}(b). Such fits are typically very delicate, as including data points at larger $L$ can substantially alter the extrapolated value. Indeed, at $U/t=3.80$, we observe clear shifts in $Z_{\infty}$, from $Z_{\infty}\simeq 0.234$ to $0.188$ and $0.172$, when the largest $L$ used in the fit increases from $L_{\rm max}=30$ to $36$ and $42$, respectively. We note that, at $U/t=3.80$, our $Z_{\infty}$ value with $L_{\rm max}=36$ agrees with that reported in the previous ground-state AFQMC study~\cite{Otsuka2016} using the same $L_{\rm max}$, within statistical uncertainties. This suggests that the above contradiction is not due to methodological differences but rather to the intrinsic finite-size effects in $Z_L$. The same extrapolation issue should persist at $U/t=3.75$ and $3.70$ from the similar downward curvatures. Therefore, the extrapolated values in Fig.~\ref{fig:fig11Qpwght}(b) are unreliable, and one should be cautious about claiming the QCP location based on these results.

\begin{figure}[t]
\includegraphics[width=0.873\columnwidth]{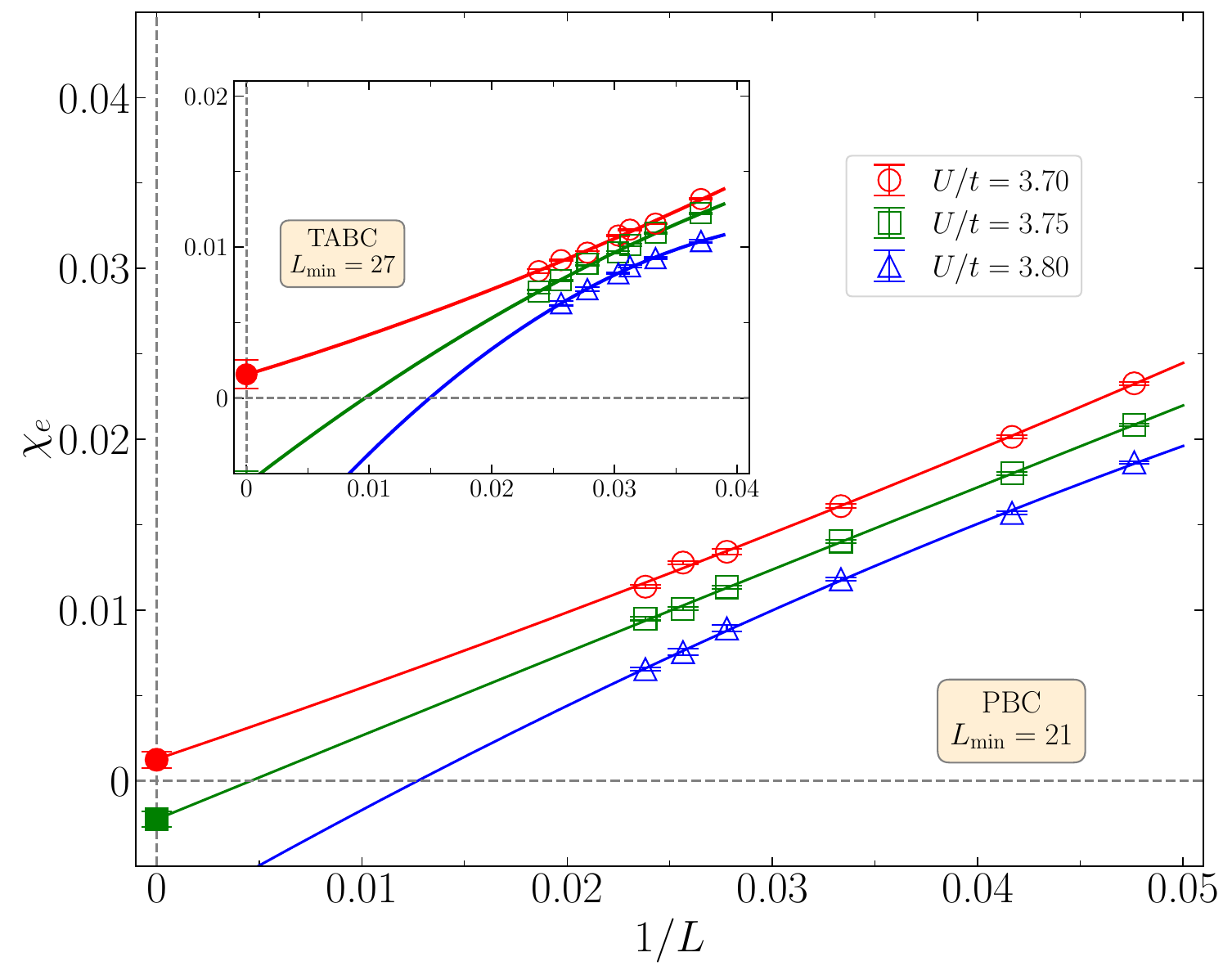}
\caption{
Extrapolation of the charge compressibility $\chi_e$ to the TDL using quadratic polynomial fits in $1/L$ (solid curves). $\chi_e$ is computed by Eq.~(\ref{eq:ChiCharge}). The main panel displays AFQMC data obtained under PBC with $L\ge21$, while the inset shows results under TABC with $L\ge27$. The AFQMC simulations are performed at $\beta t=L$. 
}
\label{fig:fig13chie}
\end{figure}

An alternative way to analyze the $n(\varepsilon_{\mathbf{k}})$ results is to compute its slope with respect to $\varepsilon_{\mathbf{k}}$ at the Fermi level. This slope is quantified by the Fermi-liquid parameter $Q_{\rm FL}$ defined in Eq.~(\ref{eq:FLP}). It is closely related to the quasiparticle weight, since the jump of $n(\varepsilon_{\mathbf{k}})$ in the DSM phase implies an infinite slope and hence $Q_{\rm FL}=0$, whereas in the AFMI phase $Q_{\rm FL}$ remains finite. To accurately evaluate $Q_{\rm FL}$ in a finite-size system, we combine AFQMC data of $n(\varepsilon_{\mathbf{k}})$ under PBC and TBC with at least two sets of specially chosen twist angles, to achieve a relatively uniform and high-resolution momentum grid around $\varepsilon_{\mathbf{k}}=0$. As illustrated in Fig.~\ref{fig:fig12FLparamt}(a), we perform cubic-spline interpolation for the combined data of $n(\varepsilon_{\mathbf{k}})$, extract the slope, and compute $Q_{\rm FL}$ from the fitting function. Then we extrapolate $Q_{\rm FL}$ to the TDL, as shown in Fig.~\ref{fig:fig12FLparamt}(b). Interestingly, $Q_{\rm FL}$ exhibits a much better finite-size scaling behavior than $Z_L$ in Fig.~\ref{fig:fig11Qpwght}(b), although the fitting curves still show a slight downward curvature. Its extrapolated values conform with our estimate $U_c/t=3.731(4)$, as $Q_{\rm FL}(L=\infty)=0$ at $U/t=3.70$ and $Q_{\rm FL}(L=\infty)>0$ at $U/t=3.75$ and $3.80$. Hence, our results suggest that $Q_{\rm FL}$ is more reliable than $Z_L$ in locating the QCP of the Dirac quantum criticality. 

The charge compressibility $\chi_e$ serves as another sensitive probe of the metal-Mott insulator transition. The DSM phase is gapless and compressible, giving rise to a finite $\chi_e$, while $\chi_e=0$ signals the opening of a charge gap and the associated incompressibility in the AFMI phase. In Fig.~\ref{fig:fig13chie}, we present the $\chi_e$ results computed via Eq.~(\ref{eq:ChiCharge}) in AFQMC simulations under both PBC and TABC, along with their extrapolations to the TDL. The two data sets show good consistency, both giving a finite $\chi_e(L=\infty)$ at $U/t=3.70$ and vanishing $\chi_e(L=\infty)$ at $U/t=3.75$ and $3.80$ (the negative intercepts are expected to vanish once the data with larger $L$ are included). Moreover, the almost linear fitting curves near the QCP (e.g., at $U/t=3.70$ and $3.75$) indicate well-behaved finite-size scaling behavior in $\chi_e$, ensuring the reliability of the extrapolated $\chi_e(L=\infty)$. These results are consistent with our estimate $U_c/t=3.731(4)$, demonstrating that $\chi_e$ successfully resolves the QCP. 

\begin{figure}[t]
\includegraphics[width=1.00\columnwidth]{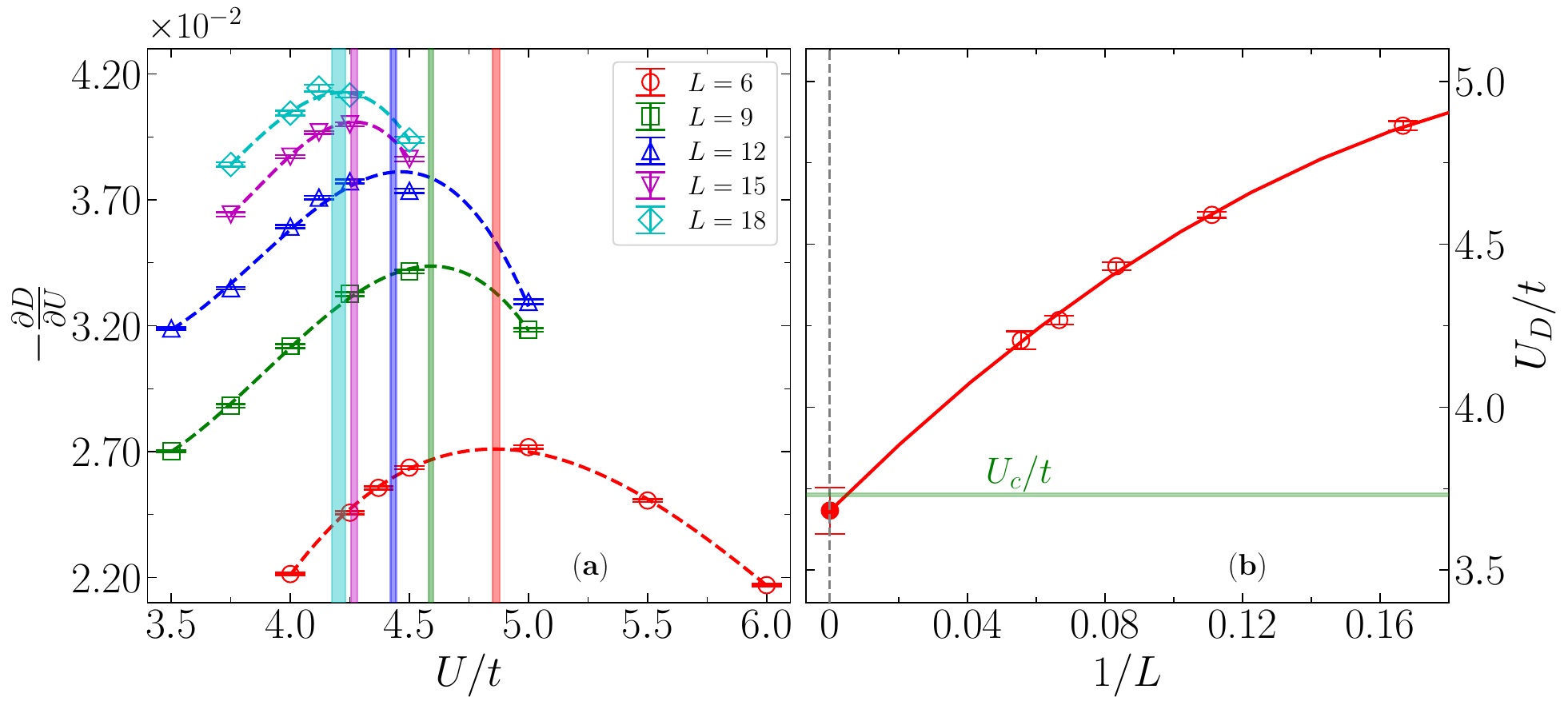}
\caption{
Examination for the QCP location $U_c/t$ via the most rapid suppression of double occupancy $D$. Panel (a) plots $-\partial D/\partial U$, computed by Eq.~(\ref{eq:DouOccDerive}), near the QCP for $L=6,9,12,15,18$. Dashed lines represent cubic-spline fits to the data. Vertical shading bands mark the peak position $U_{\rm D}/t$ of the curves (the most rapid suppression of $D$), with the values included. Panel (b) shows the extrapolation of $U_{\rm D}/t$ to the TDL using a quadratic polynomial fit in $1/L$ (red solid curve), yielding an intercept of $3.68(7)$. The green shading band represents $U_c/t=3.731(4)$ obtained in Fig.~\ref{fig:fig05m2CrF}. The AFQMC simulations are performed at $\beta t=L$ with PBC.
}
\label{fig:fig14dDoudU}
\end{figure}

In the Hubbard model, double occupancy $D$ is a direct and quantitative metric of local charge fluctuations~\cite{Lu2026CPL}. As a result, $D$ decreases monotonically as $U/t$ increases at fixed temperature~\cite{Song2025L,*Song2025B,Lu2026}, reflecting the progressively suppressed charge fluctuations. Across a metal-Mott insulator transition, $D$ typically displays the most rapid suppression at the critical point~\cite{Marcelo1999,Parcollet,Ohashi2008}, which manifests as a peak behavior in the derivative $-\partial{D}/\partial{U}$. We directly compute $-\partial{D}/\partial{U}$ using Eq.~(\ref{eq:DouOccDerive})~\cite{dDdUCompt} for the model~(\ref{eq:HubbModel}) near the QCP, using AFQMC simulations with PBC. As shown in Fig.~\ref{fig:fig14dDoudU}(a), $-\partial{D}/\partial{U}$ indeed displays a prominent peak, and we extract its peak position $U_{\rm D}/t$ from the fitting curve. An evident shift of $U_{\rm D}/t$ towards smaller values and toward the QCP with increasing $L$ can be observed. However, the result $U_{\rm D}/t=4.20(3)$ at $L=18$ remains well above our estimate $U_c/t=3.731(4)$, indicating strong finite-size effects. In Fig.~\ref{fig:fig14dDoudU}(b), we further extrapolate the finite-size values of $U_{\rm D}/t$ to the TDL and reach an intercept of $3.68(7)$, which is roughly consistent with our $U_c/t$ value within uncertainties. However, in AFQMC simulations, obtaining high-precision results of $-\partial{D}/\partial{U}$ [via Eq.~(\ref{eq:DouOccDerive})] in large systems is typically prohibited by the heavy computational cost of imaginary-time dynamics. This severely limits the ability of resolving the QCP through the peak of $-\partial{D}/\partial{U}$. Furthermore, the peak values of $-\partial{D}/\partial{U}$ are quite small and tend to converge as $L$ increases, implying a smooth evolution of $D$ with no discontinuity across the Dirac quantum criticality in the TDL. 

\begin{figure}[t]
\includegraphics[width=0.920\columnwidth]{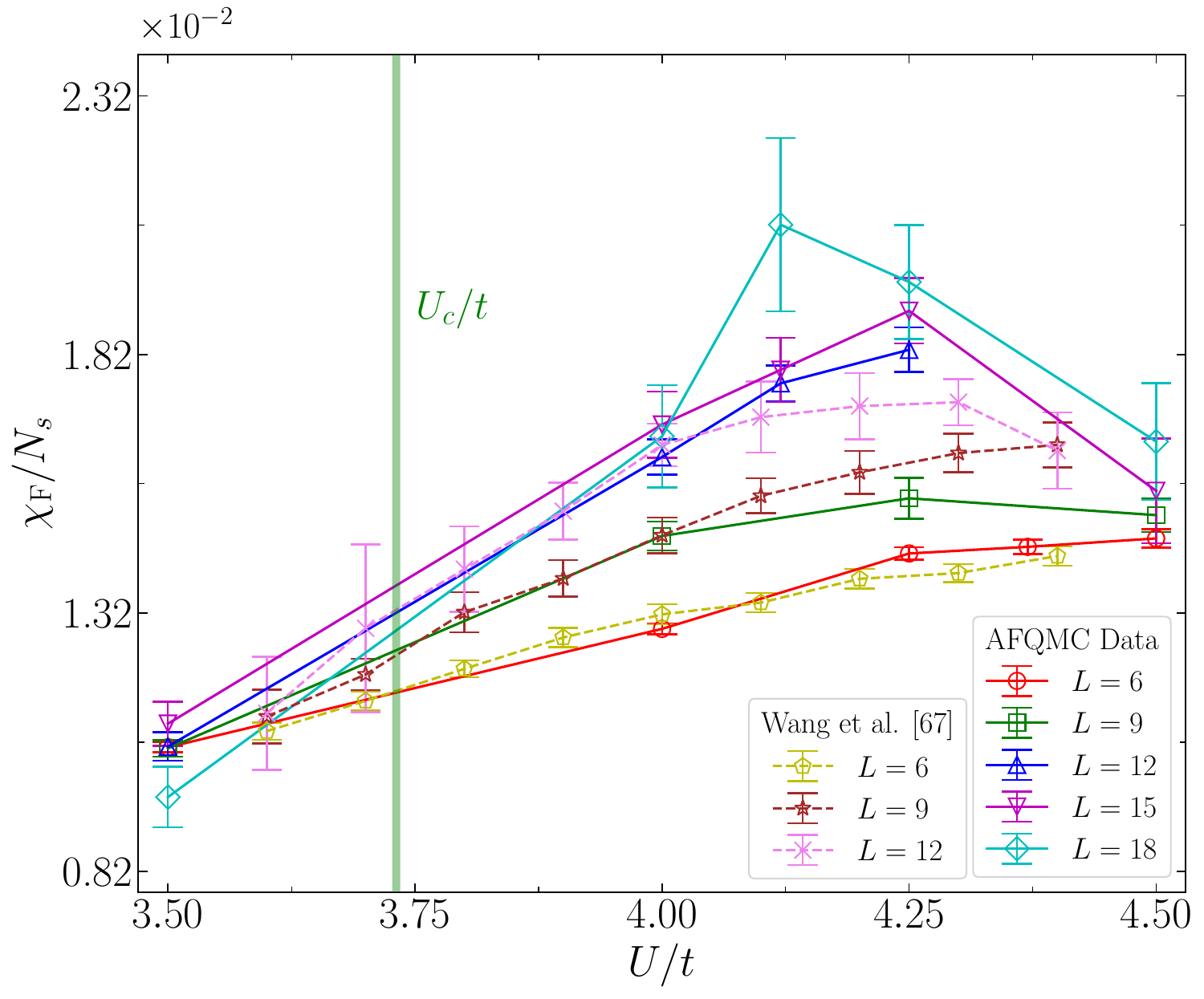}
\caption{
Examination for the QCP location $U_c/t$ via the peak of fidelity susceptibility $\chi_{\mathrm{F}}$, computed by Eq.~(\ref{eq:fidelitysusp}). The results of $\chi_{\mathrm{F}}/N_s$ near the QCP, with $L=6,9,12,15,18$. The data from a previous study, Wang et al.~\cite{WangLei2015}, are also included as benchmark. The vertical green shading band represents $U_c/t=3.731(4)$ obtained in Fig.~\ref{fig:fig05m2CrF}. The AFQMC simulations are performed at $\beta t=L$ with PBC.
}
\label{fig:fig15Fidelity}
\end{figure}

In the fidelity approach~\cite{Paolo2006,Amico2008,Gu2010}, the fidelity, defined as the overlap between ground-state wave functions, develops a dip at the QCP. Its second-order derivative, i.e., the fidelity susceptibility $\chi_{\rm F}$, therefore diverges and exhibits critical scaling behavior~\cite{You2007,Venuti2007,Gu2009,Schwandt2009,Albuquerque2010,WangLei2015,HuangLi2016}, analogous to other relevant observables. This renders $\chi_{\rm F}$ an effective probe for detecting the quantum criticality. For the model~(\ref{eq:HubbModel}), we compute $\chi_{\rm F}$ via Eq.~(\ref{eq:fidelitysusp}), using the same imaginary-time dynamics as for $-\partial{D}/\partial{U}$. In Fig.~\ref{fig:fig15Fidelity}, we present the results of $\chi_{\rm F}/N_s$ near the QCP from AFQMC simulations under PBC with up to $L=18$. Although these finite-size data are rather noisy, they agree within uncertainties with those reported in Ref.~\cite{WangLei2015}, which are included as a reference in the plot. Rather than a sharp peak, we observe only a broad peak in $\chi_{\rm F}/N_s$, located above our estimate $U_c/t$. Such a distinction from previous studies~\cite{Schwandt2009,Albuquerque2010,WangLei2015} may be attributed to the gapless nature of the Dirac quantum criticality, which smears the critical signature in $\chi_{\rm F}$. As $L$ increases, the peak of $\chi_{\rm F}/N_s$ becomes more pronounced and shifts toward smaller interaction strength. Nevertheless, the large error bars in the data prevent a reliable extrapolation of the peak position to the TDL, thereby making it difficult to verify the QCP location $U_c/t$. Further algorithmic improvements are required to achieve precise $\chi_{\rm F}$ results for the model~(\ref{eq:HubbModel}), so that the critical point and even the critical exponents of the Dirac quantum criticality can be determined from this quantity.

From the numerical results in Figs.~\ref{fig:fig11Qpwght}-\ref{fig:fig15Fidelity} and the associated discussions, we conclude that the Fermi-liquid parameter and charge compressibility can resolve the QCP location of the model~(\ref{eq:HubbModel}) in a relatively reliable manner. Our AFQMC simulations can achieve precise results for both quantities in large systems up to $L=42$ using reasonable computational effort. Moreover, these results corroborate our estimate of $U_c/t$ from the perspective of the metal-Mott insulator transition, beyond spin-spin correlations, which has been much less explored in previous studies of the Dirac quantum criticality.

\subsection{Benchmark data of total energy and double occupancy}
\label{sec:EnergyDouble}

\renewcommand{\arraystretch}{1.15}
\begin{table}[t]
	\centering
	\caption{Summary of the total energy per site, denoted as $e/t$, for the model~(\ref{eq:HubbModel}) with various $U/t$ and $L$, obtained from AFQMC simulations under TABC with $\beta t=L$, using the HS-$\hat{n}$ transformation. The extrapolated values at $L=\infty$ represent the ground-state results.}
	\begin{tabular*}{\columnwidth}{@{\extracolsep{\fill}} c c c c c }
		\hline\hline
		\diagbox{$L$}{$U/t$} & 3.65 & 3.70 & 3.75 & 3.80 \\
		\hline
		24 & -0.8336(2) & -0.8259(3) & -0.8188(2) & -0.8112(2) \\
		27 & -0.8338(3) & -0.8263(4) & -0.8189(6) & -0.8117(9) \\
		30 & -0.8332(3) & -0.8264(4) & -0.8195(7) & -0.8105(3) \\
		33 & -0.8339(5) & -0.8256(2) & -0.8192(6) & -0.8112(5) \\
		36 & -0.8334(4) & -0.8266(5) & -0.8184(5) & -0.8122(8) \\
		39 & -0.8340(4) & -0.8258(5) & -0.8191(8) & -0.8108(5) \\
		42 &            & -0.8259(3) & -0.8187(6) &            \\
		\hline
		$\infty$ & -0.8336(4) & -0.8261(5) & -0.8190(5) & -0.8113(8) \\
		\hline\hline
	\end{tabular*}
	\label{table:ETABC}
\end{table}

\renewcommand{\arraystretch}{1.15}
\begin{table}[t]
	\centering
	\caption{Summary of the double occupancy $D$ for the model~(\ref{eq:HubbModel}) with various $U/t$ and $L$, obtained from AFQMC simulations under TABC with $\beta t=L$, using the HS-$\hat{n}$ transformation. The extrapolated values at $L=\infty$ represent the ground-state results.}
	\begin{tabular*}{\columnwidth}{@{\extracolsep{\fill}} c c c c c }
		\hline\hline
		\diagbox{$L$}{$U/t$} & 3.65 & 3.70 & 3.75 & 3.80 \\
		\hline
		24 & 0.15307(4) & 0.15153(5) & 0.14988(4) & 0.14839(5) \\
		27 & 0.15311(5) & 0.15158(8) & 0.15003(7) & 0.14842(8) \\
		30 & 0.15324(6) & 0.15162(5) & 0.15003(7) & 0.14854(6) \\
		33 & 0.15321(9) & 0.15166(5) & 0.15002(9) & 0.14836(9) \\
		36 & 0.15332(7) & 0.15157(9) & 0.15020(6) & 0.14834(9) \\
		39 & 0.15327(6) & 0.15186(6) & 0.15026(8) & 0.14857(6) \\
		42 &            & 0.15164(7) & 0.15009(8) &            \\
		\hline
		$\infty$ & 0.15321(11) & 0.15169(17) & 0.15006(20) & 0.14845(12) \\
		\hline\hline
	\end{tabular*}
	\label{table:DTABC}
\end{table}

In this subsection, we present high-precision AFQMC results for the total energy and double occupancy in the model~(\ref{eq:HubbModel}) near the QCP, which we expect to serve as useful benchmark for future studies.

We define the total energy as $E=\langle\hat{H}_0\rangle+N_sUD$, with $D$ denoting the double occupancy, and $e=E/N_s$ is the total energy per site. We summarize the finite-size data of $e/t$ and $D$, together with their TDL values, in Tables~\ref{table:ETABC} and~\ref{table:DTABC}, respectively, for $U/t=3.65,3.70,3.75,3.80$ with $L\ge 24$. These results are obtained via finite-temperature AFQMC simulations at $\beta t=L$ under TABC using HS-$\hat{n}$ transformation [Eq.~(\ref{eq:DensityHS})]. Upon extrapolation to $L=\infty$, the data correspond to the true ground-state results. The relative errors of all these results are comparable to or below $0.1\%$. Furthermore, both observables show rather weak finite-size effects, as reflected in their consistency within one to two error bars from $L=24$ to $L=42$. For comparison, we also present similar results for $e/t$ and $D$ from AFQMC simulations at $\beta t=L$ under PBC using HS-$\hat{s}^z$ transformation [Eq.~(\ref{eq:SzHS})] in Appendix~\ref{sec:AppEnDouOcc}. Those results achieve even higher precision for both quantities, since the HS-$\hat{s}^z$ suppresses fluctuations in density-related quantities, as discussed in Sec.~\ref{sec:AFQMCmethod}. More importantly, as illustrated in Appendix~\ref{sec:AppEnDouOcc}, the two sets of AFQMC simulations using different HS transformations and boundary conditions yield consistent TDL results for both $e/t$ and $D$ at $U/t=3.75$. This consistency provides further validation of our AFQMC simulations in this work.

\section{Summary and discussion}
\label{sec:Summary}

In summary, we have employed the numerically unbiased finite-temperature AFQMC method to investigate the Dirac quantum criticality in the honeycomb Hubbard model. By setting $\beta t=L$ in the simulations, the ground state and TDL are approached simultaneously for the model. We have resolved the quantum critical properties via combined finite-size scaling analysis for the spin-spin correlations and off-diagonal single-particle Green's function, and have further checked our estimated QCP location using several relevant quantities.

On the methodological side, we have established a highly efficient AFQMC framework for multi-sublattice systems. It incorporates several key ingredients, including the FFT technique for the path propagation, the delayed update algorithm for configuration updates, and the TABC to reduce finite-size effects. This framework enables us to access unprecedented system sizes up to $L=42$ (and $\beta t=42$) for the honeycomb Hubbard model. We have systematically tested the speedup achieved by the FFT technique and the delayed update algorithm relative to the conventional matrix-matrix multiplication and the local update, respectively. We expect this AFQMC framework to be broadly applicable to a wide range of multi-sublattice systems in future studies. Notably, the FFT technique has also been implemented in a very recent work~\cite{Zhang2026} on the Kagome-lattice Hubbard model within the ground-state AFQMC method. 

On the physical aspect, we determine the QCP location $U_c/t=3.731(4)$ and the critical exponents $\nu=1.097(7)$, $\eta_{\phi}=0.72(2)$, $\eta_{\psi}=0.155(6)$ through the finite-size scaling analysis, for the Dirac quantum criticality in the honeycomb Hubbard model. Unlike previous ground-state studies, our results from finite-temperature AFQMC calculations provide an independent benchmark for the chiral Heisenberg Gross-Neveu-Yukawa criticality. We have further examined the QCP location from the perspective of metal-Mott insulator transition and found that Fermi-liquid parameter and charge compressibility reliably corroborate our estimated QCP location. These results, beyond spin-spin correlations, offer a perspective that has rarely been addressed in previous studies of Dirac quantum criticality. Our work also serves as a foundation for the subsequent studies of the finite-temperature properties of the honeycomb Hubbard model, including quantum critical thermodynamics, the semimetal-to-Mott insulator crossover, and the interplay between them. These open opportunities are left for future work.

\begin{figure}[t]
\includegraphics[width=0.963\columnwidth]{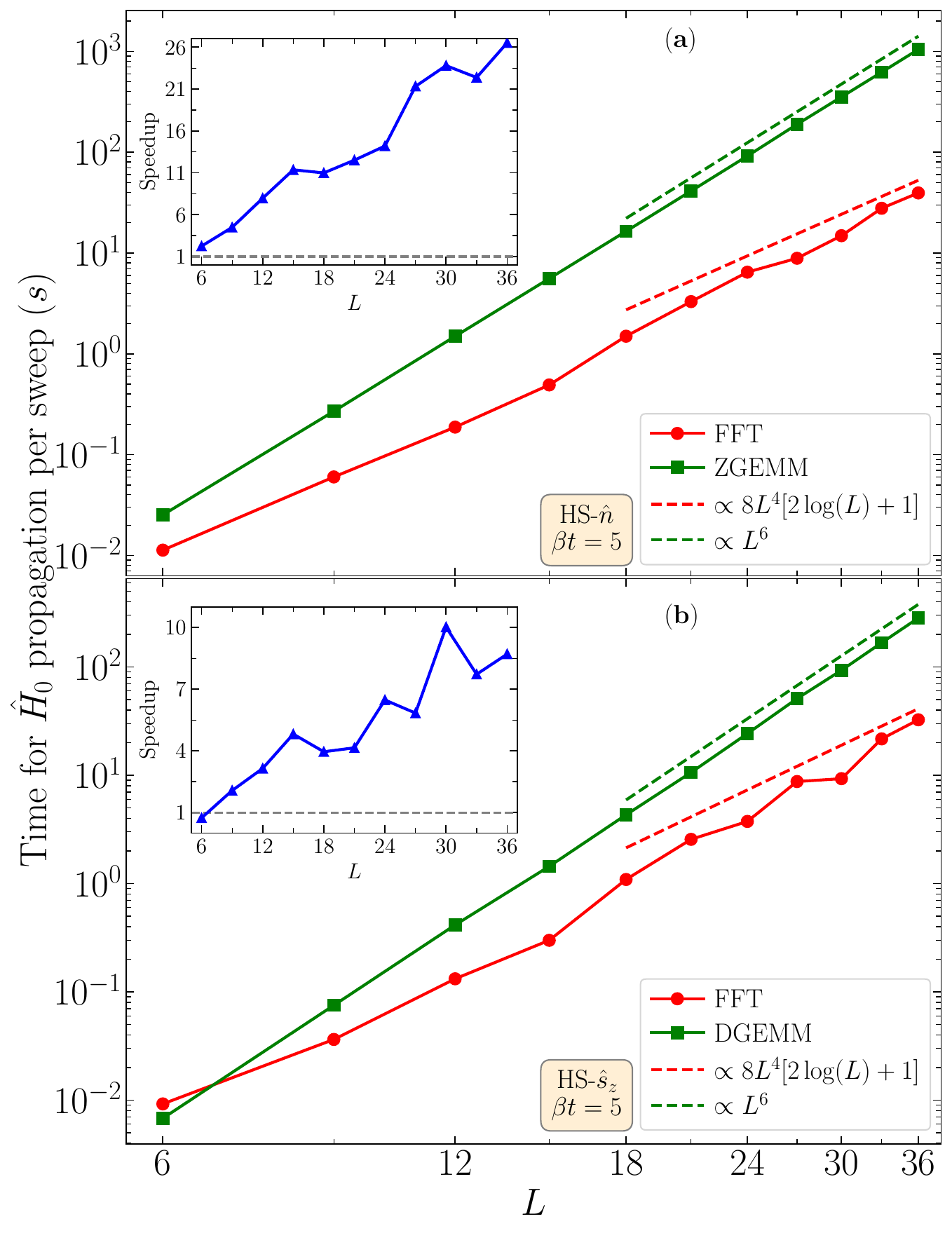}
\caption{
Comparison of the average time for propagating the noninteracting part per sweep (in seconds) between the FFT and ZGEMM/DGEMM methods in AFQMC simulations of the honeycomb Hubbard model with $U/t=1$ and $\beta t=5$. Panels (a) and (b) show the results applying HS-$\hat{n}$ [Eq.~(\ref{eq:DensityHS})] and HS-$\hat{s}^z$ [Eq.~(\ref{eq:SzHS})], respectively. The dashed lines plot the corresponding theoretical scalings. The insets in both panels illustrate the corresponding speedups achieved by the FFT compared to ZGEMM/DGEMM for the propagation.
}
\label{fig:figS1FftZgemm}
\end{figure}

\begin{figure}[t]
\includegraphics[width=0.91\columnwidth]{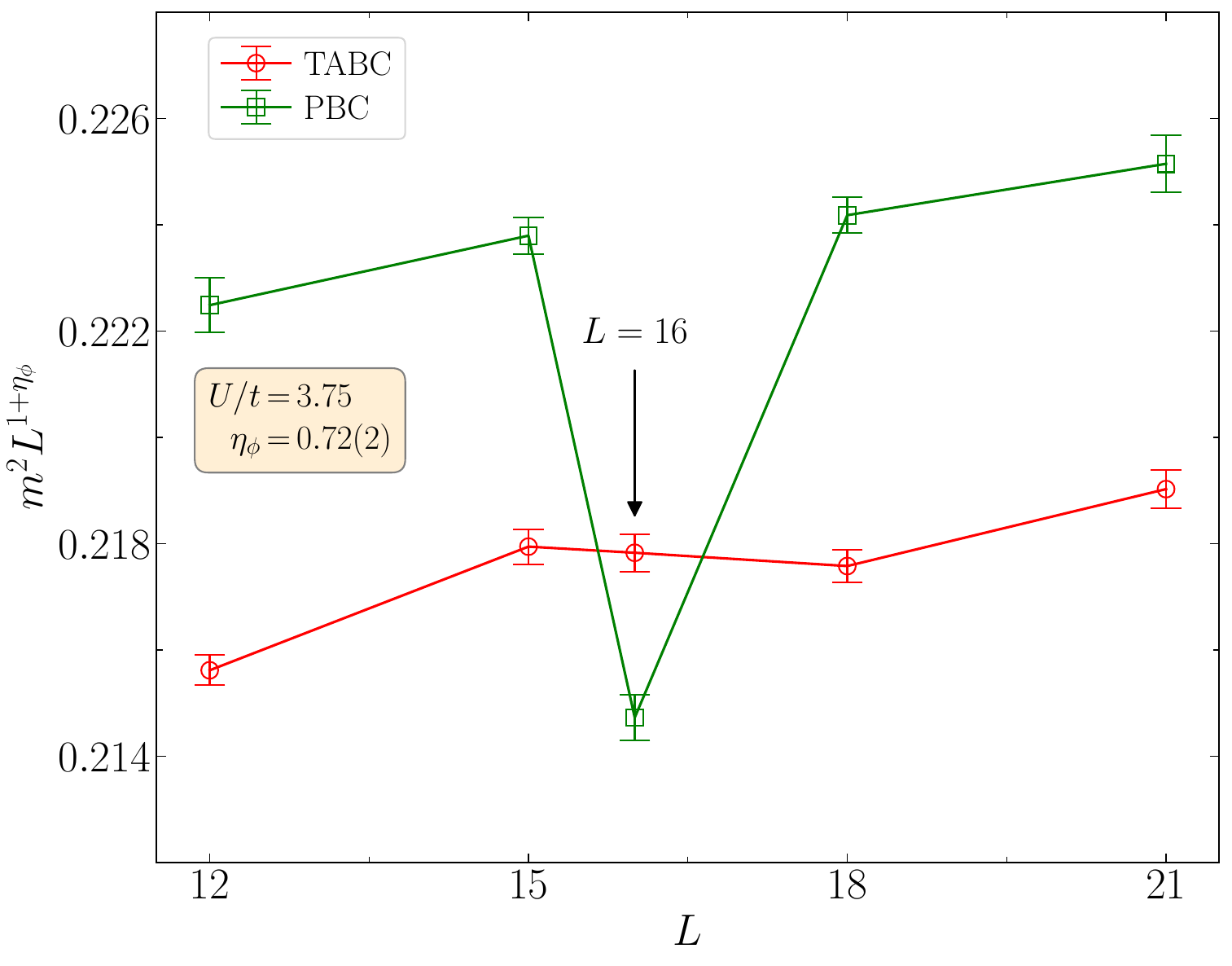}
\caption{
Comparison of the numerical results for the rescaled mean-squared magnetization $m^2L^{1+\eta_{\phi}}$ as a function of $L$ for the model~(\ref{eq:HubbModel}) at $U/t=3.75$, obtained from AFQMC simulations under TABC and PBC (with $\beta t=L$). Here $\eta_{\phi}=0.72$ is adopted. The arrow marks the data point at $L=16$.
}
\label{fig:figS2m2Oscil}
\end{figure}

\begin{acknowledgments}
We thank Xiao Yan Xu for valuable discussions. This work was supported by the National Natural Science Foundation of China (Grants No. 12247103, No. 12204377, and No. 12275263), the Quantum Science and Technology-National Science and Technology Major Project (Grant No. 2021ZD0301900), the Natural Science Foundation of Fujian province of China (Grant No. 2023J02032), and the Youth Innovation Team of Shaanxi Universities.
\end{acknowledgments}

\section*{Data Availability}
The data that support the findings of this article are not publicly available. The data are available from the authors upon reasonable request.

\begin{table*}[t]
	\centering
	\caption{Summary of the total energy per site $e/t$ for the model~(\ref{eq:HubbModel}) with various $U/t$ and $L$, obtained from AFQMC simulations under PBC with $\beta t=L$, using the HS-$\hat{s}^z$ transformation. }
	\renewcommand{\arraystretch}{1.15}
	\begin{tabular*}{\textwidth}{@{\extracolsep{\fill}} c c c c c c c }
		\hline\hline
		\diagbox{$L$}{$U/t$} & 3.50 & 3.75 & 4.00 & 4.12 & 4.25 & 4.50 \\
		\hline
		9  & -0.85397(4) & -0.81593(3) & -0.77998(4) & -0.76345(4) & -0.74621(3) & -0.71456(5) \\
		12 & -0.85604(2) & -0.81793(2) & -0.78179(2) & -0.76520(2) & -0.74792(2) & -0.71605(3) \\
		15 & -0.85681(1) & -0.81855(3) & -0.78241(1) & -0.76578(2) & -0.74837(1) & -0.71646(2) \\
		18 & -0.85711(2) & -0.81882(1) & -0.78266(2) & -0.76599(2) & -0.74857(2) & -0.71659(2) \\
		24 & -0.85739(1) & -0.81906(1) & -0.78277(2) & -0.76614(2) & -0.74864(2) & -0.71663(2) \\
		\hline\hline
	\end{tabular*}
	\label{table:EPBC}
\end{table*}

\renewcommand{\arraystretch}{1.15}
\begin{table*}[t]
	\centering
	\caption{Summary of the double occupancy $D$ for the model~(\ref{eq:HubbModel}) with various $U/t$ and $L$, obtained from AFQMC simulations under PBC with $\beta t=L$, using the HS-$\hat{s}^z$ transformation. }
	\begin{tabular*}{\textwidth}{@{\extracolsep{\fill}} c c c c c c c }
		\hline\hline
		\diagbox{$L$}{$U/t$} & 3.50 & 3.75 & 4.00 & 4.12 & 4.25 & 4.50 \\
		\hline
		9  & 0.156872(6) & 0.149188(9) & 0.140990(9) & 0.136936(9) & 0.132484(9) & 0.123942(9) \\
		12 & 0.157293(5) & 0.149491(5) & 0.141255(9) & 0.137175(8) & 0.132681(8) & 0.124101(9) \\
		15 & 0.157412(3) & 0.149612(6) & 0.141356(4) & 0.137249(8) & 0.132771(6) & 0.124166(7) \\
		18 & 0.157469(4) & 0.149680(6) & 0.141420(6) & 0.137283(6) & 0.132791(7) & 0.124178(6) \\
		24 & 0.157509(3) & 0.149735(4) & 0.141451(6) & 0.137314(5) & 0.132826(5) & 0.124232(5) \\
		\hline\hline
	\end{tabular*}
	\label{table:DPBC}
\end{table*}

\appendix

\section{The speedup by FFT with \texorpdfstring{$\beta t=5$}{}}
\label{sec:AppFFTSpeedUp}

In Sec.~\ref{sec:FFT} of the main text, we have presented the details for implementing the FFT technique in the path propagation in AFQMC simulations, and have shown the speedup by FFT compared to the conventional matrix-matrix multiplication in Fig.~\ref{fig:fig02FFTGEMM}. Those results are for the simulations with $\beta t=L$. In this appendix, we further present similar speedup results at fixed $\beta t=5$. 

In Fig.~\ref{fig:figS1FftZgemm}, we plot the average time per sweep consumed by propagating the noninteracting part, using FFT and matrix-matrix multiplication. The results for both HS-$\hat{n}$ and HS-$\hat{s}^z$ are shown, with the corresponding speedup depicted in the insets. We find that the achieved speedups are close to those obtained in Fig.~\ref{fig:fig02FFTGEMM} for $\beta t=L$ simulations. This illustrates that the acceleration by FFT is independent of the specific parameter $\beta t$ in AFQMC simulations. Moreover, for fixed $\beta t=5$, the consumed time by ZGEMM/DGEMM (matrix-matrix multiplication) conforms well with the theoretical scaling of $O(L^6)$. Similar agreement is also observed for the FFT, which follows the computational scaling of $O[8L^4(2\ln L + 1)]$. The speedup achieved with HS-$\hat{s}^z$ is relatively smaller than that with HS-$\hat{n}$, again due to the real-version AFQMC calculations, as discussed in Sec.~\ref{sec:FFT}. 

\section{Comparison of the TABC and PBC results for \texorpdfstring{$m^2$}{}}
\label{sec:AppTABCPBC}

In Sec.~\ref{sec:TABC} of the main text, we have discussed in detail the implementation of TABC, and mentioned that TABC simulations can eliminate the oscillating results versus $L$. Such results were previously reported in Ref.~\cite{Song2025B}. Here, in this appendix, we show similar results in the honeycomb Hubbard model~(\ref{eq:HubbModel}). 

In Fig.~\ref{fig:figS2m2Oscil}, we plot AFQMC results of $m^2 L^{1+\eta_{\phi}}$ versus $L$ at $U/t=3.75$, from the simulations under both PBC and TABC. Evidently, the PBC results show a significant dip at $L=16$, while the remaining data at $L=12,15,18,21$ follow a consistent trend. This anomaly can be attributed to the accessible momentum points in the Brillouin zone. Under PBC, when $L$ is a multiple of $3$, the Dirac points on the Fermi level are available, and the noninteracting energy spectrum is gapless. Otherwise, the Dirac points are absent, and the corresponding spectrum is gapped. Such a mismatch gives rise to markedly different low-energy properties, and hence to the oscillatory behavior of $m^2 L^{1+\eta_{\phi}}$ (and also $m^2$) with respect to $L$. When TABC is employed, the twist angles are generated using quasi-random numbers, and the averaged results eliminate the shell effect present in the noninteracting energy spectrum. As a result, the TABC results vary smoothly with $L$, as shown in Fig.~\ref{fig:figS2m2Oscil}. In Figs.~\ref{fig:fig07Crossing} and~\ref{fig:fig08Ratio} in the main text, we have indeed incorporated the results of $L=16$ and $L=32$, and observed smooth results of the crossing points versus $1/(2L)$. These results demonstrate the effectiveness of the TABC implementation in the AFQMC simulations for the model~(\ref{eq:HubbModel}). 

\begin{figure}[h!]
\includegraphics[width=0.985\columnwidth]{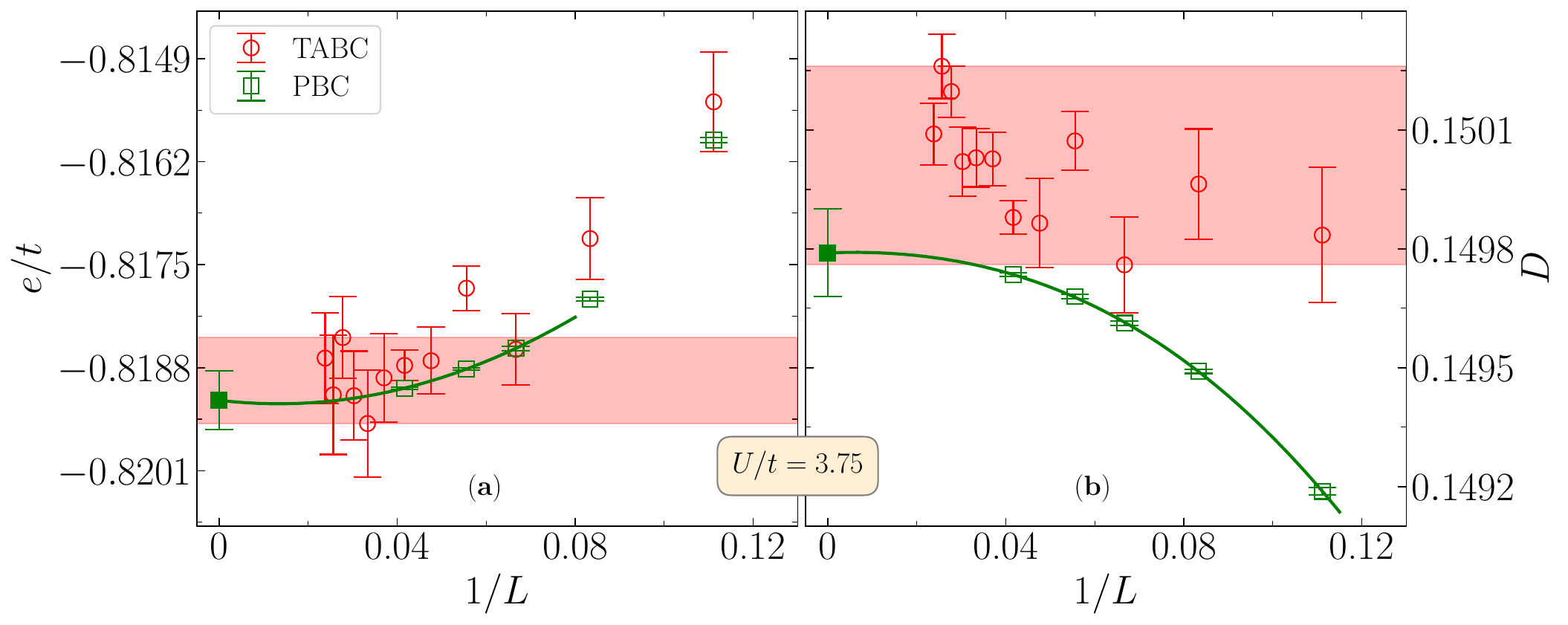}
\caption{Comparison of the total energy per site $e/t$ and double occupancy $D$ obtained from the AFQMC calculations under TABC and PBC with $\beta t=L$, for the model~(\ref{eq:HubbModel}) at $U/t=3.75$. Panels (a) and (b) plot the results for $e/t$ and $D$, respectively. The TABC data are obtained using the HS-$\hat{n}$ transformation, while the PBC data are obtained using the HS-$\hat{s}^z$ transformation. Red shading bands mark the TABC results at TDL, and solid green curves represent polynomial fits of the PBC data.}
\label{fig:figS3compare}
\end{figure}

\section{Benchmark data of total energy and double occupancy under PBC}
\label{sec:AppEnDouOcc}

In Sec.~\ref{sec:EnergyDouble} of the main text, we have presented the benchmark data of total energy per site $e/t$ and double occupancy $D$ for the model~(\ref{eq:HubbModel}), from AFQMC simulations under TABC using the HS-$\hat{n}$ transformation. Here in this appendix, we further summarize the similar data set under PBC using HS-$\hat{s}^z$ transformation, in Tables~\ref{table:EPBC} and~\ref{table:DPBC}. 

As discussed in Sec.~\ref{sec:AFQMCmethod}, applying HS-$\hat{s}^z$ in AFQMC can substantially suppress the statistical fluctuations in density-related quantities, such as the total energy and double occupancy. This is confirmed by the higher precision of both $e/t$ and $D$ results in Tables~\ref{table:EPBC} and~\ref{table:DPBC}, compared with the results in Tables~\ref{table:ETABC} and~\ref{table:DTABC}. At $U/t=3.75$, we carry out a careful comparison between the two data sets, which are plotted in Fig.~\ref{fig:figS3compare}. Because the TABC data exhibit negligible finite-size effects, we simply average the results over $24\le L\le 42$ to obtain the TDL values for both observables. For the PBC data, given their tiny error bars, we perform further extrapolations of the finite-size data to reach the TDL values. As illustrated in Fig.~\ref{fig:figS3compare}, the final ground-state results from the two sets of AFQMC simulations agree within statistical uncertainties for both $e/t$ and $D$.

\bibliography{HoneycombRef.bib}

\end{document}